\documentclass[12pt,draftclsnofoot,onecolumn]{IEEEtran}
\ifCLASSINFOpdf
\else
\fi
\usepackage{amsfonts}
\usepackage{cite}
\usepackage{graphicx}
\usepackage{amssymb}
\usepackage{caption}
\usepackage{epsfig}
\usepackage{subfigure}
\usepackage{amsmath}
\usepackage{epstopdf}
\newtheorem{theorem}{Theorem}
\newtheorem{lemma}{Lemma}

\usepackage{array}
\usepackage{color}
\usepackage{amsmath}
\usepackage{algorithmic}
\usepackage{algorithm}
\usepackage{cases}
\usepackage{mathrsfs}
\usepackage{psfrag}
\usepackage{url}
\usepackage{setspace}

\usepackage{etoolbox}
\makeatletter
\patchcmd{\@makecaption}
  {\scshape}
  {}
  {}
  {}
\makeatletter
\patchcmd{\@makecaption}
  {\\}
  {.\ }
  {}
  {}
\makeatother

\newcommand{\dis}{\stackrel{d}{\sim}}
\newcommand{\eqla}{\stackrel{(a)}{=}}
\newcommand{\eqlb}{\stackrel{(b)}{=}}
\newcommand{\eqlc}{\stackrel{(c)}{=}}
\newcommand{\eqld}{\stackrel{(d)}{=}}
\newcommand{\eqle}{\stackrel{(e)}{=}}
\newcommand{\eqlf}{\stackrel{(f)}{=}}

\newtheorem{Prob}{Problem}

\newtheorem{remark}{Remark}

\IEEEoverridecommandlockouts

\newcommand{\blue}[1]{\textcolor{black}{#1}}

\begin{document}
%
\title{Robust Optimization of Instantaneous Beamforming and Quasi-static Phase Shifts in an IRS-assisted Multi-Cell Network}


\author{Yuhang Jia, Ying Cui, and Wuyang Jiang
\footnote{Y. Jia, Y. Cui are with Shanghai Jiao Tong University, Shanghai, China.
Wuyang Jiang is with Shanghai University of Engineering Science, Shanghai, China. The paper has been submitted in part to 2021 IEEE Global Communications Conference.
This work is under minor revision.
}
}

\maketitle

\begin{abstract}
\blue{The impacts of channel estimation errors, inter-cell interference, phase adjustment cost, and computation cost on an intelligent reflecting surface (IRS)-assisted system are severe in practice but have been ignored for simplicity in most existing works.} In this paper, we \blue{investigate} a multi-antenna base station (BS) serving a single-antenna user with the help of a multi-element \blue{IRS} in a multi-cell network with inter-cell interference. We consider imperfect channel state information (CSI) at the BS\blue{, i.e., imperfect CSIT,} and focus on the robust optimization of the BS's instantaneous CSI-adaptive beamforming and the IRS's quasi-static phase shifts in two scenarios. In the scenario of coding over many slots, we formulate a robust optimization problem to maximize the user's ergodic rate. In the scenario of coding within each slot, we formulate a robust optimization problem to maximize the user's average goodput under \blue{the} successful transmission probability constraints. \blue{The robust optimization problems are challenging two-timescale stochastic non-convex problems.} In both scenarios, we obtain closed-form robust beamforming designs for any given phase shifts and more tractable stochastic non-convex approximate problems only for the phase shifts. Besides, we propose an iterative algorithm to obtain a Karush-Kuhn-Tucker (KKT) point of each of the stochastic problems for the phase shifts. It is worth noting that the proposed methods offer closed-form robust \blue{instantaneous CSI-adaptive} beamforming designs which can promptly adapt to \blue{rapid CSI} changes \blue{over slots} and robust quasi-static phase shift designs of low computation and phase adjustment costs in the presence of imperfect CSIT and inter-cell interference. Numerical results further demonstrate the notable gains of the proposed robust joint designs over existing \blue{ones and reveal the practical values of the proposed solutions}.
\end{abstract}


%
\IEEEpeerreviewmaketitle

\section{Introduction}
With the deployment of the fifth-generation (5G) wireless network, the urgent requirement for network capacity is gradually being achieved. Nevertheless, the increasingly demanding need for energy efficiency remains unaddressed. Recently, intelligent reflecting surface (IRS), consisting of nearly passive, low-cost, reflecting elements with reconfigurable parameters, has been envisioned to serve as a promising solution for improving spectrum and energy efficiency and has received more and more attention \cite{QingqingWu1,EBasar},\blue{\cite{R7,SJin1}}. 
Specifically, the IRS's phase shifts can be determined by a smart controller attached to the IRS. A nearby base station (BS) can communicate to the IRS controller to configure the IRS's phase shifts, based on its knowledge of channel characteristics.

Most existing works consider an IRS-assisted system with one multi-antenna BS serving one or multiple users with the help of one multi-element IRS and optimize BS beamforming and \blue{IRS} phase shifts \cite{HGuo1,HShen1,XYu,CHuang,8855810,QingqingWu2,SJin2,hu2020statistical,YuhangJia,zhao2019intelligent,CGuo,QTao}. In \cite{HGuo1,HShen1,XYu,CHuang,8855810,QingqingWu2}, the authors formulate the maximization of the weighted sum rate \cite{HGuo1}, secrecy rate \cite{HShen1,XYu}, and energy efficiency \cite{CHuang,8855810} and the minimization of the transmit power \cite{QingqingWu2} as \blue{(deterministic)} non-convex problems and propose iterative algorithms to obtain locally optimal solutions or nearly optimal solutions \blue{using block-wise coordinate descent (BCD) \cite{HGuo1}, successive convex approximation (SCA) \cite{CHuang}, conjugate gradient \cite{8855810}, and semidefinite relaxation \cite{QingqingWu2}, etc.} Notice that in \cite{HGuo1,HShen1,XYu,CHuang,8855810,QingqingWu2}, the \blue{BS's} beamformer and \blue{IRS's} phase shifts \blue{both} adapt to instantaneous channel state information (CSI). In \cite{SJin2,hu2020statistical,zhao2019intelligent,YuhangJia,CGuo,QTao}, the authors formulate the maximization of the ergodic rate \cite{SJin2,hu2020statistical,YuhangJia} and average rate \cite{zhao2019intelligent} and the minimization of the outage probability \cite{CGuo,QTao} as \blue{stochastic} non-convex problems. \blue{The problems in \cite{SJin2,hu2020statistical,YuhangJia,QTao} are first converted to deterministic non-convex problems using Jensen's inequality and then solved, whereas the problems in \cite{zhao2019intelligent,CGuo} are directly tackled.} Specifically, the authors in \cite{SJin2,hu2020statistical,CGuo,QTao} obtain closed-form optimal phase shifts, and the authors in \cite{YuhangJia,zhao2019intelligent} propose iterative algorithms to obtain locally optimal solutions or nearly optimal solutions \blue{using BCD \cite{YuhangJia} and stochastic successive convex approximation (SSCA) \cite{zhao2019intelligent}}. In \cite{SJin2,hu2020statistical,zhao2019intelligent,YuhangJia,CGuo,QTao}, the BS beamformer adapts to instantaneous CSI as in \cite{HGuo1,HShen1,XYu,CHuang,8855810,QingqingWu2}, while the \blue{IRS's} phase shifts are adaptive to the statistics of CSI, which remains unchanged over several slots \cite{SJin2,hu2020statistical,YuhangJia,zhao2019intelligent,CGuo,QTao}, \blue{contrary to} \cite{HGuo1,HShen1,XYu,CHuang,8855810,QingqingWu2}. We refer to the phase shift designs in \cite{HGuo1,HShen1,XYu,CHuang,8855810,QingqingWu2} and \cite{SJin2,hu2020statistical,YuhangJia,zhao2019intelligent,CGuo,QTao} as instantaneous CSI-adaptive phase shift designs and quasi-static phase shift designs, respectively. Compared with an instantaneous CSI-adaptive phase shift design, a quasi-static phase shift design yields a low phase adjustment cost at the sacrifice of some performance. Considering the practical implementation issue, a quasi-static phase shift design may be more valuable.

\blue{Robust beamforming designs under imperfect CSI at the transmitter (CSIT) have been widely studied for conventional systems without IRSs \cite{R8,R9}. Nevertheless, very few works investigate robust designs for IRS-assisted systems where channel estimation is more challenging and estimation errors are more inevitable.} The authors in \cite{xu2020resource,yu2020robust,hong2020robust,ZhouGui,JWang,TALe,deng2020robust,Czhong1} investigate the robust optimization of \blue{instantaneous CSI-adaptive} beamformers and phase shifts to maximize the worst-case average sum rate \cite{xu2020resource} and secrecy rate \cite{yu2020robust} and to minimize the transmit power \cite{hong2020robust,ZhouGui,JWang,TALe,deng2020robust} and the average mean square error (MSE) \cite{Czhong1}, under imperfect instantaneous CSI at the BS, i.e., imperfect instantaneous CSIT. The resulting non-convex robust optimization problems are generally more challenging than their counterparts with perfect instantaneous CSIT. Rather than directly tackling the challenging robust optimization problems, the authors in \cite{xu2020resource,yu2020robust,hong2020robust,ZhouGui,JWang,TALe,deng2020robust,Czhong1} consider their simplified versions and propose iterative algorithms to obtain locally optimal solutions or Karush-Kuhn-Tucker (KKT) points. The convergence speeds of the iterative algorithms \blue{with} imperfect instantaneous CSIT in \cite{xu2020resource,yu2020robust,hong2020robust,ZhouGui,JWang,TALe,deng2020robust,Czhong1} are lower than those \blue{with} perfect instantaneous CSIT in \cite{HGuo1,HShen1,XYu,CHuang,8855810,QingqingWu2,SJin2,hu2020statistical,YuhangJia,zhao2019intelligent,CGuo,QTao}. Thus, they may not \blue{provide} effective robust \blue{instantaneous CSI-adaptive} beamforming designs before CSI changes, by noting that the channel coherence time is in the order of milliseconds. Furthermore, the robust \blue{instantaneous CSI-adaptive} phase shift designs in \cite{xu2020resource,yu2020robust,hong2020robust,ZhouGui,JWang,TALe,deng2020robust,Czhong1} have higher phase adjustment costs and are less practical. Therefore, it is critical to obtain robust instantaneous CSI-adaptive beamforming designs with highly efficient methods and robust quasi-static phase shift designs with low phase adjustment costs.

Furthermore, notice that most existing works, including the abovementioned ones under perfect CSIT \cite{HGuo1,HShen1,XYu,CHuang,8855810,QingqingWu2,SJin2,hu2020statistical,zhao2019intelligent,CGuo,QTao} and imperfect CSIT \cite{xu2020resource,yu2020robust,hong2020robust,ZhouGui,JWang,TALe,deng2020robust,Czhong1}, consider single-cell networks and ignore interference from other BSs. However, in practice, inter-cell interference usually has a severe impact, especially for dense networks or cell-edge users. It is thus critical to take into account the influence of interference when designing practical IRS-assisted systems. In \cite{YongjunXu,CPan1,WangQun,QingqingWu3,YuanweiLiu,jia2020reconfigurable,YuhangJia}, the authors propose instantaneous CSI-adaptive beamforming designs and instantaneous CSI-adaptive \cite{YongjunXu,CPan1,WangQun,QingqingWu3,YuanweiLiu,jia2020reconfigurable} or quasi-static\blue{\cite{YuhangJia}} phase shift designs for IRS-assisted multi-cell networks with inter-cell interference. As \cite{YongjunXu,CPan1,WangQun,QingqingWu3,YuanweiLiu,jia2020reconfigurable,YuhangJia} assume perfect instantaneous CSIT, the proposed solutions for multi-cell networks are not robust against channel estimation errors. Thus, it is highly desirable to obtain robust beamforming and phase shift design for IRS-assisted multi-cell networks with inter-cell interference.

In this paper, we shall address the above issues. Specifically, we consider a multi-antenna BS serving a single-antenna user with the help of a multi-element IRS in a multi-cell network with inter-cell interference. The antennas at the BS and the reflecting elements at the IRS are arranged in uniform rectangular arrays (URAs). The indirect signal and interference links passing the IRS are modeled with Rician fading, whereas the direct signal and interference links follow Rayleigh fading. We assume that the \blue{line-of-sight} (LoS) components do not change \blue{over slots}, and the \blue{non-line-of-sight} (NLoS) components vary from slots to slots. Furthermore, we suppose that the BS has imperfect CSI, \blue{i.e., imperfect CSIT}. We focus on the robust optimization of instantaneous CSI-adaptive beamforming and quasi-static phase shifts for the IRS-assisted network with imperfect CSIT and inter-cell interference in two scenarios, \blue{which is more challenging than the robust instantaneous CSI-adaptive joint design optimization for each slot in
\cite{HGuo1,HShen1,XYu,CHuang,8855810,QingqingWu2,SJin2,hu2020statistical,zhao2019intelligent,CGuo,QTao}.}
\begin{itemize}
\item Firstly, we consider coding over a large number of slots. In this scenario, we formulate a robust optimization problem to maximize the ergodic rate of the user. The problem is a very challenging two-timescale stochastic non-convex problem since the beamforming design and phase shift design are in different \blue{timescales} and the number of random variables involved is prohibitively huge. Based on \blue{Jensen's inequality and} the analysis of the expectations of the received signal power and interference power, we obtain a closed-form \blue{instantaneous CSI-adaptive} beamforming design for any given phase shifts and a more tractable stochastic non-convex approximate problem only for the phase shifts.

\item Secondly, we consider coding within each slot and adopt transmission rate adaptation over slots. In this scenario, we formulate a robust optimization problem to maximize the average goodput of the user under \blue{the} successful transmission probability constraints. The problem is a more challenging two-timescale stochastic non-convex optimization problem than the one in the first scenario, due to infinitely many constraints and an extra \blue{rate adaptation} function to be optimized. By constructing a deterministic channel error set and using the Bernstein-type inequality \blue{and total probability theorem}, we obtain two closed-form \blue{instantaneous CSI-adaptive} beamforming \blue{and rate adaptation} designs for any given phase shifts and two more tractable stochastic non-convex approximate problems only for the phase shifts.

\item Thirdly, using SSCA, we propose an iterative algorithm to obtain a KKT point of each of the three approximate stochastic problems for the phase shifts. The proposed algorithm has low computational complexity, as it solves a sequence of approximate convex problems analytically \blue{by carefully analyzing the KKT conditions}.
\end{itemize}

In sum, we propose one method for the robust maximization of \blue{the} ergodic rate and two methods for the robust maximization of \blue{the} average goodput. The proposed methods \blue{for both scenarios} offer closed-form robust \blue{instantaneous CSI-adaptive} beamforming designs which can promptly adapt to \blue{rapid CSI changes over slots} and robust quasi-static phase shift designs of low computation and phase adjustment costs in the presence of imperfect CSIT and inter-cell interference. Besides, the two proposed robust joint designs in the scenario of average goodput maximization have different preferable system parameters and complement each other to provide a higher average goodput. Numerical results further demonstrate notable gains of the proposed robust joint designs over existing instantaneous CSI-adaptive beamforming and quasi-static phase shift designs.
\subsection*{\textbf{Notation}}
We represent vectors by boldface lowercase letters (e.g., $\mathbf{x}$), matrices by boldface uppercase letters (e.g., $\mathbf{X}$), scalar constants by non-boldface letters (e.g., $x$), and sets by calligraphic letters (e.g., $\mathcal{X}$).  The notation $X(i,j)$ denotes the $(i,j)$-th element of matrix $\mathbf{X}$, and $x_i$ represents the $i$-th element of vector $\mathbf{x}$. $\mathbf{X}^H$, $\mathbf{X}^T$\blue{, and} $\text{tr}\left(\mathbf{X}\right)$ denote the conjugate transpose, transpose, \blue{and}  trace of a matrix, respectively. rvec$(\mathbf{X})$ and vec$(\mathbf{X})$ denote the row and column vectorization of a matrix, respectively. $\left\lVert\cdot\right\rVert_2$ denotes the Euclidean norm of a vector. diag$\left(\blue{\mathbf{x}}\right)$ is a diagonal matrix with the entries of $\mathbf{x}$ on its main diagonal. $Re\left\{\cdot\right\}$, \blue{$\angle(\cdot)$}, and $\lvert \cdot \rvert$ denote the real part, \blue{phase}, and modulus of a complex number, respectively. $\text{Pr}[x] \in [0,1]$ represents the probability of the event $x$. The complex field and real field are denoted by $\mathbb{C}$ and $\mathbb{R}$, respectively. \blue{$\mathbf{I}_{N}$ represents the $N\times N$ identity matrix. $x \dis y$ represents that $x$ and $y$ follow the same distribution.}
\section{System Model}\label{sec:system}
\begin{figure}[t]
\begin{center}
 \includegraphics[width=6cm]{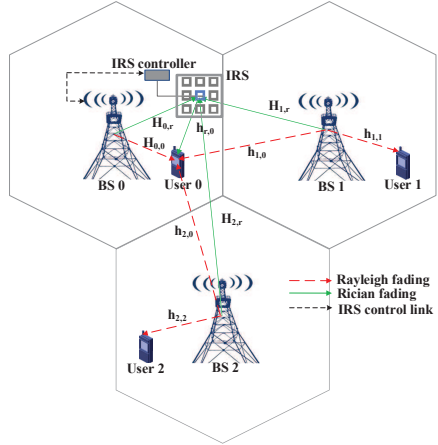}
  \end{center}
 \vspace*{-3mm}
     \caption{\small{System Model.}}
\label{fig:system model}
\vspace*{-3mm}
\end{figure}

As shown in Fig. \ref{fig:system model}, one multi-antenna BS, i.e., BS $0$, serves one single-antenna user, i.e., user $0$, with the help of one multi-element IRS in its cell, in the presence of $K$ interference BSs, i.e., BS $1$, $...\blue{,}$ BS $K$. Denote $\mathcal{K} \triangleq \{0,1,...,K\}$. For all $k \in \mathcal{K}\backslash\{0\}$, BS $k$ has one user, i.e., user $k$. Suppose that for all $k \in \mathcal{K}\backslash\{0\}$, the IRS is far from either BS $k$ or user $k$. Thus, each BS $k \in \mathcal{K}\backslash\{0\}$ serves user $k$, ignoring the effect of \blue{the} IRS. We do not consider cooperation or coordination among the $K+1$ BSs. \blue{We consider a time period consisting of $S$ slots (coherence blocks) during which} the locations of the BSs and IRS are fixed, and the users are \blue{almost} static.\blue{\footnote{\blue{The time period is on the minute time-scale. One slot is on the milliseconds time-scale. Thus, $S$ is roughly $\frac{60}{0.001}=6\times 10^4$.}}} Each BS $k\in\mathcal{K}$ is equipped with a URA of $M_k\times N_k$ antennas, and the IRS is equipped with a URA of $M_r\times N_r$ reflecting elements. For notation simplicity, define $\mathcal M_k \triangleq \{1,2,...,M_k\}$, $\mathcal N_k \triangleq \{1,2,...,N_k\}$, $\mathcal M_r \triangleq \{1,2,...,M_r\},$ and $\mathcal N_r \triangleq \{1,2,...,N_r\}$, where $k \in \mathcal{K}$. The phase shifts of the IRS's reflecting elements can be determined by a smart controller attached to the IRS. BS $0$ communicates to the IRS controller to configure the IRS's phase shifts so that the IRS can assist its communication to user $0$.

We consider a narrow-band system and adopt the block-fading model for small-scale fading. As scattering is often rich near the ground, we adopt the Rayleigh fading model for the small-scale fading channels between the BSs and the users\blue{\cite{SJin2,YuhangJia}}. Let $\mathbf{h}^H_{k,j} \in \mathbb{C}^{ 1 \times M_kN_k}$ denote the random channel vector for the channel between BS $k\in\mathcal{K}$ and user $j\in\mathcal{K}$ in each slot. Specifically,
\begin{align*}
\mathbf{h}^H_{k,j}=\sqrt{{\alpha_{k,j}}}\tilde{\mathbf{h}}^H_{k,j},\
\quad k \in \mathcal{K}, j \in \mathcal{K},
\end{align*}
where $\alpha_{k,j}>0$ represents the large-scale fading power, and the elements of $\tilde{\mathbf{h}}^H_{k,j}$ are \blue{i.i.d.} according to $C\mathcal N(0,1)$. 
As scattering is much weaker far from the ground, we adopt the Rician fading model for the small-scale fading channels between the BSs and the IRS and the small-scale fading channel between the IRS and user $0$ \blue{\cite{SJin2,YuhangJia,zhao2019intelligent}}. Let $\mathbf{H}_{k,r} \in \mathbb{C}^{M_rN_r \times M_kN_k}$ and $\mathbf{h}^{H}_{r,0} \in \mathbb{C}^{1 \times M_rN_r}$ denote the channel matrices for the channel between BS $k\in\mathcal{K}$ and the IRS and the channel between the IRS and user $0$, respectively, in each slot. \blue{Specifically, we have \cite{SJin2,YuhangJia,zhao2019intelligent}:}
\begin{align*}
\mathbf{H}_{k,r}=&\sqrt{\alpha_{k,r}} \left(\sqrt{\frac{K_{k,r}}{K_{k,r}+1}} \bar{\mathbf{H}}_{k,r}+ \sqrt{\frac{1}{K_{k,r}+1}} \tilde{\mathbf{H}}_{k,r}\!\right), \quad k\in\mathcal{K},\\
\mathbf{h}^{H}_{r,0} = & \sqrt{\alpha_{r,0}} \left(\sqrt{\frac{K_{r,0}}{K_{r,0}+1}} \bar{\mathbf{h}}^{H}_{r,0} + \sqrt{\frac{1}{K_{r,0}+1}} \tilde{\mathbf{h}}^{H}_{r,0}\right),
\end{align*}
where $\alpha_{k,r}$, $\alpha_{r,0}>0$ represent the large-scale fading powers, $K_{k,r}$, $K_{r,0}\geq 0$ denote the Rician factors,\footnote{If $K_{k,r}=0$ or $K_{r,0}=0$, the corresponding Rician fading reduces down to Rayleigh fading. If $K_{k,r} \rightarrow \infty$ or $K_{r,0} \rightarrow \infty$, only the LoS components exist.} \blue{$\bar{\mathbf{H}}_{k,r} \in \mathbb{C}^{M_rN_r \times M_kN_k}$, $\bar{\mathbf{h}}^{H}_{r,0}\in \mathbb{C}^{1\times M_rN_r}$ represent the normalized LoS components with unit-modulus elements, and $\tilde{\mathbf{H}}_{k,r} \in \mathbb{C}^{M_rN_r \times M_kN_k}$, $\tilde{\mathbf{h}}^{H}_{r,0} \in \mathbb{C}^{1 \times M_rN_r}$ represent the random normalized NLoS components in each slot with elements i.i.d. according to $C\mathcal N(0,1)$.} \blue{We suppose that the channel model, i.e., all large-scale fading powers, Rician factors, LoS components, and distributions of all random NLoS components for each slot, remain unchanged during the considered time period.}

Let $\lambda$ and $d$ $(\leq \frac{\lambda}{2})$ denote the wavelength of transmit signals and the distance between adjacent elements or antennas  in each row and each column of the URAs. Define:
\begin{align*}
 f(\theta^{(h)},\theta^{(v)},m,n)\triangleq  & 2\pi\frac{d}{\lambda}\sin\theta^{(v)}((m-1)\cos\theta^{(h)}+(n-1)\sin\theta^{(h)}), 
 \\
 \mathbf{A}_{m,n}(\theta^{(h)},\theta^{(v)},M,N) \triangleq & \left(e^{jf(\theta^{(h)},\theta^{(v)},m,n)}\right)_{m=1,...,M,n=1,...,N}    \in \mathbb{C}^{M \times N},\\
 \mathbf{a}(\theta^{(h)},\theta^{(v)},M,N) \triangleq & \text{rvec}\left( \mathbf{A}_{m,n}(\theta^{(h)},\theta^{(v)},M,N) \right) \in \mathbb{C}^{1\times MN}.
 \end{align*}
Note that $f(\theta^{(h)},\theta^{(v)},m,n)$ represents the difference of the corresponding phase change over the LoS component.
Then, $\bar{\mathbf{H}}_{k,r}$
and $\bar{\mathbf{h}}^{H}_{r,0}$ are modeled as \cite{IEEEexample:van2004optimum}:
\begin{align*}
\bar{\mathbf{H}}_{k,r} = & \mathbf{a}^H(\delta^{(h)}_{k,r},\delta^{(v)}_{k,r},M_r,N_r) \mathbf{a}(\varphi^{(h)}_{k,r},\varphi^{(v)}_{k,r},M_k,N_k),\
\quad k \in \mathcal{K}, \\ \bar{\mathbf{h}}_{r,0}^H = &  \mathbf{a}(\varphi^{(h)}_{r,0},\varphi^{(v)}_{r,0},M_r,N_r),
\end{align*}
where $\delta^{(h)}_{k,r}$ $\left(\delta^{(v)}_{k,r}\right)$ represents the azimuth (elevation) angle between the direction of a row (column) of the URA at the IRS and the projection of the signal from  BS $k$ to the IRS on the plane of the URA at the IRS; $\varphi^{(h)}_{k,r}$ $\left(\varphi^{(v)}_{k,r}\right)$ represents the azimuth (elevation) angle between the direction of a row (column) of the URA at BS $k$ and the projection of the signal from BS $k$ to the IRS on the plane of the URA at BS $k$; $\varphi^{(h)}_{r,0}$ $\left(\varphi^{(v)}_{r,0}\right)$ represents the azimuth (elevation) angle between the direction of a row (column) of the URA at the IRS and the projection of the signal from the IRS to user $0$ on the plane of the URA at the IRS.

\blue{We consider a quasi-static phase shift design where the phase shifts do not change with the fast varying NLoS components to reduce phase adjustment cost.}
Let $\phi_{m,n} \in [0,2\pi)$ denote the phase shift of the $(m,n)$-th element of the IRS. For notation convenience, we introduce
$\mathbf{v}\triangleq\text{vec}\left(\left(e^{j\phi_{m,n}}\right)_{m \in \mathcal M_r, n \in \mathcal N_r}\right)\in \mathbb{C}^{M_rN_r\times 1}$
to represent the phase shifts. Denote $\mathcal{N}\triangleq \{1,...,M_rN_r\}$. $\mathbf{v}$ can also be expressed as $\mathbf{v}=(v_n)_{n \in \mathcal{N}}$, where $v_n\in\mathbb{C}$ satisfies:
\begin{align}
\lvert v_n \rvert= 1,\ n \in \mathcal{N}. \label{eq:phi}
\end{align}
Then, the channel of the indirect link between BS $k\in\mathcal{K}$ and user $0$ via the IRS is given by:
\begin{align}
\mathbf{h}_{r,0}^H\text{diag}\left(\mathbf{v}\right)
\mathbf{H}_{k,r}=\mathbf{v}^H\mathbf{G}_{k,0},\quad k \in \mathcal{K},
\end{align}
where $\mathbf{G}_{k,0}\triangleq \text{diag}\left(\mathbf{h}_{r,0}^H\right)\mathbf{H}_{k,r} \in \mathbb{C}^{M_rN_r\times M_kN_k}$ is referred to as the cascaded channel between BS $k\in\mathcal{K}$ and user $0$. We express the channel of each indirect link in terms of the corresponding cascaded channel, for ease of analysis. Accordingly, the LoS components of the cascaded channel are given by:
\begin{align}
\bar{\mathbf{G}}_{k,0}\triangleq \sqrt{\alpha_{k,r}\alpha_{r,0}\tau_{k}}\text{ diag}
\left(\bar{\mathbf{h}}_{r,0}^H\right)\bar{\mathbf{H}}_{k,r},\quad k \in \mathcal{K},
\end{align}
where $$\tau_{k} \triangleq \frac{K_{k,r}K_{r,0}}{(K_{k,r}+1)(K_{r,0}+1)}.$$
Considering both the indirect and direct links, the equivalent channel between BS $k\in\mathcal{K}$ and user $0$ in the IRS-assisted system is expressed as  $\mathbf{h}_{k,0}^H+\mathbf{G}_{k,0}^H\mathbf{v}$.

For all $k\in\mathcal{K}$, we consider linear beamforming at BS $k$ for serving user $k$. Let $\mathbf{w}_k \in \mathbb{C}^{M_kN_k \times 1}$ denote the corresponding normalized beamforming vector, where ${\left\lVert \mathbf{w}_k \right\rVert}^2_2=1$.
Thus, the signal received at user $0$ is expressed as:
\begin{align}
Y_0 \triangleq & \sqrt{P_0}\left(\mathbf{v}^H\mathbf{G}_{0,0} + \mathbf{h} ^H_{0,0}\right) \mathbf{w}_0 X_0 + \sum_{k \in \mathcal{K}\backslash \{0\} }\sqrt{P_k}\left(\mathbf{v}^H\mathbf{G}_{k,0} + \mathbf{h}^H_{k,0}\right) \mathbf{w}_{k} X_{k} + Z,
\end{align}
where $P_k$ denotes the transmit power of BS $k\in\mathcal{K}$, $X_k\in\mathbb{C}$ is the information symbol for user $k\in\mathcal{K}$, with $\mathbb{E}\left[{\lvert X_k\rvert}^2\right] = 1$, and $Z \sim C\mathcal{N}(0,\sigma^2)$ is the additive white Gaussian noise (AWGN).

In this paper, we focus on downlink transmission. We assume that BS $0$ has perfect knowledge
of all large-scale fading powers, Rician factors, LoS components, and distributions of all random NLoS components, \blue{as they change relatively slowly and can be estimated with high accuracy}. Furthermore, BS $0$ estimates the cascaded channel $G_{0,0}$ and direct channel $h_{0,0}$ \blue{in each slot with certain estimation errors, as instantaneous NLoS components change from slots to slots and cannot be estimated very accurately with a very limited number of
pilot symbols \cite{hong2020robust,ZhouGui,Czhong1}.}
Let $\hat{\mathbf{G}}_{0,0} \in \mathbb{C}^{M_rN_r\times M_0N_0}$ and $\hat{\mathbf{h}}_{0,0}\in\mathbb{C}^{M_0N_0\times 1}$ denote the estimated imperfect CSI for $\mathbf{G}_{0,0}$ and $\mathbf{h}_{0,0}$, respectively, and let $\Delta{\mathbf{G}}_{0,0} \in\mathbb{C}^{M_rN_r\times M_0N_0}$ and $\Delta{\mathbf{h}}_{0,0} \in\mathbb{C}^{M_0N_0\times 1}$ denote the corresponding estimation errors. Thus, we have:
\begin{align}\label{eq:observedG}
\mathbf{G}_{0,0}=\hat{\mathbf{G}}_{0,0}+\Delta{\mathbf{G}}_{0,0},\quad
\mathbf{h}_{0,0}=\hat{\mathbf{h}}_{0,0}+\Delta{\mathbf{h}}_{0,0}.
\end{align}
As in \cite{hong2020robust,ZhouGui,Czhong1}, we adopt the statistical CSI error model and assume that all elements of $\Delta{\mathbf{G}}_{0,0}$ and $\Delta{\mathbf{h}}_{0,0}$ are i.i.d. according to $C\mathcal N(0,\delta_{1}^2)$ and $C\mathcal N(0,\delta_{2}^2)$, respectively. \blue{Note that it has been shown that widely used  estimation methods yield zero-mean complex Gaussian distributed estimation errors \cite{DMishra,Que}.}
Thus, the elements of $\hat{\mathbf{G}}_{0,0}$ and $\hat{\mathbf{h}}_{0,0}$ are independent, the $(m,n)$-th element of $\hat{\mathbf{G}}_{0,0}$, denoted by $\hat{G}_{0,0}(m,n)$, follows $C\mathcal{N}\left(\bar{G}_{0,0}(m,n),1-\delta_1^2\right)$, and the $n$-th element of $\hat{\mathbf{h}}_{0,0}$ follows $C\mathcal{N}\left(0,1-\delta_2^2\right)$. Assume that these distributions are known to BS $0$. Denote $\hat{\mathbf{H}}_0 \triangleq \left[\hat{\mathbf{h}}_{0,0},
\hat{\mathbf{G}}_{0,0}\right]\in\mathbb{C}^{M_0N_0\times (M_rN_r+1)}$ and $\Delta{\mathbf{H}}_0\triangleq \left[\Delta{\mathbf{h}}_{0,0}, \Delta{\mathbf{G}}_{0,0}\right]\in\mathbb{C}^{M_0N_0\times (M_rN_r+1)}$. Assume that in each slot user $0$ perfectly estimates the effective channel $\left(\mathbf{v}^H\mathbf{G}_{k,0}+\mathbf{h}_{k,0}^H\right)\mathbf{w}_0 \in \mathbb{C}$ and does not know  $\left(\mathbf{v}^H\mathbf{G}_{k,0}+\mathbf{h}_{k,0}^H\right)\mathbf{w}_0, k \in \mathcal{K}\backslash\{0\}$.
For all $k \in \mathcal{K}\backslash\{0\}$, assume that in each slot BS $k$ perfectly estimates $\mathbf{h}_{k,k}$.

We consider instantaneous CSI-adaptive beamforming design at BS $0$. Let
$\mathbf{w}_{0}\left(\hat{\mathbf{H}}_0\right) \in \mathbb{C}^{M_0N_0\times 1}$ denote the normalized  beamformer for BS $0$ given the  imperfectly estimated CSI $\hat{\mathbf{H}}_0$, where
\begin{align}
\left\lVert\mathbf{w}_0(\hat{\mathbf{H}}_0)\right\rVert^2_2=1.\label{eq:w}
\end{align}
We can view $\mathbf{w}_0:\mathbb{C}^{M_0N_0\times(M_rN_r+1)}\rightarrow
\mathbb{C}^{M_0N_0\times 1}$ as a vector-valued beamforming function from imperfect CSI to a beamforming vector for BS $0$.
Recall that the IRS is far from either BS $k$ or user $k$, for all $k \in \mathcal{K}\backslash\{0\}$. Thus, for all $k \in \mathcal{K}\backslash\{0\}$, to enhance the signal received at user $k$, we consider the instantaneous  CSI-adaptive maximum ratio transmission (MRT) at BS $k$ in each slot, i.e., $\frac{{\mathbf{h}}_{k,k}}{\left\lVert {\mathbf{h}}_{k,k} \right\rVert_2}$, which relies on the perfectly estimated CSI $\mathbf{h}_{k,k}$ at BS $k$.

\begin{figure}[t]
\begin{center}
\includegraphics[width=13cm]{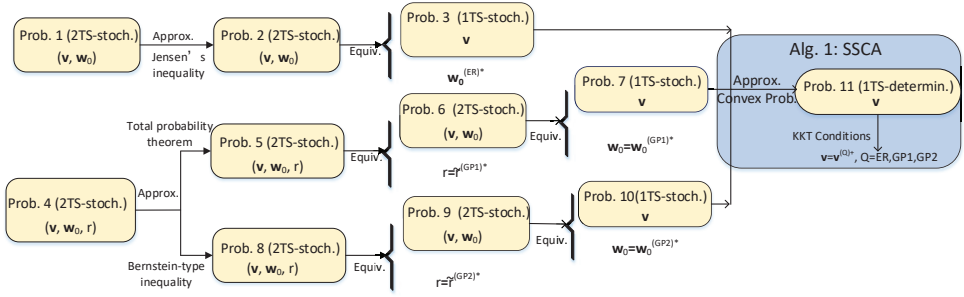}
\end{center}
\vspace{-3mm}
\caption{\small{\blue{Illustration of proposed solution framework. ``1TS" and ``2TS" are short for ``single-timescale" and ``two-timescale", respectively. ``stoch." and ``determin." are short for ``stochastic" and ``deterministic", respectively. ``Approx", ``Equiv.", and ``Prob." represent ``Approximation", ``Equivalent", and ``Problem", respectively.}}}
\label{fig:transform}
\vspace{-4mm}
\end{figure}

In Section~\ref{sec:averagerate} and Section~\ref{sec:Non-ergodicrate}, we consider robust optimization of instantaneous CSI-adaptive beamforming and quasi-static phase shifts for the IRS-assisted multi-cell network with imperfect CSIT and inter-cell interference in two scenarios. We obtain closed-form \blue{robust instantaneous CSI-adaptive} beamforming design\blue{s} for any given phase shifts and more tractable stochastic non-convex approximate problems only for the phase shifts. In Section~\ref{alg:SSCA}, we propose a low-complexity algorithm, conducted at BS $0$\blue{,} to obtain \blue{robust quasi-static phase shifts which correspond to KKT points of the approximate problems.}\blue{\footnote{\blue{When the channel model is not analytically tractable, artificial intelligence techniques are effective for robust design based on massive channel samples.}}} \blue{The proposed solution framework is illustrated in Fig.~\ref{fig:transform}.}
\section{Maximization of Ergodic Rate}\label{sec:averagerate}
In this section, we consider coding over a large number of slots. First, we formulate a robust optimization problem to maximize the ergodic rate of user $0$, which is a challenging two-timescale stochastic non-convex problem. Then, we obtain a closed-form \blue{robust instantaneous CSI-adaptive} beamforming design for any given phase shifts and a more tractable single-timescale stochastic non-convex approximate problem only for the \blue{robust quasi-static} phase shifts.
\subsection{Problem Formulation}
We consider coding over a large number of slots. \blue{Then,} the ergodic rate \blue{of user $0$}, $C^{(ER)}\left(\mathbf{v},\mathbf{w}_0\right)$ (bit/s/Hz), is given by:\footnote{For all $k \in\mathcal{K}\backslash\{0\}$, by treating $\left(\mathbf{v}^H\mathbf{G}_{k,0} + \mathbf{h}^H_{k,0}\right) \frac{{\mathbf{h}}_{k,k}}{\left\lVert {\mathbf{h}}_{k,k} \right\rVert_2} X_{k} \sim C\mathcal{N}\left(0,\mathbb{E}\left[\left\lvert\left(\mathbf{v}^H\mathbf{G}_{k,0} +\mathbf{h}^H_{k,0}\right) \frac{{\mathbf{h}}_{k,k}}{\left\lVert {\mathbf{h}}_{k,k} \right\rVert_2}\right\rvert^2\right]\right)$, which corresponds to the worst-case noise, $C^{(ER)}\left(\mathbf{v},\mathbf{w}_0\right)$ in \eqref{eq:sinr} can be achieved.}
\begin{align}
C^{(ER)}(\mathbf{v},\mathbf{w}_0)=\mathbb{E}\left[\log_2\left(1+
\frac{{P_0}\left\lvert\left(\mathbf{v}^H\mathbf{G}_{0,0}+ \mathbf{h}^H_{0,0}\right)\mathbf{w}_{0}\left(\hat{\mathbf{H}}_0\right)  \right\rvert^2}{\sum\limits_{k \in \mathcal{K}\backslash\{0\}} {P_{k}}\mathbb{E} \left[ {\left\lvert \left(\mathbf{v}^H\mathbf{G}_{k,0} + \mathbf{h}^H_{k,0}\right)  \frac{{\mathbf{h}}_{k,k}}{\left\lVert {\mathbf{h}}_{k,k} \right\rVert_2}\right\rvert}^2 \right]+ {\sigma}^2}\right)\right].\label{eq:sinr}
\end{align}
We aim to maximize  $C^{(ER)}\left(\mathbf{v},\mathbf{w}_0\right)$ by optimizing the phase shifts $\mathbf{v}$ and beamforming function $\mathbf{w}_0$ subject to the phase shift constraints in \eqref{eq:phi} and the normalized beamforming constraints in \eqref{eq:w}.
\begin{Prob}[Robust Ergodic Rate Maximization]\label{prob:eq_delta}
\begin{align*}
\begin{split}
\mathop{\max}_{\mathbf{v},\mathbf{w}_0}\quad &
C^{(ER)}(\mathbf{v},\mathbf{w}_0) \\
s.t. \quad
& \eqref{eq:phi},\quad \eqref{eq:w}.
\end{split}
\end{align*}
\end{Prob}
\begin{remark}[Robust Design in Problem~\ref{prob:eq_delta}]
An optimal solution of Problem~\ref{prob:eq_delta} adapts to the variances of the Gaussian estimation errors $\delta_1^2$ and $\delta_2^2$ and hence is robust against CSI estimation errors.
\end{remark}
\begin{remark}[Challenge for Solving Problem~\ref{prob:eq_delta}]
As the objective function does not have an analytical expression, Problem~\ref{prob:eq_delta} has to be treated as a stochastic optimization problem. As $\mathbf{v}$ is constant, $\mathbf{w}_0(\hat{\mathbf{H}}_0)$ adapts to $\hat{\mathbf{H}}_0$, and $C^{(ER)}(\mathbf{v},\mathbf{w}_0)$ is non-convex in $\left(\mathbf{v},\mathbf{w}_0\right)$, Problem~\ref{prob:eq_delta} is actually a two-timescale stochastic non-convex optimization problem. Moreover, the number of random variables in Problem~\ref{prob:eq_delta} is $2M_0N_0(M_rN_r+1)$, which is usually quite large. Therefore, Problem~\ref{prob:eq_delta} is very challenging.
\end{remark}
\begin{table}
\centering
\begin{tabular}{|c|c|c|c|}
\hline
Problem & Problem~\ref{prob:eq_delta} & Problem~\ref{prob:approximation} & Problem~\ref{prob:theta}  \\
\hline
Random Variables  & $\hat{\mathbf{H}}_0, \Delta{\mathbf{H}}_0$ & $\hat{\mathbf{H}}_0$ & $\hat{\mathbf{H}}_0$ \\
\hline
Optimization Variables & $\left(\mathbf{v},\mathbf{w}_0\right)$ & $\left(\mathbf{v},\mathbf{w}_0\right)$ & $\mathbf{v}$
\\
\hline
Constraints  & \eqref{eq:phi}, \eqref{eq:w}& \eqref{eq:phi}, \eqref{eq:w} & \eqref{eq:phi}
\\
\hline
Timescale& Two-timescale & Two-timescale & Single-timescale \\
\hline
\end{tabular}
\caption{Comparisons of the problems in the ergodic rate maximization.}
\label{tab:prob}
\vspace{-3mm}
\end{table}
\subsection{Closed-form Beamforming and Approximate Phase Shift Optimization }\label{sec:erB}
First, we simplify the objective function of Problem~\ref{prob:eq_delta}.
As in \cite{SJin2,YuhangJia}, we can obtain an upper bound of $C^{(ER)}(\mathbf{v},\mathbf{w}_0)$ using Jensen's inequality \blue{and channel statistics} \cite{cover1991elements}.
\begin{theorem}[Upper Bound of $C^{(ER)}(\mathbf{v},\mathbf{w}_0)$]\label{lem:ergodic Case 1 with reflector}
\begin{align*}
 &C^{(ER)}(\mathbf{v},\mathbf{w}_0) \leq  \log_2\left(1+
 \frac{P_0\mathbb{E}\left[{g}_0
 ^{(ER)}\left(\mathbf{v},\mathbf{w}_0
\left(\hat{\mathbf{H}}_0\right),\hat{\mathbf{H}}_0\right)\right]}
{\sum\limits_{k \in \mathcal{K}\backslash \{0\}}P_k{g}_{k}\left(\mathbf{v}\right)+\sigma^2}\right),
\end{align*}
where ${g}_0^{(ER)}\left(\mathbf{v},\mathbf{w}_0\left(\hat{\mathbf{H}}_0\right),\hat{\mathbf{H}}_0\right)$ and ${g}_{k}\left(\mathbf{v}\right)$ are given by:
\begin{align}
&g_0^{(ER)}\left(\mathbf{v},\mathbf{w}_0\left(\hat{\mathbf{H}}_0\right),\hat{\mathbf{H}}_0\right)
\triangleq
\left\lvert\left(\mathbf{v}^H\hat{\mathbf{G}}_{0,0}+ \hat{\mathbf{h}}^H_{0,0}\right)\mathbf{w}_0
\left(\hat{\mathbf{H}}_0\right)\right\rvert^2 +
\delta_{2}^2+M_rN_r\delta_1^2,\label{eq:h_sphi}\\
 & g_{k}\left(\mathbf{v}\right)\triangleq
\blue{\frac{1}{M_kN_k}\left\lVert\mathbf{v}^H
\bar{\mathbf{G}}_{k,0}
\right\rVert_2^2} +\alpha_{k,r}\alpha_{r,0}M_rN_r\left(1-\tau_{k}\right)
+\alpha_{k,0},\quad k \in \mathcal{K}\backslash\{0\}.\label{eq:h_kphi}
\end{align}
\end{theorem}\begin{IEEEproof}
Please refer to Appendix A.
\end{IEEEproof}

As the upper bound in Theorem~\ref{lem:ergodic Case 1 with reflector} is a good approximation of $C^{(ER)}(\mathbf{v},\mathbf{w}_0)$, which will be seen in Fig.~\ref{fig:ergodic}, we can consider the maximization of the upper bound instead of Problem~\ref{prob:eq_delta} \cite{YuhangJia,CPan1}. $\log_2(\cdot)$ is an increasing function, so the optimization is equivalent to the following problem, which is simpler than Problem~\ref{prob:eq_delta}, as shown in Table~\ref{tab:prob}.
\begin{Prob}[Approximate Problem of Problem \ref{prob:eq_delta}]\label{prob:approximation}
\begin{equation*}
\begin{split}
\mathop{\max}_{\mathbf{v},\mathbf{w}_0}\quad & \frac{P_0\mathbb{E}\left[{g}_0^{(ER)}\left(\mathbf{v},\mathbf{w}_0
\left(\hat{\mathbf{H}}_0\right),\hat{\mathbf{H}}_0\right)\right]}
{\sum\limits_{k \in \mathcal{K}\backslash \{0\}}P_k{g}_{k}\left(\mathbf{v}\right)+\sigma^2} \\
s.t. \quad
& \eqref{eq:phi},\quad \eqref{eq:w}.
\end{split}
\end{equation*}
\end{Prob}

Next, \blue{using Cauchy-Schwartz inequality and the structural property of Problem~\ref{prob:approximation},} we obtain a closed-form \blue{robust} beamforming design for any given phase shifts and equivalently transform Problem~\ref{prob:approximation}, a two-timescale non-convex problem, to a single-timescale stochastic non-convex problem only for the phase shifts, as shown in Table~\ref{tab:prob}.
\begin{Prob}[Equivalent Problem of Problem~\ref{prob:approximation}]\label{prob:theta}
\begin{align*}
\begin{split}
\mathop{\max}_{\mathbf{v}}\quad &
 \frac{P_0\mathbb{E}_{\hat{\mathbf{H}}_0}\left[
 {g}_0^{(ER)}\left(\mathbf{v},\frac{\left(\mathbf{v}^H\hat{\mathbf{G}}_{0,0}+ \hat{\mathbf{h}}^H_{0,0}\right)^H}{\left\lVert\mathbf{v}^H\hat{\mathbf{G}}_{0,0}+ \hat{\mathbf{h}}^H_{0,0}\right\rVert_2},\hat{\mathbf{H}}_0\right)\right]}
{\sum\limits_{k \in \mathcal{K}\backslash \{0\}}P_k{g}_{k}\left(\mathbf{v}\right)+\sigma^2}
 \\
s.t. \quad
& \eqref{eq:phi}.
\end{split}
\end{align*}
\end{Prob}
\begin{theorem}[Equivalence between Problem~\ref{prob:approximation} and Problem~\ref{prob:theta}]\label{theorem:eq}
If $\mathbf{v}^{(ER)*}$ is an optimal solution of Problem~\ref{prob:theta}, then
$\left(\mathbf{v}^{(ER)*},\mathbf{w}_0^{(ER)*}\right)$ is an optimal solution of  Problem~\ref{prob:approximation}, where
\begin{align} \mathbf{w}_0^*\left(\hat{\mathbf{H}}_0\right)=\frac{\left(\left(\mathbf{v}^{(ER)*}\right)^H\hat{\mathbf{G}}_{0,0}+ \hat{\mathbf{h}}^H_{0,0}\right)^H}{\left\lVert\left(\mathbf{v}^{(ER)*}\right)^H\hat{\mathbf{G}}_{0,0}+ \hat{\mathbf{h}}^H_{0,0}\right\rVert_2},\quad
\hat{\mathbf{H}}_0 \in \mathbb{C}^{M_0N_0\times (M_rN_r+1)}. \label{eq:wequivalence}
\end{align}
\end{theorem}
\begin{IEEEproof}
Please refer to Appendix B.
\end{IEEEproof}

\blue{The} closed-form robust \blue{instantaneous CSI-adaptive} beamforming design \blue{(for given $\mathbf{v}^{(ER)*}$)} in \eqref{eq:wequivalence} \blue{has computational complexity $\mathcal{O}(M_0N_0M_rN_r)$ and hence} can promptly adapt to \blue{rapid CSI} changes \blue{over slots}. In Section~\ref{alg:SSCA}, we will obtain a robust quasi-static phase shift design by solving Problem~\ref{prob:theta}.
\section{Maximization of Average Goodput}\label{sec:Non-ergodicrate}
In this section, we consider coding within each slot and adopt transmission rate adaptation over slots. First, we formulate a robust optimization problem to maximize the average goodput of user $0$, which is a \blue{more} challenging two-timescale stochastic non-convex problem. Then, we obtain two closed-form robust \blue{instantaneous CSI-adaptive} beamforming \blue{and rate adaptation} designs for any given phase shifts and two more tractable single-timescale stochastic non-convex approximate problems only for the \blue{robust quasi-static} phase shifts. Later in Section~\ref{sec:simulation}, we will see that the two \blue{resulting} robust joint designs have average goodput advantages at different system parameters and well complement each other.
\subsection{Problem Formulation}
We consider coding within one slot. In a slot with imperfect CSIT $\hat{\mathbf{H}}_0$ and actual CSI $\mathbf{H}_0=\hat{\mathbf{H}}_0+\Delta{\mathbf{H}}_0$, the channel capacity \blue{of} user $0$, $C\left(\mathbf{v},
\mathbf{w}_0\left(
\hat{\mathbf{H}}_0\right),
\hat{\mathbf{H}}_0,\Delta{\mathbf{H}}_0\right)$, is given by:
\begin{align*}
C\left(\mathbf{v},
\mathbf{w}_0\left(
\hat{\mathbf{H}}_0\right),
\hat{\mathbf{H}}_0,\Delta{\mathbf{H}}_0\right)=
\log_2\left(1+
\frac{{P_0}\left\lvert\left(\mathbf{v}^H\mathbf{G}_{0,0}+ \mathbf{h}^H_{0,0}\right)\mathbf{w}_{0}\left(\hat{\mathbf{H}}_0\right)  \right\rvert^2}{\sum\limits_{k \in \mathcal{K}\backslash\{0\}} {P_{k}}\mathbb{E} \left[ {\left\lvert \left(\mathbf{v}^H\mathbf{G}_{k,0} + \mathbf{h}^H_{k,0}\right)  \frac{{\mathbf{h}}_{k,k}}{\left\lVert {\mathbf{h}}_{k,k} \right\rVert_2}\right\rvert}^2 \right]+ {\sigma}^2}\right).
\end{align*}
As $\Delta\mathbf{H}_0$ is unknown, $C\left(\mathbf{v},
\mathbf{w}_0\left(
\hat{\mathbf{H}}_0\right),
\hat{\mathbf{H}}_0,\Delta{\mathbf{H}}_0\right)$ is unknown to BS $0$. Thus, we adopt transmission rate adaptation according to $\hat{\mathbf{H}}_0$. Let $r\left(\hat{\mathbf{H}}_0\right)\in\mathbb{R}$ denote the transmission rate of user $0$ at $\hat{\mathbf{H}}_0$. We can view $r:\mathbb{C}
^{M_0N_0\times(M_rN_r+1)}\rightarrow \mathbb{R}$ as a rate adaption function from imperfect CSI to a transmission rate for user $0$. The transmission from BS $0$ to user $0$ is successful if $C\left(\mathbf{v},
\mathbf{w}_0\left(
\hat{\mathbf{H}}_0\right),
\hat{\mathbf{H}}_0,\Delta{\mathbf{H}}_0\right)\geq r\left(\hat{\mathbf{H}}_0\right)$. We consider the following successful transmission probability constraints:
\begin{equation}\label{eq:proconstraints}
\text{Pr}\left[C\left(\mathbf{v},
\mathbf{w}_0\left(
\hat{\mathbf{H}}_0\right),\hat{\mathbf{H}}_0,\Delta{\mathbf{H}}_0\right)
\geq r\left(\hat{\mathbf{H}}_0\right) \right]\geq \rho,\quad \hat{\mathbf{H}}_0 \in \mathbb{C}^{M_0N_0\times(M_rN_r+1)},
\end{equation}
where $\rho \in (0,1)$ is the successful transmission probability requirement. The average goodput of user $0$ in a slot with $\hat{\mathbf{H}}_0$ is expressed as $\text{Pr}\left[C\left(\mathbf{v},
\mathbf{w}_0\left(
\hat{\mathbf{H}}_0\right),\hat{\mathbf{H}}_0,\Delta{\mathbf{H}}_0\right)
\geq r\left(\hat{\mathbf{H}}_0\right) \right] r\left(\hat{\mathbf{H}}_0\right)$, which is greater than or equal to $\rho r\left(\hat{\mathbf{H}}_0\right)$ under \eqref{eq:proconstraints}. Then, the average goodput of user $0$ over slots, denoted by $C^{(GP)}\left(\mathbf{v},\mathbf{w}_0,
r\right)$, can be defined as:
\begin{equation}\label{eq:averagegoodput}
C^{(GP)}\left(\mathbf{v},\mathbf{w}_0,
r\right)=\rho
\mathbb{E}\left[r\left(
\hat{\mathbf{H}}_0\right)\right].
\end{equation}
We aim to maximize the average goodput of user $0$, $C^{(GP)}\left(\mathbf{v},\mathbf{w}_0,r\right)$, by optimizing the phase shifts $\mathbf{v}$, beamforming function $\mathbf{w}_0$, and rate adaption function $r$ subject to the phase shift constraints in \eqref{eq:phi}, the normalized  beamforming constraints in \eqref{eq:w}, and the successful transmission probability constraints in \eqref{eq:proconstraints}.
\begin{Prob}[Robust Average Goodput Maximization]\label{prob:eq_sta_ergodic}
\begin{align*}
\begin{split}
C^{(GP)*}\triangleq\mathop{\max}_{\mathbf{v},\mathbf{w}_0,
r} \quad &
C^{(GP)}\left(\mathbf{v},\mathbf{w}_0,
r\right)\\
s.t. \quad
& \eqref{eq:phi}, \quad \eqref{eq:w}, \quad \eqref{eq:proconstraints}.
\end{split}
\end{align*}
\end{Prob}
\begin{remark}[Robust Design in Problem~\ref{prob:eq_sta_ergodic}]\label{rem:Robust_nonergodic}
An optimal solution of Problem~\ref{prob:eq_sta_ergodic} adapts to the variances of the Gaussian estimation errors $\delta_1^2$ and $\delta_2^2$ and hence is robust against CSI estimation errors.
\end{remark}
\begin{remark}[Challenge for Solving Problem~\ref{prob:eq_sta_ergodic}]
As $\mathbf{v}$ is constant, $\mathbf{w}_0\left(
\hat{\mathbf{H}}_0\right)$ and $r\left(\hat{\mathbf{H}}_0\right)$ both adapt to $\hat{\mathbf{H}}_0$, and $C^{(GP)}\left(\mathbf{v},\mathbf{w}_0,
r\right)$ is non-convex in $\left(\mathbf{v},\mathbf{w}_0,
r\right)$, Problem~\ref{prob:eq_sta_ergodic} is actually a two-timescale stochastic non-convex optimization problem with $2M_0N_0(M_rN_r+1)$  random variables, which is usually quite large. Besides, Problem~\ref{prob:eq_sta_ergodic} has infinitely many successfully transmission constraints, one for each $\hat{\mathbf{H}}_0$, and an extra function $r$ to be optimized. Thus, Problem~\ref{prob:eq_sta_ergodic} is even more challenging than Problem~\ref{prob:eq_delta}.
\end{remark}
\begin{table}
\centering
\begin{tabular}{|c|c|c|c|c|}
\hline
Problem & Problem~\ref{prob:eq_sta_ergodic} & Problem~\ref{prob:eq_sta_determin} & Problem~\ref{prob:eq_sta_determin_2} & Problem~\ref{prob:2} \\
\hline
Random Variables  & $\hat{\mathbf{H}}_0, \Delta{\mathbf{H}}_0$ & $\hat{\mathbf{H}}_0,\Delta{\mathbf{H}}_0$ & $\hat{\mathbf{H}}_0$ & $\hat{\mathbf{H}}_0$ \\
\hline
Optimization Variables & $\left(\mathbf{v},\mathbf{w}_0,r\right)$ & $\left(\mathbf{v},\mathbf{w}_0,r\right)$ & $\left(\mathbf{v},\mathbf{w}_0\right)$ & $\mathbf{v}$
\\
\hline
Constraints  & \eqref{eq:phi}, \eqref{eq:w}, \eqref{eq:proconstraints}& \eqref{eq:phi}, \eqref{eq:w}, \eqref{eq:probability} & \eqref{eq:phi}, \eqref{eq:w}& \eqref{eq:phi} \\
\hline
Timescale& Two-timescale & Two-timescale & Two-timescale & Single-timescale\\
\hline
\end{tabular}
\caption{Comparisons of the problems in the average goodput maximization based on a deterministic bounded error set.}
\label{tab:prob1}
\vspace{-3mm}
\end{table}
\subsection{Closed-form Beamforming and Approximate Phase Shift Optimization }\label{sec:GPB}
\subsubsection{Approximation based on a  deterministic bounded error set}
First, we tackle the challenge caused by the infinitely many constraints in \eqref{eq:proconstraints} via constructing a deterministic bounded set of $\Delta\mathbf{H}_0$:
\begin{align}
\mathcal{E} \triangleq\left\{\left(
\Delta{\mathbf{G}}_{0,0},\Delta{\mathbf{h}}_{0,0}\right)
\big|\left\lVert\Delta{\mathbf{G}}_{0,0}\right\rVert_2 \leq \varepsilon_{1},
\left\lVert\Delta{\mathbf{h}}_{0,0}\right\rVert_2 \leq \varepsilon_{2}\right\},\label{eq:bound}
\end{align}
where $\varepsilon_{1}$ and $\varepsilon_{2}$ are given by:
\begin{align*}
\varepsilon_{1}=\sqrt{\frac{\delta_{1}^2}{2}
F_{2M_0N_0M_rN_r}^{-1}(\rho)}, \quad
\varepsilon_{2}=\sqrt{\frac{\delta_{2}^2}{2}
F_{2M_0N_0}^{-1}(\rho)}.
\end{align*}
Here, $F_{2M_0N_0}^{-1}(\cdot)$ and $F_{2M_0N_0M_rN_r}^{-1}(\cdot)$ denote the inverse cumulative distribution functions (CDFs) of the Chi-square distributions with $2M_0N_0$ and $2M_0N_0M_rN_r$ degrees of freedom, respectively.
Based on the deterministic bounded error set $\mathcal{E}$ in \eqref{eq:bound} \blue{and the total probability theorem,} we obtain Problem~\ref{prob:eq_sta_determin}, which is an approximate problem of Problem~\ref{prob:eq_sta_ergodic}.
\begin{Prob}[Approximate Problem of Problem~\ref{prob:eq_sta_ergodic}]\label{prob:eq_sta_determin}
\begin{align}
C^{(GP1)*}\triangleq\mathop{\max}_{\mathbf{v},\mathbf{w}_0,r}\quad
& \rho \mathbb{E}\left[r\left(
\hat{\mathbf{H}}_0\right)\right]\nonumber \\
s.t. \quad & \eqref{eq:phi},\quad \eqref{eq:w},\nonumber\\ &\text{Pr}\left[C\left(\mathbf{v},
\mathbf{w}_0\left(
\hat{\mathbf{H}}_0\right),\hat{\mathbf{H}}_0,\Delta{\mathbf{H}}_0\right)
\geq r\left(\hat{\mathbf{H}}_0\right) \big|\Delta\mathbf{H}_0\in \mathcal{E}\right]=  1. \label{eq:probability}
\end{align}
\end{Prob}

The relationship between Problem~\ref{prob:eq_sta_ergodic} and Problem~\ref{prob:eq_sta_determin} is summarized in the following \blue{lemma}.
\begin{lemma}[Relationship between Problem~\ref{prob:eq_sta_ergodic} and Problem~\ref{prob:eq_sta_determin}]\label{lem:lemma1}
$C^{(GP1)*}\leq C^{(GP)*}$, where $C^{(GP)*}$ and $C^{(GP1)*}$ denote the optimal values of Problem~\ref{prob:eq_sta_ergodic} and Problem~\ref{prob:eq_sta_determin}, respectively. Furthermore,
a feasible solution of Problem~\ref{prob:eq_sta_determin} is feasible for Problem~\ref{prob:eq_sta_ergodic}.
\end{lemma}
\begin{IEEEproof}
Please refer to Appendix C.
\end{IEEEproof}

Define:
\begin{align}
g_0^{(GP1)}\left(\mathbf{v},\mathbf{w}_0\left(
\hat{\mathbf{H}}_0\right),\hat{\mathbf{H}}_0\right)
\triangleq \left(\left\lvert \left(\mathbf{v}^H\hat{\mathbf{G}}_{0,0}+
\hat{\mathbf{h}}_{0,0}^H\right)
\mathbf{w}_0\left(\hat{\mathbf{H}}_0\right)
\right\rvert-\varepsilon_1\sqrt{M_rN_r}-
\varepsilon_2\right)^2. \label{eq:g_0_GP1}
\end{align}\blue{Using Cauchy-Schwartz inequality and the structural property of Problem~\ref{prob:eq_sta_determin},} we equivalently convert Problem~\ref{prob:eq_sta_determin} to the following problem.
\begin{Prob}[Equivalent Problem of Problem~\ref{prob:eq_sta_determin}]\label{prob:eq_sta_determin_2}
\begin{align}
\begin{split}
\mathop{\max}_{\mathbf{v}}\quad & \rho\mathbb{E}
\left[\tilde{r}^{(GP1)*}\left(\mathbf{v},\hat{\mathbf{H}}_0\right)
\right]\\ s.t. \quad
& \eqref{eq:phi},
\end{split}\label{eq:vvptimalGP}
\end{align}
where
\begin{align}
\begin{split}
\tilde{r}^{(GP1)*}\left(\mathbf{v},\hat{\mathbf{H}}_0\right)=
\mathop{\max}\limits_{\mathbf{w}_0} \quad & \log_2\left(1+\frac{g_0^{(GP1)}\left(\mathbf{v},\mathbf{w}_0
\left(\hat{\mathbf{H}}_0\right),
\hat{\mathbf{H}}_0\right)}{\sum\limits_{k \in \mathcal{K}\backslash \{0\}}P_k{g}_{k}\left(\mathbf{v}\right)+\sigma^2}\right)
\\ s.t. \quad & \eqref{eq:w}.
\end{split} \label{eq:woptimalGP}
\end{align}
Let $\tilde{\mathbf{v}}^{(GP1)*}$ and $\tilde{\mathbf{w}}_{0}^{(GP1)*}\left(\mathbf{v},
\hat{\mathbf{H}}_0\right)$ denote the optimal solutions of the problems in \eqref{eq:vvptimalGP} and \eqref{eq:woptimalGP}, respectively.
\end{Prob}
\begin{lemma}[Equivalence between Problem~\ref{prob:eq_sta_determin} and Problem~\ref{prob:eq_sta_determin_2}
]\label{theorem:GPeq}
The optimal solution and optimal value of the problem in \eqref{eq:woptimalGP} are given by:
\begin{align}
\tilde{\mathbf{w}}_{0}^{(GP1)*}\left(\mathbf{v},
\hat{\mathbf{H}}_0\right)
=&\frac{\left(\mathbf{v}^H\hat{\mathbf{G}}_{0,0}+ \hat{\mathbf{h}}^H_{0,0}\right)^H}{\left\lVert
\mathbf{v}^H\hat{\mathbf{G}}_{0,0}+ \hat{\mathbf{h}}^H_{0,0}\right\rVert_2},\label{eq:GP1tildew}\\
\tilde{r}^{(GP1)*}\left(\mathbf{v},\hat{\mathbf{H}}_0\right)=&\log_2\left(1+
\frac{g_0^{(GP1)}\left(\mathbf{v},\tilde{\mathbf{w}}_{0}
^{(GP1)*}\left(\mathbf{v},\hat{\mathbf{H}}_0\right),
\hat{\mathbf{H}}_0\right)}{\sum\limits_{k \in \mathcal{K}\backslash \{0\}}P_k{g}_{k}\left(\mathbf{v}\right)+\sigma^2}\right),\label{eq:GP1tilder}
\end{align}
\blue{where $g_0^{(GP1)}\left(\cdot\right)$ is given by \eqref{eq:g_0_GP1}.}
Furthermore, $\left(\tilde{\mathbf{v}}^{(GP1)*},\tilde{\mathbf{w}}_{0}^{(GP1)*}\left(\tilde{\mathbf{v}}^{(GP1)*},
\hat{\mathbf{H}}_0\right),
\tilde{r}^{(GP1)*}\left(\tilde{\mathbf{v}}^{(GP1)*},\hat{\mathbf{H}}_0\right)\right)$ is an optimal solution of Problem~\ref{prob:eq_sta_determin}.
\end{lemma}\begin{IEEEproof}
Please refer to Appendix D.
\end{IEEEproof}

By Lemma~\ref{theorem:GPeq},
we equivalently transform Problem~\ref{prob:eq_sta_determin_2} to
a single-timescale stochastic non-convex problem only for the phase shifts, as shown in Table~\ref{tab:prob1}.
\begin{Prob}[Equivalent Problem of Problem~\ref{prob:eq_sta_determin_2}]\label{prob:2}
\begin{align*}
\begin{split}
\mathop{\max}_{\mathbf{v}}\quad &
 \frac{P_0\mathbb{E}\left[
 {g}_0^{(GP1)}\left(\mathbf{v},\frac{\left(\mathbf{v}^H\hat{\mathbf{G}}_{0,0}+ \hat{\mathbf{h}}^H_{0,0}\right)^H}{\left\lVert\mathbf{v}^H\hat{\mathbf{G}}_{0,0}+ \hat{\mathbf{h}}^H_{0,0}\right\rVert_2},\hat{\mathbf{H}}_0\right)\right]}
{\sum\limits_{k \in \mathcal{K}\backslash \{0\}}P_k{g}_{k}\left(\mathbf{v}\right)+\sigma^2} \\
s.t. \quad
& \eqref{eq:phi}.
\end{split}
\end{align*}
\end{Prob}
\begin{theorem}[Equivalence between Problem~\ref{prob:eq_sta_determin} and Problem~\ref{prob:2}]\label{theorem:2}
If $\tilde{\mathbf{v}}^{(GP1)*}$ is an optimal solution of Problem~\ref{prob:2}, then $\left(\mathbf{v}^{(GP1)*},\mathbf{w}_0^{(GP1)*},
r^{(GP1)*}\right)$ is an optimal solution of Problem~\ref{prob:eq_sta_determin}, where \blue{$\mathbf{v}^{(GP1)*}=\tilde{\mathbf{v}}^{(GP1)*}$ and}
\begin{align}
\mathbf{w}_0^{(GP1)*}\left(\hat{\mathbf{H}}_0\right)= &
\tilde{\mathbf{w}}^{(GP1)*}_{0}
\left(\tilde{\mathbf{v}}^{(GP1)*},
\hat{\mathbf{H}}_0\right), \label{eq:equivalencew} \\
r^{(GP1)*}\left(\hat{\mathbf{H}}_0\right)=&
\tilde{r}^{(GP1)*}
\left(\tilde{\mathbf{v}}^{(GP1)*},\hat{\mathbf{H}}_0\right),\quad
\hat{\mathbf{H}}_0 \in \mathbb{C}^{M_0N_0\times (M_rN_r+1)}, \label{eq:equivalencer}
\end{align}
\blue{where $\mathbf{w}_0^{(GP1)*}(\cdot)$ and $r^{(GP1)*}(\cdot)$ are given by \eqref{eq:GP1tildew} and \eqref{eq:GP1tilder}, respectively.}
\end{theorem}
\begin{IEEEproof}
Please refer to Appendix E.
\end{IEEEproof}

Similarly, the closed-form robust \blue{instantaneous CSI-adaptive} beamforming \blue{and rate adaptation} design \blue{for given $\mathbf{v}^{(GP1)*}$} in \eqref{eq:equivalencew} \blue{and \eqref{eq:equivalencer} has computational complexity $\mathcal{O}(M_0N_0M_rN_r)$ and hence} can promptly adapt to \blue{rapid CSI} changes \blue{over slots}. In Section~\ref{alg:SSCA}, we will obtain a robust quasi-static phase shift design by solving Problem~\ref{prob:2}.
\begin{table}
\centering
\begin{tabular}{|c|c|c|c|c|}
\hline
Problem & Problem~\ref{prob:eq_sta_ergodic} & Problem~\ref{prob:appBernstein}& Problem~\ref{prob:eq_sta_determin_xx} & Problem~\ref{prob:3}  \\
\hline
Random Variables  & $\hat{\mathbf{H}}_0, \Delta{\mathbf{H}}_0$ & $\hat{\mathbf{H}}_0, \Delta{\mathbf{H}}_0$ & $\hat{\mathbf{H}}_0$ & $\hat{\mathbf{H}}_0$ \\
\hline
Optimization Variables & $\left(\mathbf{v},\mathbf{w}_0,r\right)$ & $\left(\mathbf{v},\mathbf{w}_0,r\right)$ & $\left(\mathbf{v},\mathbf{w}_0\right)$ & $\mathbf{v}$
\\
\hline
Constraints  & \eqref{eq:phi}, \eqref{eq:w}, \eqref{eq:proconstraints}& \eqref{eq:phi}, \eqref{eq:w}, \eqref{eq:constraint1} & \eqref{eq:phi}, \eqref{eq:w} & \eqref{eq:phi}
\\
\hline
Timescale& Two-timescale & Two-timescale & Two-timescale& Single-timescale \\
\hline
\end{tabular}
\caption{Comparisons of the problems in the average goodput maximization based on the Bernstein-type inequality.}
\label{tab:prob2}
\vspace{-3mm}
\end{table}\subsubsection{Approximation based on the Bernstein-type inequality}
First, we address the challenge caused by the infinitely many constraints in \eqref{eq:proconstraints} using the Bernstein-type inequality.
Based on the Bernstein-type inequality \cite{KWang} \blue{and the structural property of Problem~\ref{prob:eq_sta_ergodic},} \eqref{eq:proconstraints} can be approximated to \blue{(please refer to Appendix F for details)}:
\begin{align}
&\delta_1^2M_rN_r+\delta_{2}^2
+\left\lvert\left(\hat{\mathbf{h}}_{0,0}^H+
\mathbf{v}^H\hat{\mathbf{G}}_{0,0}\right)
\mathbf{w}_0\left(\hat{\mathbf{H}}_0\right)
\right\rvert^2-\frac{1}{P_0}\left(2^{r
\left(\hat{\mathbf{H}}_0\right)}-1\right)
\left(\sum\limits_{k \in \mathcal{K}\backslash\{0\}}P_kg_k(\mathbf{v})
+\sigma^2\right)\nonumber \\ &-\sqrt{2\ln\frac{1}{1-\rho}\left(\left(\delta_1^2M_rN_r+\delta_{2}^2\right)^2+2\left(\delta_1^2M_rN_r+\delta_{2}^2\right)\left\lvert
\left(\hat{\mathbf{h}}_{0,0}^H+\mathbf{v}^H\hat{\mathbf{G}}_{0,0}\right)\mathbf{w}_0
\left(\hat{\mathbf{H}}_0\right)\right\rvert^2\right)}\geq 0. \label{eq:constraint1}
\end{align}Thus, Problem~\ref{prob:eq_sta_ergodic} can be approximated to the following problem.
\begin{Prob}[Approximate Problem of Problem~\ref{prob:eq_sta_ergodic}]\label{prob:appBernstein}
\begin{align*}
\mathop{\max}_{\mathbf{v},\mathbf{w}_0,r}\quad & \rho\mathbb{E}\left[
r\left(
\hat{\mathbf{H}}_0\right)
\right]\\
s.t.\quad  &  \eqref{eq:phi}, \quad \eqref{eq:w}, \quad \eqref{eq:constraint1}.
\end{align*}
\end{Prob}

The relationship between Problem~\ref{prob:eq_sta_ergodic} and Problem~\ref{prob:appBernstein} is summarized in the following lemma.
\begin{lemma}[Relationship between Problem~\ref{prob:eq_sta_ergodic} and Problem~\ref{prob:appBernstein}]\label{lem:equivalencesta} If a feasible solution of Problem~\ref{prob:appBernstein}, $\left(\mathbf{v}^{(GP2)},\mathbf{w}_0^{(GP2)},r^{(GP2)}\right)$, satisfies \eqref{eq:proconstraints},
then $\left(\mathbf{v}^{(GP2)},\mathbf{w}_0^{(GP2)},
r^{(GP2)}\right)$ is a feasible solution of Problem~\ref{prob:eq_sta_ergodic}; otherwise, $\left(\mathbf{v}^{(GP2)},\mathbf{w}_0^{(GP2)},
\hat{r}^{(GP2)}\right)$ is feasible for Problem~\ref{prob:eq_sta_ergodic}, where $\hat{r}^{(GP2)}$ is given by:
\begin{align}
\hat{r}^{(GP2)}\left(\hat{\mathbf{H}}_0\right)=&\log_2\left(1+
\frac{g_0^{(GP1)}\left(\mathbf{v}^{(GP2)},\mathbf{w}_{0}
^{(GP2)}\left(\hat{\mathbf{H}}_0\right),
\hat{\mathbf{H}}_0\right)}{\sum\limits_{k \in \mathcal{K}\backslash \{0\}}P_k{g}_{k}\left(\mathbf{v}^{(GP2)}\right)+\sigma^2}\right).
\end{align}
\end{lemma}
\begin{IEEEproof}
Please refer to Appendix G.
\end{IEEEproof}

Define:
\begin{align}
&g_0^{(GP2)}\left(\mathbf{v},\mathbf{w}_0\left(\hat{\mathbf{H}}_0\right)
,\hat{\mathbf{H}}_0\right)=\left\lvert\left(\hat{\mathbf{h}}_{0,0}^H+\mathbf{v}^H
\hat{\mathbf{G}}_{0,0}\right)\mathbf{w}_0\left(
\hat{\mathbf{H}}_0\right)
\right\rvert^2+\delta_1^2M_rN_r+
\delta_{2}^2\nonumber \\ - & \sqrt{2\ln\frac{1}{1-\rho}\left(\left(\delta_1^2M_rN_r+\delta_{2}^2\right)^2+2\left(\delta_1^2M_rN_r+\delta_{2}^2\right)\left\lvert
\left(\hat{\mathbf{h}}_{0,0}^H+\mathbf{v}^H\hat{\mathbf{G}}_{0,0}\right)\mathbf{w}_0
\left(\hat{\mathbf{H}}_0\right)\right\rvert^2\right)}. \label{eq:g_0GP2}
\end{align}\blue{Using Cauchy-Schwartz inequality and the structural property of Problem~\ref{prob:appBernstein},} we equivalently convert Problem~\ref{prob:appBernstein} to the following problem.
\begin{Prob}[Equivalent Problem of Problem~\ref{prob:appBernstein}]\label{prob:eq_sta_determin_xx}
\begin{align}
\begin{split}
\mathop{\max}_{\mathbf{v}}\quad & \rho\mathbb{E}
\left[\tilde{r}^{(GP2)*}\left(\mathbf{v},\hat{\mathbf{H}}_0\right)
\right]\\ s.t. \quad
& \eqref{eq:phi},
\end{split}\label{eq:vvvptimalGPxx}
\end{align}
where
\begin{align}
\begin{split}
\tilde{r}^{(GP2)*}\left(\mathbf{v},\hat{\mathbf{H}}_0\right)=
\mathop{\max}\limits_{\mathbf{w}_0} \quad & \log_2\left(1+\frac{g_0^{(GP2)}\left(\mathbf{v},\mathbf{w}_0
\left(\hat{\mathbf{H}}_0\right),
\hat{\mathbf{H}}_0\right)}{\sum\limits_{k \in \mathcal{K}\backslash \{0\}}P_k{g}_{k}\left(\mathbf{v}\right)+\sigma^2}\right)
\\ s.t. \quad & \eqref{eq:w}.
\end{split} \label{eq:woptimalGPxx}
\end{align}
Let $\tilde{\mathbf{v}}^{(GP2)*}$ and
$\tilde{\mathbf{w}}_{0}^{(GP2)*}\left(\mathbf{v},
\hat{\mathbf{H}}_0\right)$ denote the optimal solutions of the problems in \eqref{eq:vvvptimalGPxx} and \eqref{eq:woptimalGPxx}, respectively.
\end{Prob}
\begin{lemma}[Equivalence between Problem~\ref{prob:appBernstein} and Problem~\ref{prob:eq_sta_determin_xx} ]\label{theorem:GPeqxx}
The optimal solution and optimal value of the problem in \eqref{eq:woptimalGPxx} are given by:
\begin{align}
\tilde{\mathbf{w}}_{0}^{(GP2)*}\left(\mathbf{v},\hat{\mathbf{H}}_0\right)
=&\frac{\left(\mathbf{v}^H\hat{\mathbf{G}}_{0,0}+ \hat{\mathbf{h}}^H_{0,0}\right)^H}{\left\lVert\mathbf{v}^H\hat{\mathbf{G}}_{0,0}+ \hat{\mathbf{h}}^H_{0,0}\right\rVert_2},\label{eq:GP2tildew}\\
\tilde{r}^{(GP2)*}\left(\mathbf{v},\hat{\mathbf{H}}_0\right)=&\log_2\left(1+
\frac{g_0^{(GP2)}\left(\mathbf{v},\tilde{\mathbf{w}}_{0}
^{(GP2)*}\left(\mathbf{v},\hat{\mathbf{H}}_0\right),
\hat{\mathbf{H}}_0\right)}{\sum\limits_{k \in \mathcal{K}\backslash \{0\}}P_k{g}_{k}\left(\mathbf{v}\right)+\sigma^2}\right),\label{eq:GP2tilder}
\end{align}
\blue{where $g_0^{(GP2)}\left(\cdot\right)$ is given by \eqref{eq:g_0GP2}.}
Furthermore, $\left(\tilde{\mathbf{v}}^{(GP2)*},\tilde{\mathbf{w}}_{0}^{(GP2)*}
\left(\tilde{\mathbf{v}}^{(GP2)*},
\hat{\mathbf{H}}_0\right),\tilde{r}^{(GP2)*}\left(\tilde{\mathbf{v}}^{(GP2)*},\hat{\mathbf{H}}_0\right)\right)$ is an optimal solution of Problem~\ref{prob:appBernstein}.
\end{lemma}
\begin{IEEEproof}
Please refer to Appendix H.
\end{IEEEproof}

By Lemma~\ref{theorem:GPeqxx}, we equivalently transform Problem~\ref{prob:eq_sta_determin_xx} to
a more tractable single-timescale stochastic non-convex problem as shown in Table~\ref{tab:prob2}.
\begin{Prob}[Equivalent Problem of Problem~\ref{prob:eq_sta_determin_xx}]\label{prob:3}
\begin{align*}
\begin{split}
\mathop{\max}_{\mathbf{v}}\quad &
 \frac{P_0\mathbb{E}\left[
 {g}_0^{(GP2)}\left(\mathbf{v},\frac{\left(\mathbf{v}^H\hat{\mathbf{G}}_{0,0}+ \hat{\mathbf{h}}^H_{0,0}\right)^H}{\left\lVert\mathbf{v}^H\hat{\mathbf{G}}_{0,0}+ \hat{\mathbf{h}}^H_{0,0}\right\rVert_2},\hat{\mathbf{H}}_0\right)\right]}
{\sum\limits_{k \in \mathcal{K}\backslash \{0\}}P_k{g}_{k}\left(\mathbf{v}\right)+\sigma^2}
 \\
s.t. \quad
& \eqref{eq:phi}.
\end{split}
\end{align*}
\end{Prob}
\begin{theorem}[Equivalence between Problem~\ref{prob:appBernstein} and Problem~\ref{prob:3}]\label{theorem:3}
If $\tilde{\mathbf{v}}^{(GP2)*}$ is an optimal solution of Problem~\ref{prob:3}, then
$\left(\mathbf{v}^{(GP2)*},\mathbf{w}_0^{(GP2)*},r^{(GP2)*}\right)$ is an optimal solution of Problem~\ref{prob:appBernstein}, where \blue{$\mathbf{v}^{(GP2)*}=\tilde{\mathbf{v}}^{(GP2)*}$ and}
\begin{align} \mathbf{w}_0^{(GP2)*}\left(\hat{\mathbf{H}}_0\right)= &
\tilde{\mathbf{w}}^{(GP2)*}_{0}
\left(\tilde{\mathbf{v}}^{(GP2)*},\hat{\mathbf{H}}_0\right), \label{eq:equivalencew2}\\
r^{(GP2)*}\left(\hat{\mathbf{H}}_0\right)=&\tilde{r}^{(GP2)*}
\left(\tilde{\mathbf{v}}^{(GP2)*},\hat{\mathbf{H}}_0\right),\quad
\hat{\mathbf{H}}_0 \in \mathbb{C}^{M_0N_0\times (M_rN_r+1)},\label{eq:equivalencew2_r}
\end{align}
\blue{where $\mathbf{w}_0^{(GP2)*}\left(\cdot\right)$ and $r^{(GP2)*}\left(\cdot\right)$ are given by \eqref{eq:GP2tildew} and \eqref{eq:GP2tilder}, respectively.}
\end{theorem}
\begin{IEEEproof}
Please refer to Appendix I.
\end{IEEEproof}

Analogously, the closed-form robust \blue{instantaneous CSI-adaptive} beamforming  \blue{and rate adaptation} design \blue{(for given $\mathbf{v}^{(GP2)*}$)} in \eqref{eq:equivalencew2} \blue{and \eqref{eq:equivalencew2_r} has computational complexity $\mathcal{O}(M_0N_0M_rN_r)$ and hence can promptly adapt to rapid CSI changes over slots}. In Section~\ref{alg:SSCA}, we will obtain a robust quasi-static phase shift design by solving Problem~\ref{prob:3}.
\section{Algorithm for Phase Shift Optimization}\label{alg:SSCA}
Recall that we have obtained closed-form robust instantaneous CSI-adaptive beamforming designs in \blue{Section}~\ref{sec:erB} and Section~\ref{sec:GPB}. It remains to obtain robust quasi-static phase shift designs by solving Problem~\ref{prob:theta}\blue{,} Problem~\ref{prob:2}\blue{, and} Problem~\ref{prob:3}. Problem~\ref{prob:theta}, Problem~\ref{prob:2}\blue{,} and Problem~\ref{prob:3} are stochastic non-convex optimization problems with similar non-convex objective functions and the same non-convex constraint functions.
This section proposes a low-complexity algorithm to obtain KKT points of each of the three stochastic non-convex problems using SSCA \cite{7412752}.

Specifically, at each iteration $t$, we randomly generate $L$ channel samples, denoted by $\hat{\mathbf{H}}_{0,l}^{(t)}, l=1,...,L$, according to the distribution of $\hat{\mathbf{H}}_0$ given in Section~\ref{sec:system}. Based on the samples, we approximate each of the three objective functions in Problem~\ref{prob:theta}, Problem~\ref{prob:2}, and Problem~\ref{prob:3} around $\mathbf{v}^{(t-1)}$ with a concave surrogate function, where $\mathbf{v}^{(t-1)}\triangleq\left(v^{(t-1)}_n\right)_{n\in\mathcal{N}}$ denotes the phase shifts at iteration $t-1$. The surrogate functions for Problem~\ref{prob:theta}, Problem~\ref{prob:2}, and Problem~\ref{prob:3}, denoted by $f^{(ER,t)}(\mathbf{v})$, $f^{(GP1,t)}(\mathbf{v})$, and $f^{(GP2,t)}(\mathbf{v})$, respectively, are given by:
\begin{align}
f^{(Q,t)}(\mathbf{v})=c_0^{(Q,t)}
+2Re\left\{\sum\limits_{n\in \mathcal{N}}c_{1,n}^{(Q,t)}\left(v_n-v_n^{(Q,t-1)}\right)\right\} -\tau\sum\limits_{n\in \mathcal{N}}\left\lvert
v_n-v^{(Q,t-1)}_n\right\rvert^2,&\nonumber \\ Q=ER,GP1,GP2,&
\label{eq:surrogate}
\end{align}
where $c_0^{(Q,t)}\in \mathbb{C}$ and $c_{1,n}^{(Q,t)}\in \mathbb{C}, n\in\mathcal{N}$ are updated according to:
\begin{align}
c_{0}^{(Q,t)}=&
\rho^{(t)}\sum\limits_{l=1}^{L}
\frac{\gamma^{(Q)}
\left(\mathbf{v}^{(Q,t-1)},\mathbf{w}_{0,l}^{(t)}
\left(\hat{\mathbf{H}}_0\right),
\hat{\mathbf{H}}_l^{(t)}\right)}{L}+
\left(1-\rho^{(t)}\right)c_{0}^{(Q,t-1)},\label{eq:average}\\
c^{(Q,t)}_{1,n}=&
\rho^{(t)}\sum\limits_{l=1}^{L}
\frac{\frac{\partial}{\partial v_n}\gamma^{(Q)}
\left(\mathbf{v}^{(Q,t-1)},\mathbf{w}_{0,l}^{(t)}
\left(\hat{\mathbf{H}}_0\right),
\hat{\mathbf{H}}_l^{(t)}\right)}{L} + \left(1-\rho^{(t)}\right)c^{(Q,t)}_{1,n},\quad n\in\mathcal{N},\label{eq:derivative}
\end{align}
with $c_{0}^{(Q,0)}=0$ and $c^{(Q,0)}_{1,n}=0, n\in\mathcal{N}$. Here, $\tau>0$ can be any constant, the term $\tau\sum\limits_{n=1}^{M_rN_r}\left\lvert
v_n-v^{(Q,t-1)}_n\right\rvert^2$ is used
to ensure strong concavity,
$\rho^{(t)}$ is a positive diminishing stepsize satisfying:
\begin{align*}
&\rho^{(t)}>0,\quad \lim_{t\to\infty}\rho^{(t)}=0,\quad \sum_{t=1}^\infty\rho^{(t)}=\infty,\quad \sum_{t=1}^\infty\left(\rho^{(t)}\right)^2<\infty,
\end{align*}
and $\gamma^{(Q)}
\left(\mathbf{v}^{(Q,t-1)},\mathbf{w}_{0,l}^{(t)}
\left(\hat{\mathbf{H}}_0\right),
\hat{\mathbf{H}}_l^{(t)}\right)$ is given by:
\begin{align*}
\gamma^{(Q)}
\left(\mathbf{v}^{(Q,t-1)},\mathbf{w}_{0,l}^{(t)}
\left(\hat{\mathbf{H}}_0\right),
\hat{\mathbf{H}}_l^{(t)}\right)\triangleq
\begin{cases}
\frac{P_0
 {g}_0^{(ER)}\left(\mathbf{v},\frac{\left(\mathbf{v}^H\hat{\mathbf{G}}_{0,0}+ \hat{\mathbf{h}}^H_{0,0}\right)^H}{\left\lVert
\mathbf{v}^H\hat{\mathbf{G}}_{0,0}+ \hat{\mathbf{h}}^H_{0,0}\right\rVert_2},\hat{\mathbf{H}}_0\right)}
{\sum\limits_{k \in \mathcal{K}\backslash \{0\}}P_k{g}_{k}\left(\mathbf{v}\right)+\sigma^2},\quad &Q=ER,\\
\frac{P_0
 {g}_0^{(GP1)}\left(\mathbf{v},\frac{\left(\mathbf{v}^H\hat{\mathbf{G}}_{0,0}+ \hat{\mathbf{h}}^H_{0,0}\right)^H}{\left\lVert
\mathbf{v}^H\hat{\mathbf{G}}_{0,0}+ \hat{\mathbf{h}}^H_{0,0}\right\rVert_2},\hat{\mathbf{H}}_0\right)}
{\sum\limits_{k \in \mathcal{K}\backslash \{0\}}P_k{g}_{k}\left(\mathbf{v}\right)+\sigma^2},\quad &Q=GP1, \\
\frac{P_0
 {g}_0^{(GP2)}\left(\mathbf{v},\frac{\left(\mathbf{v}^H\hat{\mathbf{G}}_{0,0}+ \hat{\mathbf{h}}^H_{0,0}\right)^H}{\left\lVert
\mathbf{v}^H\hat{\mathbf{G}}_{0,0}+ \hat{\mathbf{h}}^H_{0,0}\right\rVert_2},\hat{\mathbf{H}}_0\right)}
{\sum\limits_{k \in \mathcal{K}\backslash \{0\}}P_k{g}_{k}\left(\mathbf{v}\right)+\sigma^2},\quad &Q=GP2.
\end{cases}
\end{align*}
Furthermore, the non-convex constraints in \eqref{eq:phi} can be equivalently converted to the convex constraints:
\begin{align}
\lvert v_n\rvert \leq 1,\quad n\in\mathcal{N},\label{eq:cvxphi}
\end{align}
without affecting the optimality \cite{QingqingWu2}.
Then, the resulting approximate convex problem\blue{s} at iteration $t$ \blue{are} given \blue{as follows.}
\begin{Prob}[Approximation of Problem~\ref{prob:theta}, Problem~\ref{prob:2}, and Problem~\ref{prob:3} at Iteration $t$]\label{prob:ins_surrogate} For $Q=ER,GP1,GP2$,
\begin{align*}
\bar{\mathbf{v}}^{(Q,t)}=\mathop{\arg}\mathop{\max}_{\mathbf{v}}\quad &
f^{(Q,t)}(\mathbf{v}), \\
s.t. \quad & \eqref{eq:cvxphi},
\end{align*}
where $f^{(Q,t)}(\mathbf{v})$ is given by \eqref{eq:surrogate}.
\end{Prob}

As Slater's condition is satisfied, strong duality holds for Problem~\ref{prob:ins_surrogate}. Thus, based on \blue{problem decomposition and} the KKT conditions, we can obtain a closed-form optimal solution of Problem~\ref{prob:ins_surrogate}.
\begin{theorem}[Optimal Solution of Problem~\ref{prob:ins_surrogate}]\label{theoren:phi}
For $Q=ER,GP1,GP2$, the optimal solution of Problem~\ref{prob:ins_surrogate} is given by $\bar{\mathbf{v}}^{(Q,t)}=\left(\bar{v}_n^{(Q,t)}\right)_{n\in\mathcal{N}}$,  where \begin{align}
\bar{v}_n^{(Q,t)}=\frac{\tau v_n^{(Q,t-1)}+c_{1,n}^{(Q,t)}}{\left\lvert\tau v_n^{(Q,t-1)}+c_{1,n}^{(Q,t)}\right\rvert}\in \mathbb{C},\quad n\in\mathcal{N}.\label{eq:solutionv}
\end{align}
\end{theorem}
\begin{IEEEproof}
Please refer to Appendix J.
\end{IEEEproof}

Then, the phase shifts $\mathbf{v}^{(Q,t)}$ at iteration $t$ are updated according to:
\begin{align}
\mathbf{v}^{(Q,t)}=\left(1-\omega^{(t)}\right)\mathbf{v}^{(Q,t-1)}+
\omega^{(t)}\bar{\mathbf{v}}^{(Q,t)},\quad Q=ER,GP1,GP2,\label{eq:updatetheta}
\end{align}
where $\omega^{(t)}$ is a positive diminishing stepsize satisfying:
\begin{align*}
&\omega^{(t)}>0,\quad \lim_{t\to\infty}\omega^{(t)}=0,\quad \sum_{t=1}^\infty\omega^{(t)}=\infty, \quad\sum_{t=1}^\infty\left(\omega^{(t)}\right)^2<\infty,
\quad \lim_{t\to\infty}\frac{\omega^{(t)}}{\rho^{(t)}}=0.
\end{align*}

The details of the SSCA algorithm are summarized in Algorithm~\ref{alg:TTS}. Since Problem~\ref{prob:ins_surrogate} can be solved analytically, the computation complexity of Algorithm~\ref{alg:TTS} is relatively low. Specifically, the computational complexities of Step 4\blue{,} Step 5\blue{, and step 6} of Algorithm~\ref{alg:TTS} are $\mathcal{O}\left(M_0N_0M_rN_r\right)$, \blue{$\mathcal{O}\left(M_0N_0M_rN_r\right)$, and $\mathcal{O}\left(M_rN_r\right)$, respectively.}  \blue{Hence,} Algorithm~\ref{alg:TTS} \blue{has computational complexity} $\mathcal{O}\left(\blue{T}M_0N_0M_rN_r\right)$. \blue{The computational time for the \blue{robust} quasi-static phase shifts is negligible compared to the considered time period. Hence, the robust quasi-static phase shift designs have low computation and phase adjustment costs.}  Furthermore,
by Theorem~2 of \cite{A_Liu}, we know that every limit point of $\{\mathbf{v}^{(Q,t)}\}$ generated by Algorithm~\ref{alg:TTS} \blue{when $T\rightarrow\infty$}, denoted by $\mathbf{v}^{(Q)\dag}$, is a KKT point of Problem~\ref{prob:theta} \blue{($Q=ER$)}, Problem~\ref{prob:2} \blue{($Q=GP1$)}, \blue{or} Problem~\ref{prob:3} \blue{($Q=GP2$)}.

Finally, we obtain suboptimal solutions of Problem~\ref{prob:eq_delta} and Problem~\ref{prob:eq_sta_ergodic} based on the closed-form \blue{robust} instantaneous CSI-adaptive beamforming designs obtained in Section~\ref{sec:averagerate} and Section~\ref{sec:Non-ergodicrate} and the \blue{robust} quasi-static phase shift designs obtained by Algorithm~\ref{alg:TTS}.
By Theorem~\ref{theorem:eq}, we can obtain a suboptimal solution of Problem~\ref{prob:eq_delta}, i.e., $\left(\mathbf{v}^{(ER)\dag},\mathbf{w}_0^{(ER)\dag}\right)$, where $\mathbf{w}_0^{(ER)\dag}=\frac{\left(\left(\mathbf{v}^{(ER)\dag}\right)^H\hat{\mathbf{G}}_{0,0}+ \hat{\mathbf{h}}^H_{0,0}\right)^H}{\left\lVert
\left(\mathbf{v}^{(ER)\dag}\right)^H\hat{\mathbf{G}}_{0,0}+ \hat{\mathbf{h}}^H_{0,0}\right\rVert_2}$. By Lemma~\ref{theorem:GPeq}, we can obtain a suboptimal solution of Problem~\ref{prob:eq_sta_ergodic}, i.e., $\left(\mathbf{v}^{(GP1)\dag},\mathbf{w}_0^{(GP1)\dag},
r^{(GP1)\dag}\right)$, where $\mathbf{w}_0^{(GP1)\dag}=\tilde{\mathbf{w}}_{0}^{(GP1)*}\left(\mathbf{v}^{(GP1)\dag},
\hat{\mathbf{H}}_0\right)$ and $r^{(GP1)\dag}=\tilde{r}^{(GP1)*}\left(\mathbf{v}^{(GP1)\dag},\hat{\mathbf{H}}_0\right)$. Here, $\tilde{\mathbf{w}}_0^{(GP1)*}\left(\cdot\right)$ and $\tilde{r}^{(GP1)*}\left(\cdot\right)$ are given by \eqref{eq:GP1tildew} and \eqref{eq:GP1tilder}\blue{, respectively.}
By Lemma~\ref{lem:equivalencesta} and Lemma~\ref{theorem:GPeqxx}, we can obtain a suboptimal solution of Problem~\ref{prob:eq_sta_ergodic}, i.e., $\left(\mathbf{v}^{(GP2)\dag},\mathbf{w}_0^{(GP2)\dag},
r^{(GP2)\dag}\right)$, where  $\mathbf{w}_0^{(GP2)\dag}=
\tilde{\mathbf{w}}_{0}^{(GP2)*}\left(\mathbf{v}^{(GP2)\dag},
\hat{\mathbf{H}}_0\right)$ and
\begin{align*}
r^{(GP2)\dag}=
\begin{cases}
\tilde{r}^{(GP2)*}\left(\mathbf{v}^{(GP2)\dag},\hat{\mathbf{H}}_0\right),
&\text{if} \left(\mathbf{v}^{(GP2)\dag},\mathbf{w}_0^{(GP2)\dag},
r^{(GP2)\dag}\right) \text{satisfies \eqref{eq:proconstraints}},\\
\tilde{r}^{(GP1)*}\left(\mathbf{v}^{(GP2)\dag},\hat{\mathbf{H}}_0\right),
&\text{otherwise.}
\end{cases}
\end{align*}
Here, $\tilde{\mathbf{w}}_0^{(GP2)*}\left(\cdot\right)$ and $\tilde{r}^{(GP2)*}\left(\cdot\right)$ are given by \eqref{eq:GP2tildew} and \eqref{eq:GP2tilder}\blue{, respectively}. \blue{Based on the complexity analysis in Section~\ref{sec:averagerate} and Section~\ref{sec:Non-ergodicrate}, we know that the overall computational complexity for obtaining the \blue{three} proposed robust joint designs for the considered time period is $(S+T)(M_0N_0M_rN_r)$.}
\blue{\begin{remark}[Comparison with Existing Robust Joint Designs]
Let $T'$ denote the number of iterations for the algorithms in \cite{xu2020resource,yu2020robust,hong2020robust,ZhouGui,JWang,TALe,deng2020robust,Czhong1}, proposed to solve the robust instantaneous CSI-adaptive joint design problems for each slot, to meet the corresponding stopping criteria. Thus, the computational complexity of these algorithms is $\mathcal{O}(T'M_0N_0M_rN_r)$.
 Consequently, the overall computational complexity for obtaining the robust instantaneous CSI-adaptive joint designs \cite{xu2020resource,yu2020robust,hong2020robust,ZhouGui,JWang,TALe,deng2020robust,Czhong1} in the considered time period is $\mathcal{O}(ST'M_0N_0M_rN_r)$, which is much higher than $\mathcal{O}((S+T)(M_0N_0M_rN_r))$ (note that $ST'$ is usually much larger than $S+T$). Obviously, the proposed robust joint designs are more computationally efficient than those in \cite{xu2020resource,yu2020robust,hong2020robust,ZhouGui,JWang,TALe,deng2020robust,Czhong1}. Equally importantly, the phase adjustment cost of the proposed robust joint designs is only $\frac{1}{S}$ of those in \cite{xu2020resource,yu2020robust,hong2020robust,ZhouGui,JWang,TALe,deng2020robust,Czhong1}. Last but not least, unlike the proposed robust joint designs, the robust instantaneous CSI-adaptive joint designs in \cite{xu2020resource,yu2020robust,hong2020robust,ZhouGui,JWang,TALe,deng2020robust,Czhong1} fail to consider inter-cell interference and are not suitable for dense networks or cell-edge users.
\end{remark}}
\begin{algorithm}[t]\small
    \caption{SSCA Algorithm for Solving Problem~\ref{prob:theta}, Problem~\ref{prob:2} and Problem~\ref{prob:3}}
\begin{small}

        \begin{algorithmic}[1]
           \STATE \textbf{initialization}: Choose an arbitrary feasible point $\mathbf{v}^{(Q,0)}$ of Problem~\ref{prob:ins_surrogate} for $Q=ER,GP1,$ or $GP2$ as the initial point.\\
           \STATE \textbf{for} $t=1,...,T$ \textbf{do}
           \STATE \quad Randomly generate $L$ channel samples $\hat{\mathbf{H}}_{0,l}^{(t)},l=1,...,L$ according to the distribution of $\hat{\mathbf{H}}_0$.\\
           \STATE \quad Update $c_0^{(Q,t)}$ and $c^{(Q,t)}_{1,n},n\in\mathcal{N}$ according to \eqref{eq:average} and \eqref{eq:derivative}, respectively.
           \STATE \quad Calculate $\bar{\mathbf{v}}^{(Q,t)}$ according to Theorem~\ref{theoren:phi}.
           \STATE \quad Update $\mathbf{v}^{(Q,t)}$
           according to $\eqref{eq:updatetheta}$.
           \STATE \textbf{end for}
    \end{algorithmic}\label{alg:TTS}
    \end{small}

\end{algorithm}
%
\begin{figure}[t]
\begin{center}
\includegraphics[width=6.5cm]{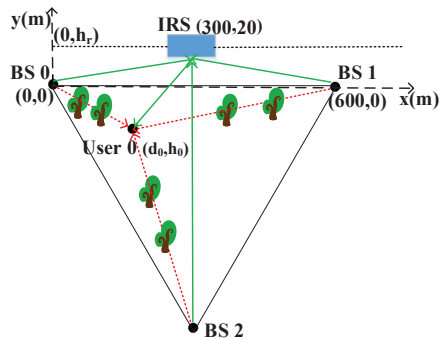}
\end{center}
\vspace{-3mm}
\caption{\small{The IRS-assisted system considered in Section \ref{sec:simulation}.}}
\label{fig:pathloss}
\vspace{-3mm}
\end{figure}
\section{Numerical Results}\label{sec:simulation}
In this section, we numerically evaluate the proposed solutions, $\left(\mathbf{v}^{(Q)\dag},\mathbf{w}_0^{(Q)\dag},
r^{(Q)\dag}\right), Q=ER,GP1,GP2$, in a multi-cell IRS-assisted system shown in Fig.~\ref{fig:pathloss}. Specifically, we  consider $K=3$. BS 0, BS 1, BS 2, user $0$, and the IRS are located at $(0,0)$, $(600,0)$, $(300,300\sqrt{3})$, $(d_{0},h_{0})$, and \blue{$(d_r,20)$} (in m), respectively, and the position of user $0$ lies on the perpendicular bisector of the line segment between BS $1$ and BS $2$. In the simulation, we set $d=\frac{\lambda}{2}$, $M_0=N_0=M_1=N_1=M_2=N_2=4$, $M_r=N_r=8$, $P_0=P_1=P_2=30$dBm, $\sigma^2=-90$dBm, $\varphi_{0,r}^{(h)}=\varphi_{0,r}^{(v)}=\pi/3$, $\varphi_{1,r}^{(h)}=\varphi_{1,r}^{(v)}=
\varphi_{2,r}^{(h)}=\varphi_{2,r}^{(v)}=\pi/8$, $\varphi_{r,0}^{(h)}=\varphi_{r,0}^{(v)}=\pi/6$, \blue{$d_r=300$m,} $d_{0,0}=d_{1,0}=d_{2,0}=200\sqrt{3}$m, $d_{r,0}=(20+100\sqrt{3})$m, $\delta_1=\delta_2=10^{-6}$, and $\rho=0.95$, if not specified otherwise. We set
$\alpha_i=1/\left(1000 d_i^{\bar\alpha_i}\right)$ $(\text{i.e.,}-30+10\bar\alpha_i\log_{10}(d_i) \text{ dB}),\ i=(0,0)$, $(1,0)$, $(2,0)$, $(0,r)$, $(1,r)$, $(2,r)$, $(r,0)$, where $\bar{\alpha}_i$ represents the corresponding path loss exponent \cite{HGuo1,CPan1,YuhangJia}.
Due to extensive obstacles and scatters, we set $\bar\alpha_{0,0}=\bar\alpha_{1,0}=\bar\alpha_{2,0}
=3.7$. As the location of the IRS is usually carefully chosen, we assume that the links between the BSs and the IRS experience free-space path loss and set $\bar\alpha_{0,r}=\bar\alpha_{1,r}
=\bar\alpha_{2,r}=2$ \cite{HGuo1}. In addition, we set $\bar\alpha_{r,0}=3$, due to few obstacles \cite{YuhangJia}.

We consider \blue{four} baseline schemes that adopt instantaneous CSI-adaptive beamforming and quasi-static\blue{\footnote{\blue{We restrict our attention to quasi-static phase shift design for a fair comparison.}}} phase shift designs, namely {\em Non-Robust Design Without Interference} \cite{SJin2}, {\em Non-Robust Design With Interference} \cite{YuhangJia}, {\em Robust Design Without Interference} \blue{\cite{HGuo1}}, \blue{{\em Robust Design With Interference} \cite{Ergodic_baseline,nonErgodic_baseline},} respectively. {\em Non-Robust Design Without Interference} and {\em Non-Robust Design With Interference} consider perfect CSIT and optimize the joint design to maximize the upper bounds on the average rates in the cases without and with inter-cell interference, respectively. Denote $\mathbf{H}_0 \triangleq \left[\mathbf{h}_{0,0},
\mathbf{G}_{0,0}\right]\in\mathbb{C}^{M_0N_0\times (M_rN_r+1)}$.  {\em Non-Robust Design Without Interference} adopts the closed-form optimal solution of the following problem:
\begin{align}
\begin{split}
\left(\mathbf{v}^{(1)*},\mathbf{w}_0^{(1)*}\left(
\mathbf{H}_0\right)\right)\triangleq \arg\mathop{\max}_{\mathbf{v},\mathbf{w}_0}\quad &
\log_2\left(1+\frac{P_0\mathbb{E}\left[
\left\lvert\left(\mathbf{v}^H
\mathbf{G}_0+\mathbf{h}_{0,0}^H\right)
\mathbf{w}_0\left(\mathbf{H}_0\right)\right\rvert^2\right]}
{\sigma^2}\right) \\
 s.t. \quad
& \eqref{eq:phi},\quad \left\lVert\mathbf{w}_0\left(\mathbf{H}_0\right)\right\rVert_2^2=1,
\end{split}\label{eq:baseline1}
\end{align}
which is given by
\cite{SJin2}: $$\mathbf{v}^{(1)*}=\angle{\left(\mathbf{a}(\delta^{(h)}_{0,r},\delta^{(v)}_{0,r},M_r,N_r) -\mathbf{a}(\varphi^{(h)}_{r,0},\varphi^{(v)}_{r,0},M_r,N_r)\right)},
\quad \mathbf{w}^{(1)*}_0\left(
\mathbf{H}_0\right)=\frac{\left(\left(\mathbf{v}^{(1)*}
\right)^H\mathbf{G}_0+\mathbf{h}_{0,0}^H\right)^H}{\left\lVert
\left(\mathbf{v}^{(1)*}\right)^H
\mathbf{G}_0+\mathbf{h}_{0,0}^H\right\rVert_2},$$
\blue{as the quasi-static phase shifts and instantaneous CSI-adaptive beamformer.}
{\em Non-Robust Design With Interference} adopts a stationary point of the following problem:
\begin{align}
\begin{split}
\mathop{\max}_{\mathbf{v},\mathbf{w}_0}\quad &
\log_2\left(1+\frac{P_0\mathbb{E}\left[
\left\lvert\left(\mathbf{v}^H
\mathbf{G}_0+\mathbf{h}_{0,0}^H\right)
\mathbf{w}_0\left(\mathbf{H}_0\right)\right\rvert^2\right]}
{P_1g_1\left(\mathbf{v}\right)+\sigma^2}\right) \\
 s.t. \quad
& \eqref{eq:phi},\quad \left\lVert\mathbf{w}_0\left(\mathbf{H}_0\right)\right\rVert_2^2=1,
\end{split}\label{eq:baseline2}
\end{align}
denoted by $\left(\mathbf{v}^{(2)*},\mathbf{w}^{(2)*}\right)$, which is obtained using Algorithm~1 in \cite{YuhangJia}, \blue{as the quasi-static phase shifts and instantaneous CSI-adaptive beamformer.} \blue{{\em Robust Design With Interference} adopts a stationary point of the following problem:
\begin{align*}
  \max_{\mathbf{v}} \quad & \frac{P_0
\left\lVert\mathbf{v}^H
\bar{\mathbf{G}}_{0,0}\right\rVert_2^2}
{\sigma^2+\sum_{k\in\mathcal{K}}P_k\left\lVert\mathbf{v}^H
\bar{\mathbf{G}}_{k,0}\right\lVert_2^2} \\
s.t. \quad & \lvert v_n \rvert= 1,\ n \in \mathcal{N},
\end{align*}
denoted by $\mathbf{v}^{(4)*}$, as the quasi-static phase shifts, and \blue{then} utilizes $\mathbf{w}^{(4)*} = \frac{\left(\left(\mathbf{v}^{(4)*}\right)^H\hat{\mathbf{G}}_{0,0}+ \hat{\mathbf{h}}^H_{0,0}\right)^H}{\left\lVert
\left(\mathbf{v}^{(4)*}\right)^H\hat{\mathbf{G}}_{0,0}+ \hat{\mathbf{h}}^H_{0,0}\right\rVert_2}$ as the robust instantaneous CSI-adaptive \blue{beamformer}.\footnote{\blue{ The quasi-static phase shifts adapt to the perfect LoS components, and the robust instantaneous CSI-adaptive beamforming design is identical to the proposed ones and
can be viewed as extensions of those in \cite{Ergodic_baseline,nonErgodic_baseline}.}} Besides, {\em Non-Robust Design Without Interference}, {\em Non-Robust Design With Interference}, and {\em Robust Design Without Interference} adopt the proposed rate adaptation design given by \eqref{eq:GP1tilder} for the scenario of goodput maximization.} {\em Robust Design Without Interference} considers imperfect CSIT but does not consider inter-cell interference. More specifically, {\em Robust Design Without Interference} adopts the proposed solutions under $P_k=0,k\in\mathcal{K}\backslash\{0\}$, denoted by \blue{$\left(\mathbf{v}^{(3,ER)*},\mathbf{w}^{(3,ER)*}\right)$ for the scenario of ergodic rate and $\left(\mathbf{v}^{(3,Q)*},\mathbf{w}^{(3,Q)*},
r^{(3,Q)*}\right), Q=GP1,GP2$ for the scenario of goodput maximization.}
The ergodic rates and average goodputs of all schemes are obtained by averaging over 10000 realizations of the random NLoS components.

\begin{figure}[t]
\begin{center}
\subfigure[\scriptsize{Ergodic rate versus $M_r(=N_r)$.
}\label{fig:Ergodic_r}]
{\resizebox{6.5cm}{!}{\includegraphics{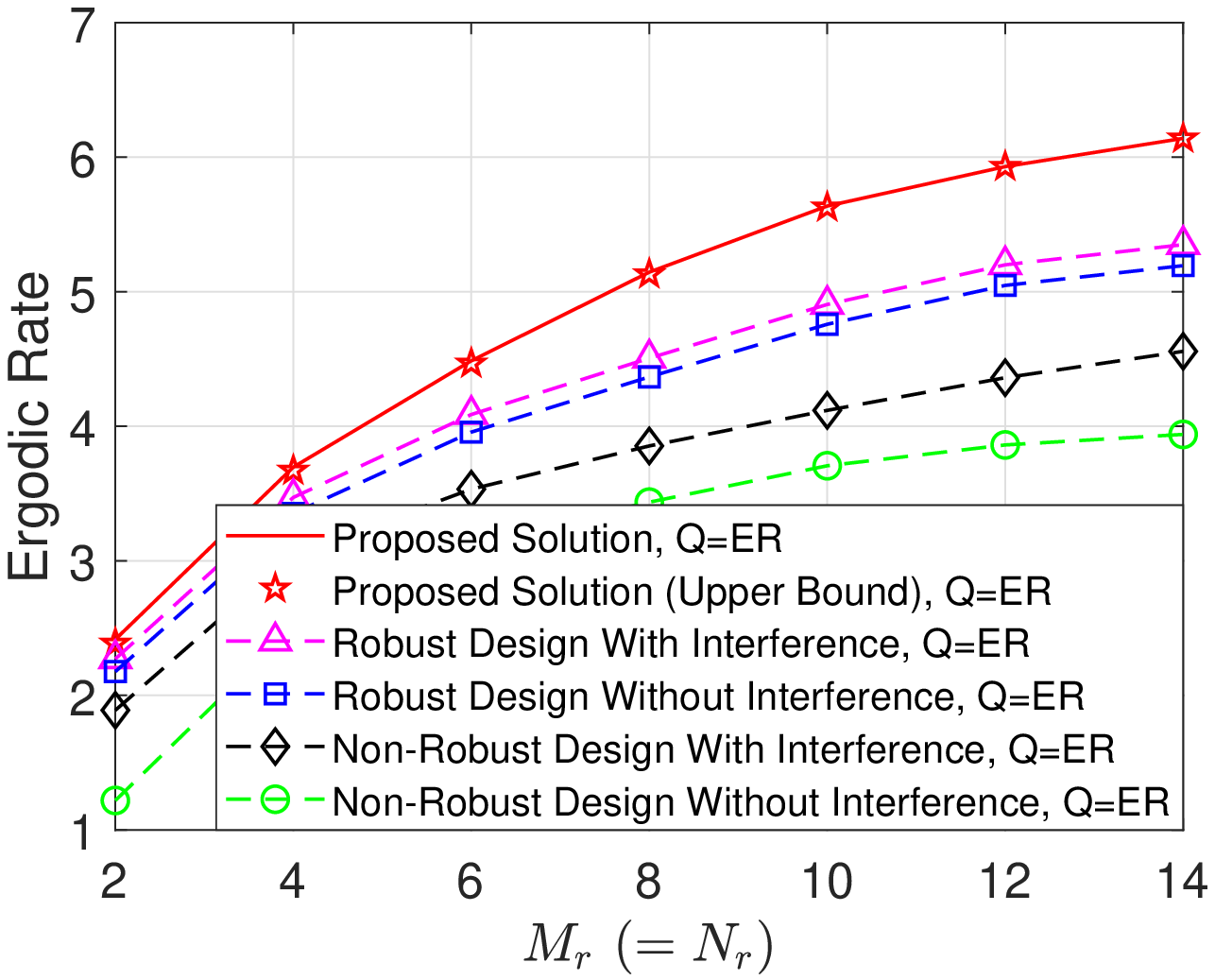}}}\quad
\subfigure[\scriptsize{Ergodic rate versus $K_{0,r}(=K_{r,0})$.
}\label{fig:Ergodic_K}]
{\resizebox{6.5cm}{!}{\includegraphics{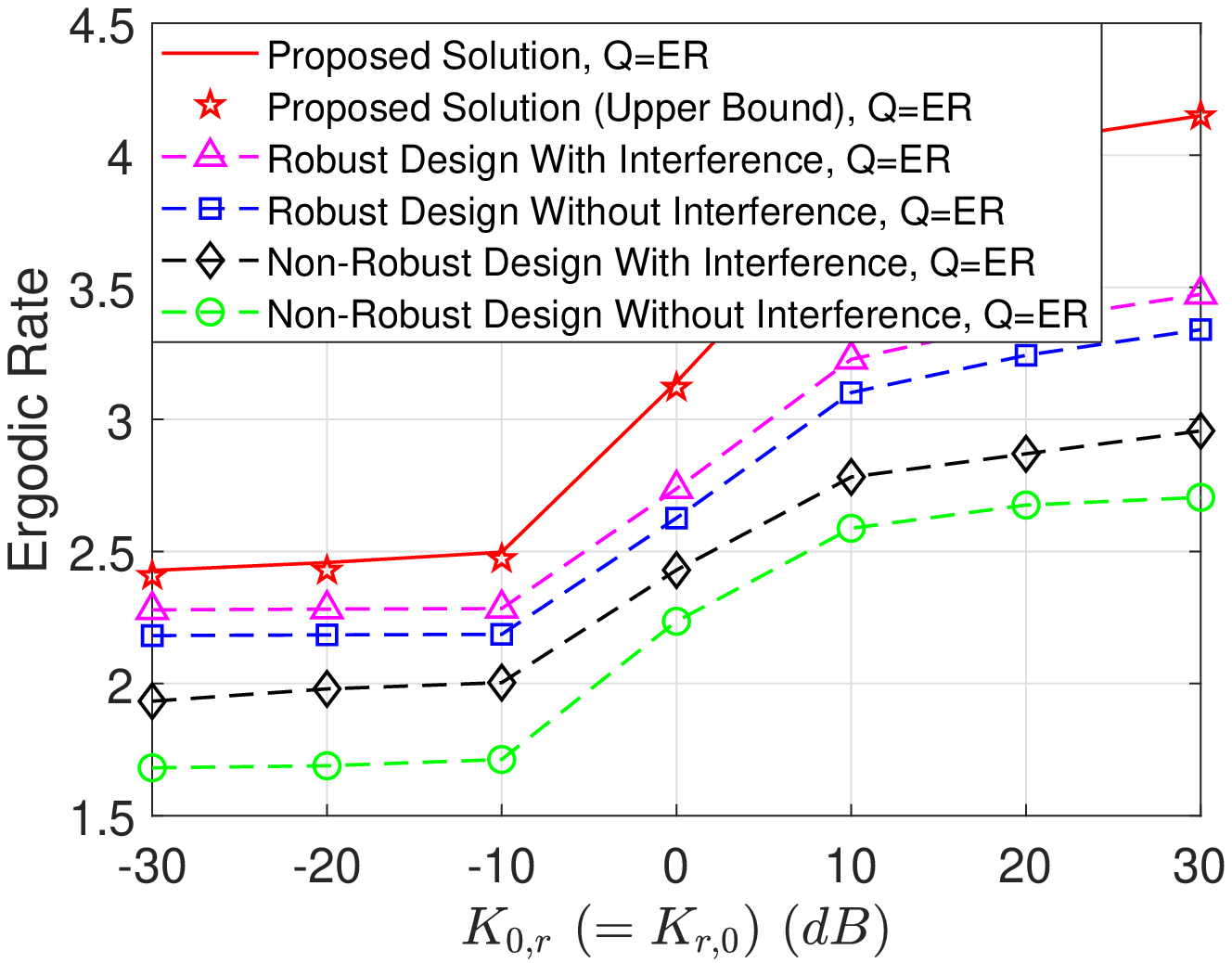}}}\quad
\subfigure[\scriptsize{Ergodic rate versus $\delta_1(=\delta_2)$.
}\label{fig:Ergodic_d_r}]
{\resizebox{6.5cm}{!}{\includegraphics{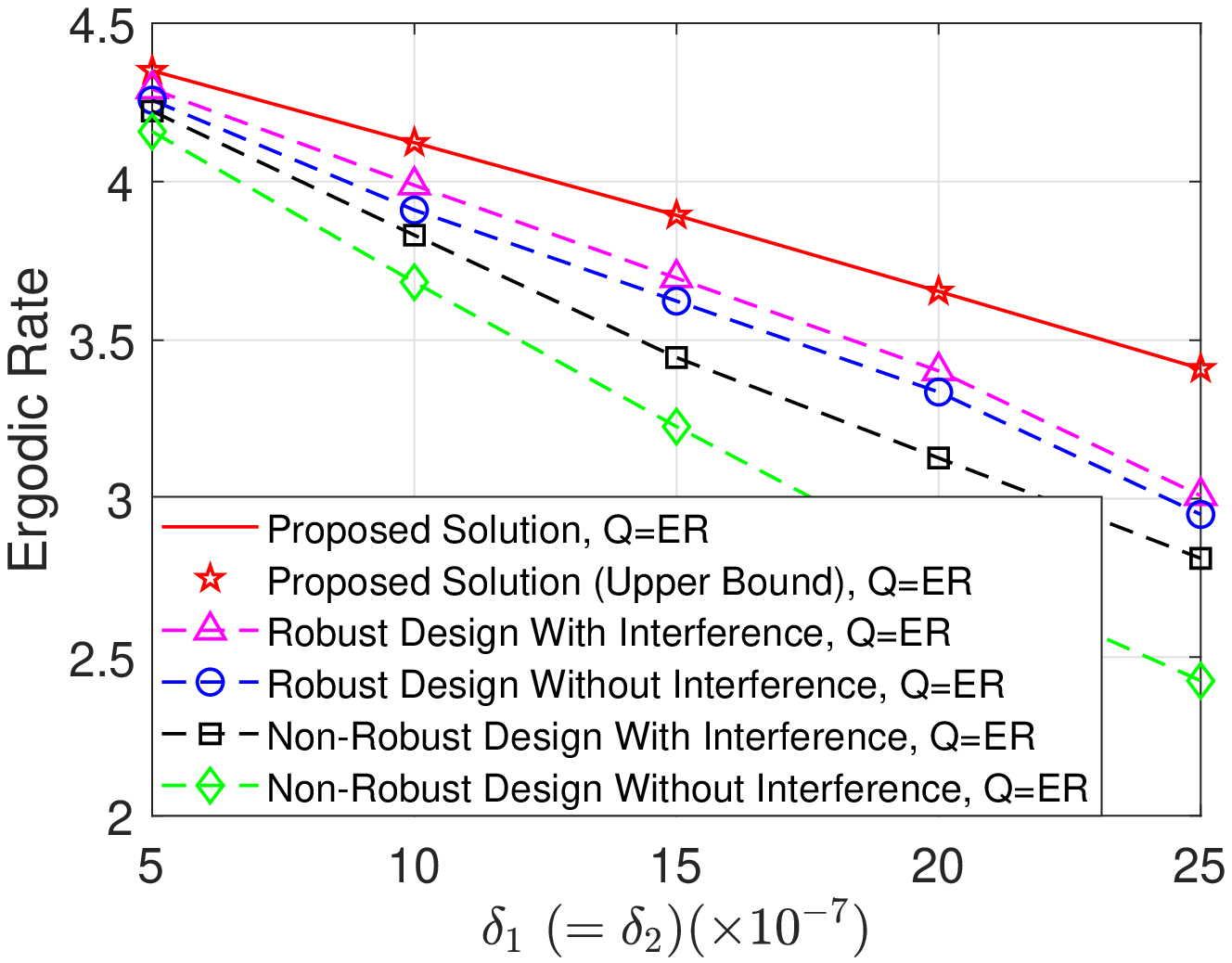}}}\quad
\subfigure[\scriptsize{Ergodic rate versus $d_{0,0}$.
}\label{fig:Ergodic_d_00}]
{\resizebox{6.5cm}{!}{\includegraphics{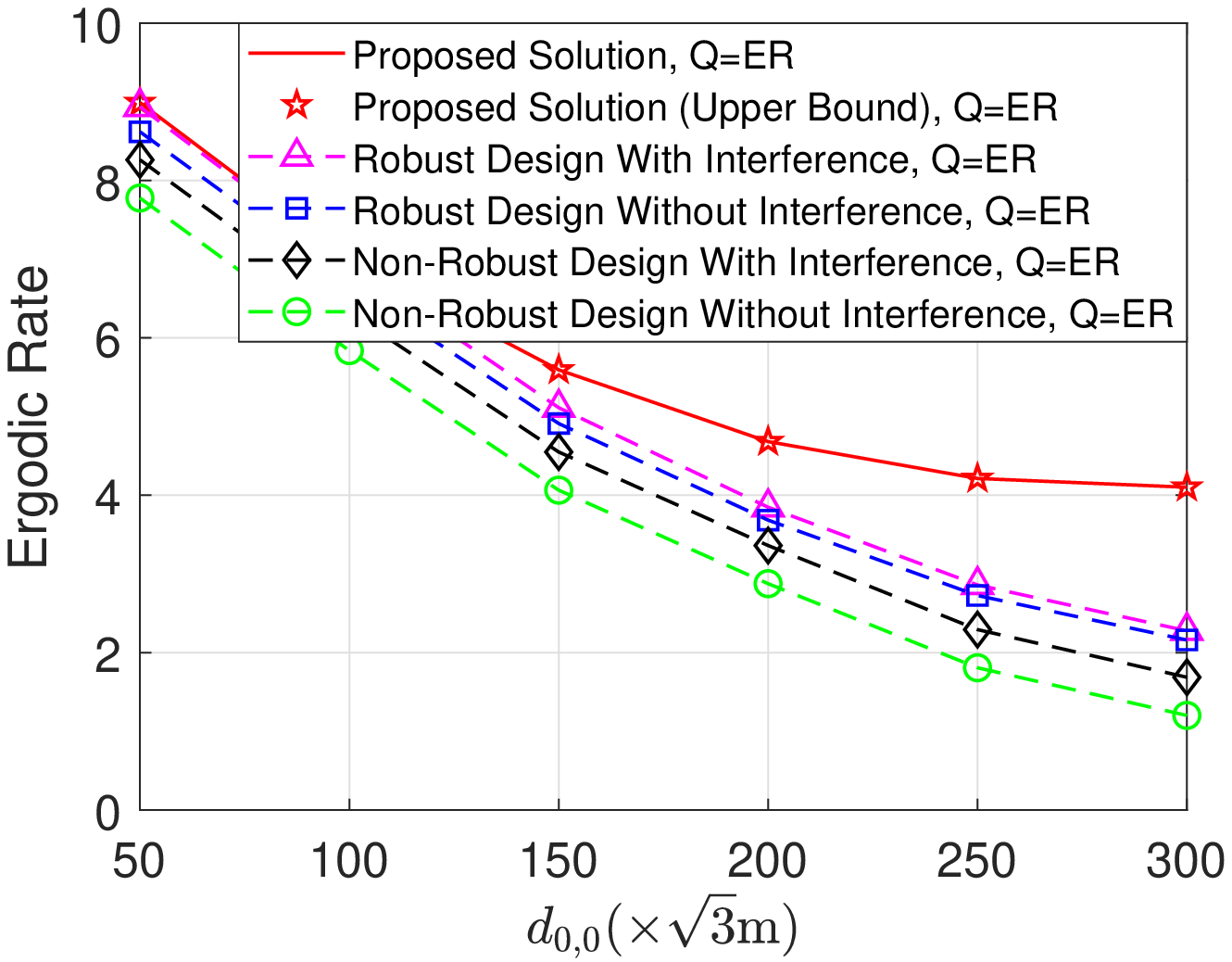}}}
\subfigure[\blue{\scriptsize{Ergodic rate versus $d_{r}$.}
}\label{fig:Ergodic_d_r_r}]
{\resizebox{6.5cm}{!}{\includegraphics{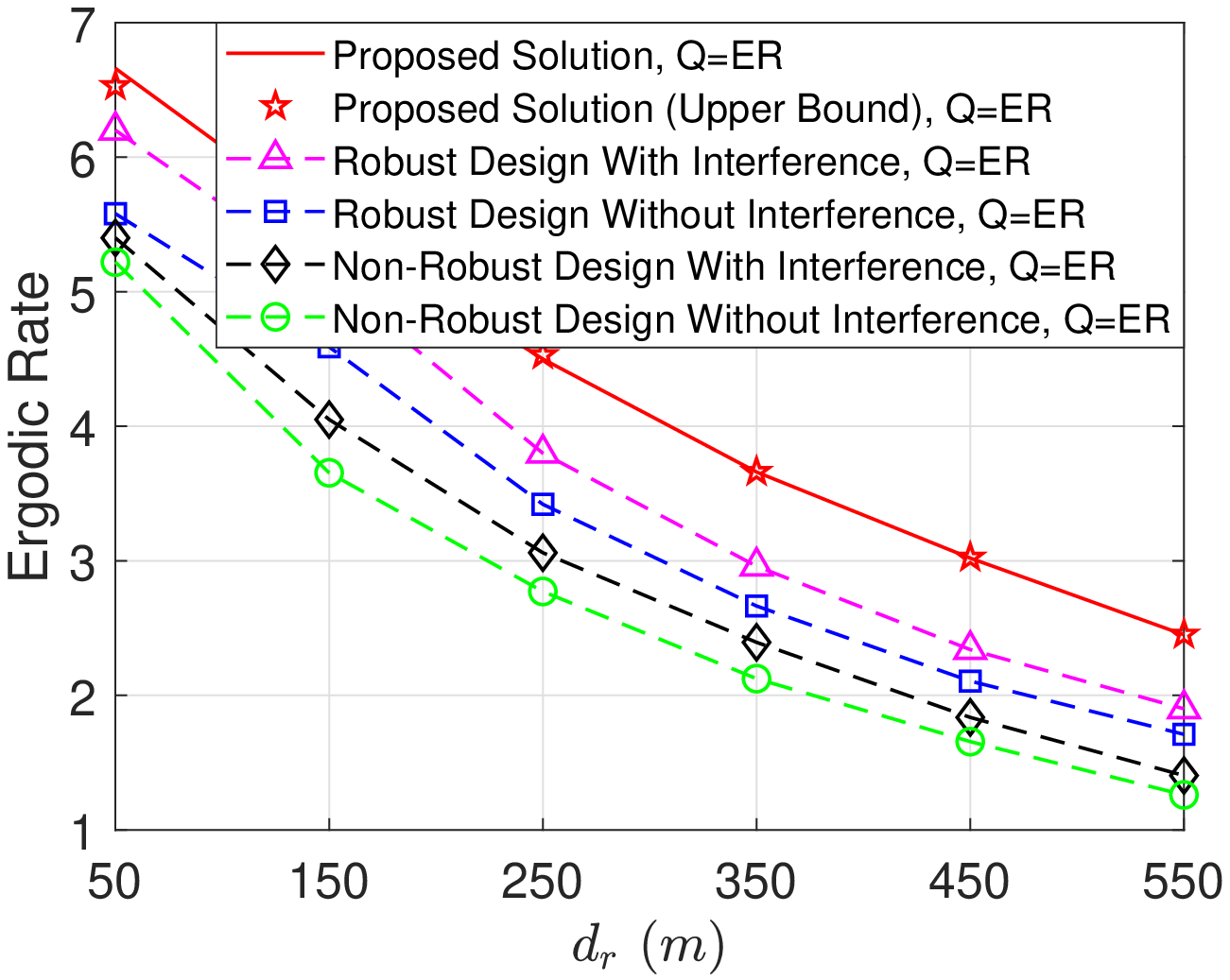}}}
\end{center}
\vspace{-2mm}
\caption{\small{Ergodic rate maximization.}}
\vspace{-2mm}
\label{fig:ergodic}
\end{figure}
\begin{figure}[t]
\begin{center}
\subfigure[\scriptsize{Average goodput versus $M_r(=N_r)$.
}\label{fig:nonErgodic_r}]
{\resizebox{6.5cm}{!}{\includegraphics{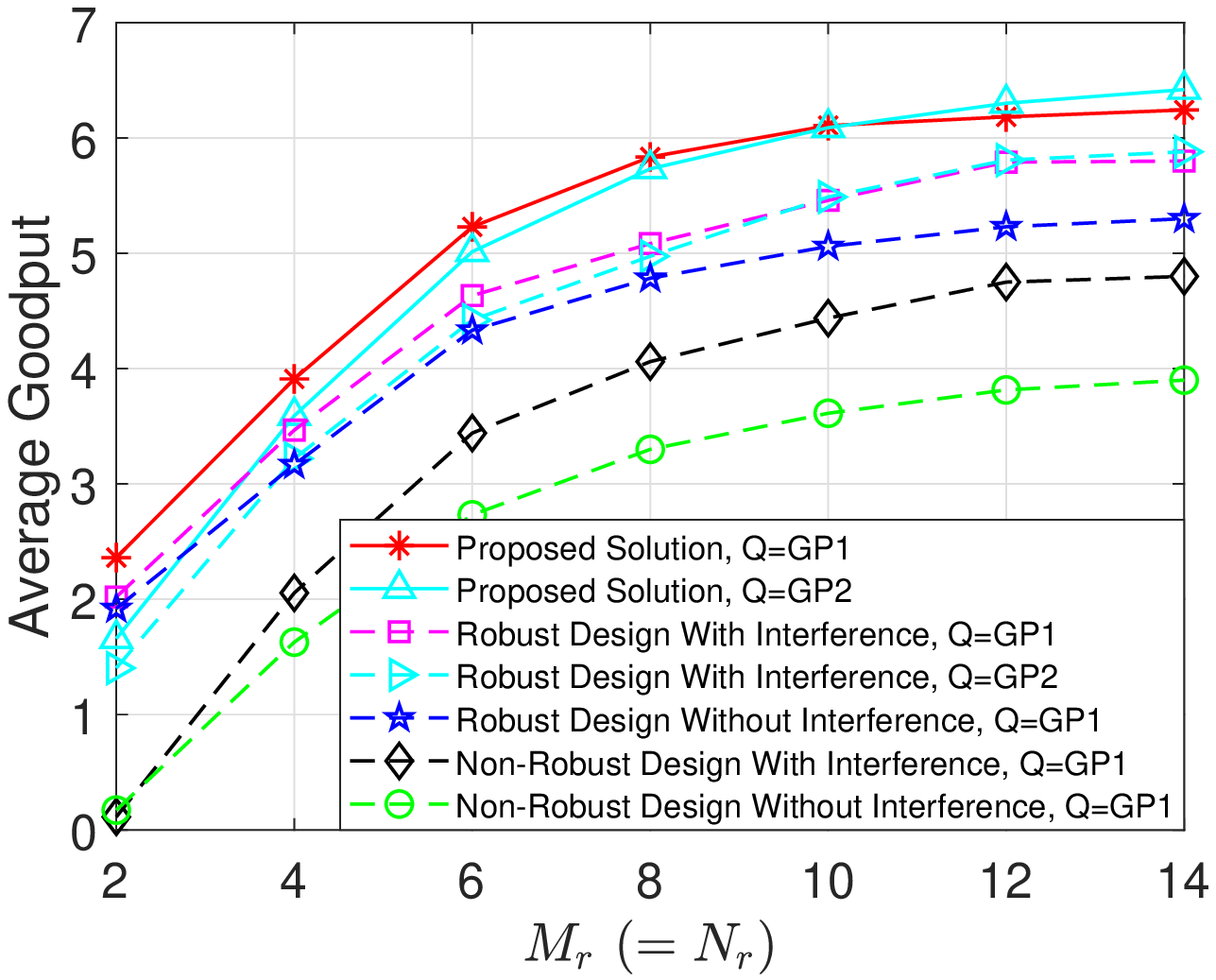}}}\quad
\subfigure[\scriptsize{Average goodput versus $K_{0,r}(=K_{r,0})$.
}\label{fig:nonErgodic_K}]
{\resizebox{6.5cm}{!}{\includegraphics{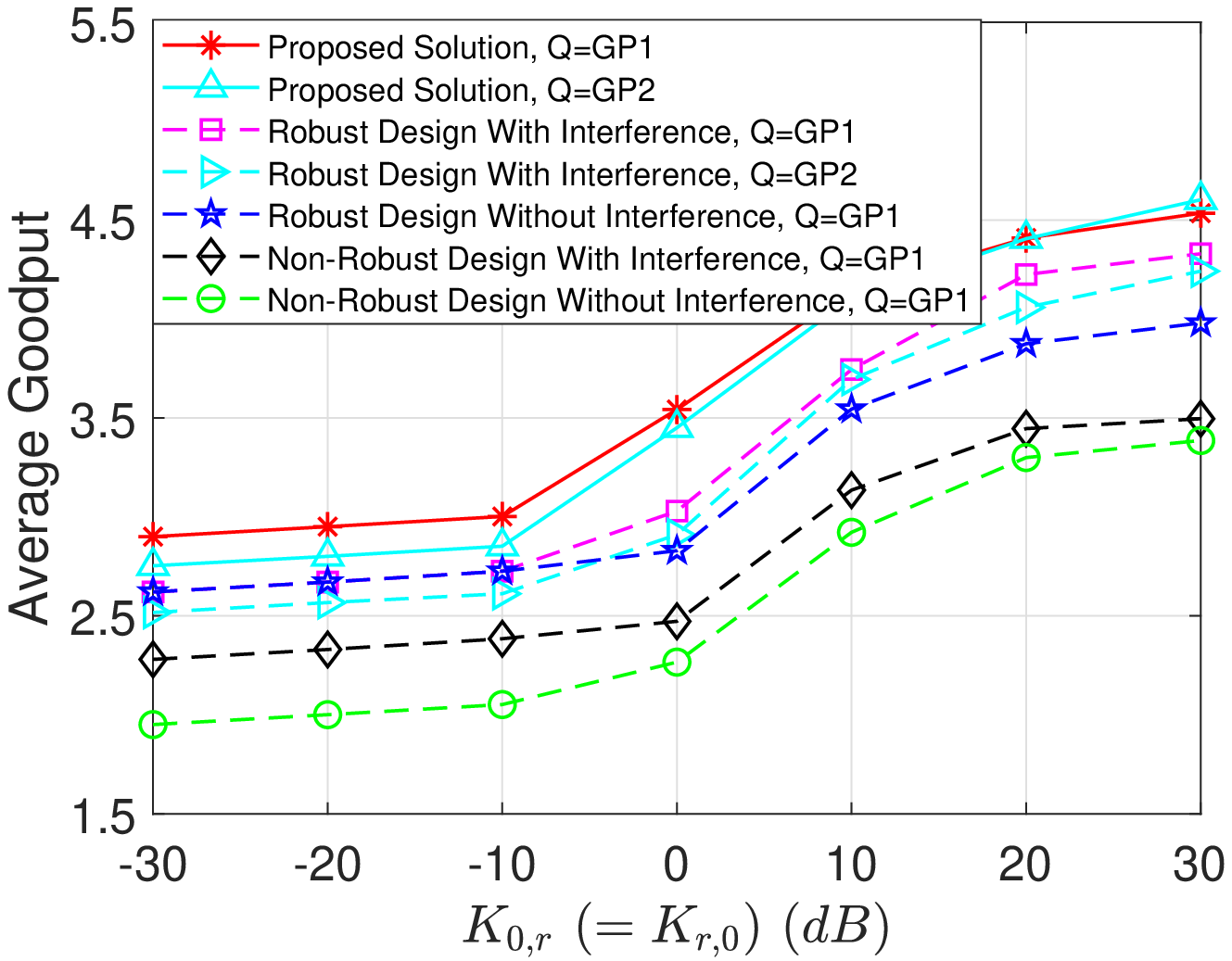}}}\quad
\subfigure[\scriptsize{Average goodput versus $\delta_1(=\delta_2)$.
}\label{fig:nonErgodic_d_r}]
{\resizebox{6.5cm}{!}{\includegraphics{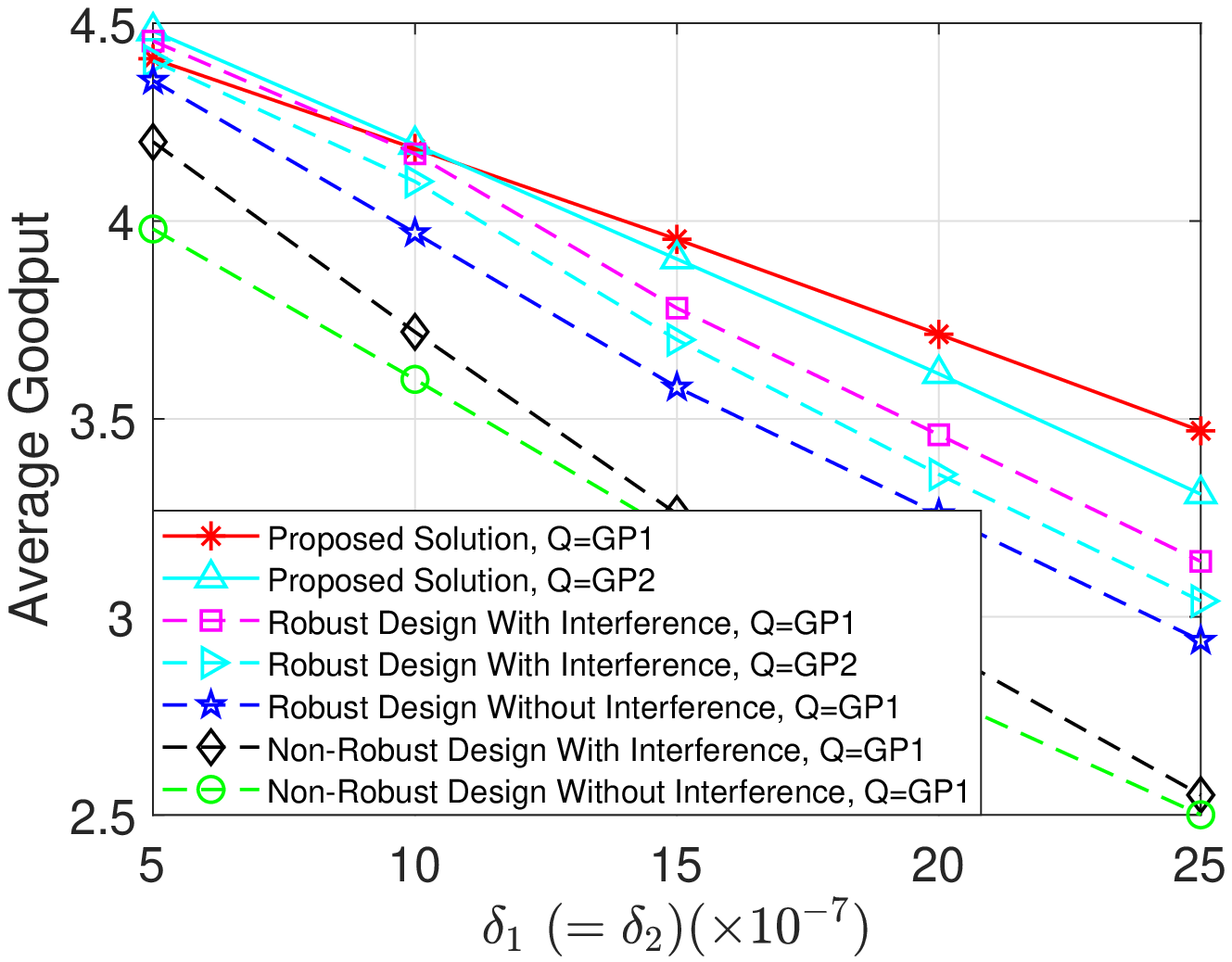}}}\quad
\subfigure[\scriptsize{Average goodput versus $d_{0,0}$.
}\label{fig:nonErgodic_d_00}]
{\resizebox{6.5cm}{!}{\includegraphics{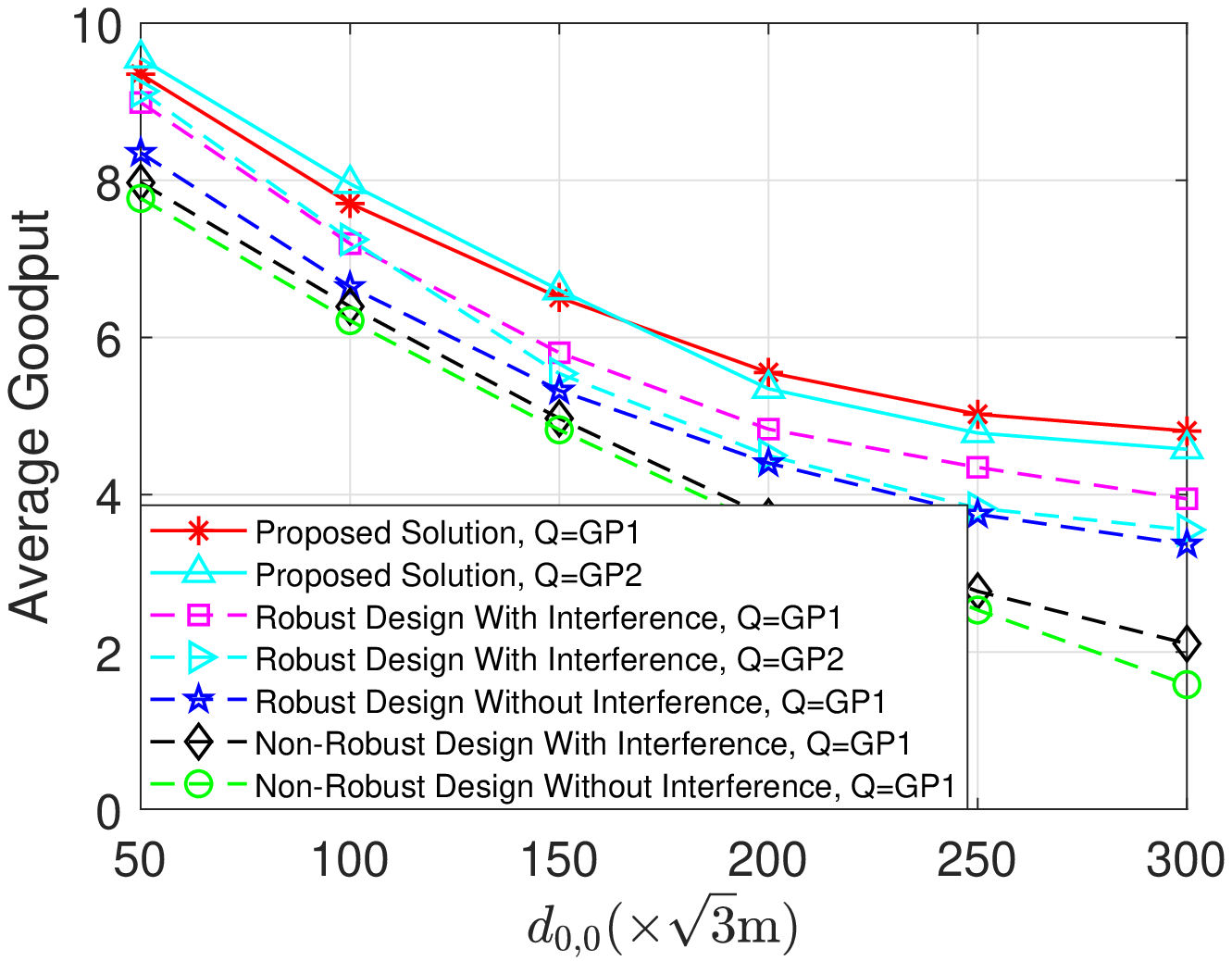}}}
\subfigure[\scriptsize{\blue{Average goodput versus $d_{r}$.}
}\label{fig:nonErgodic_d_r_r}]
{\resizebox{6.5cm}{!}{\includegraphics{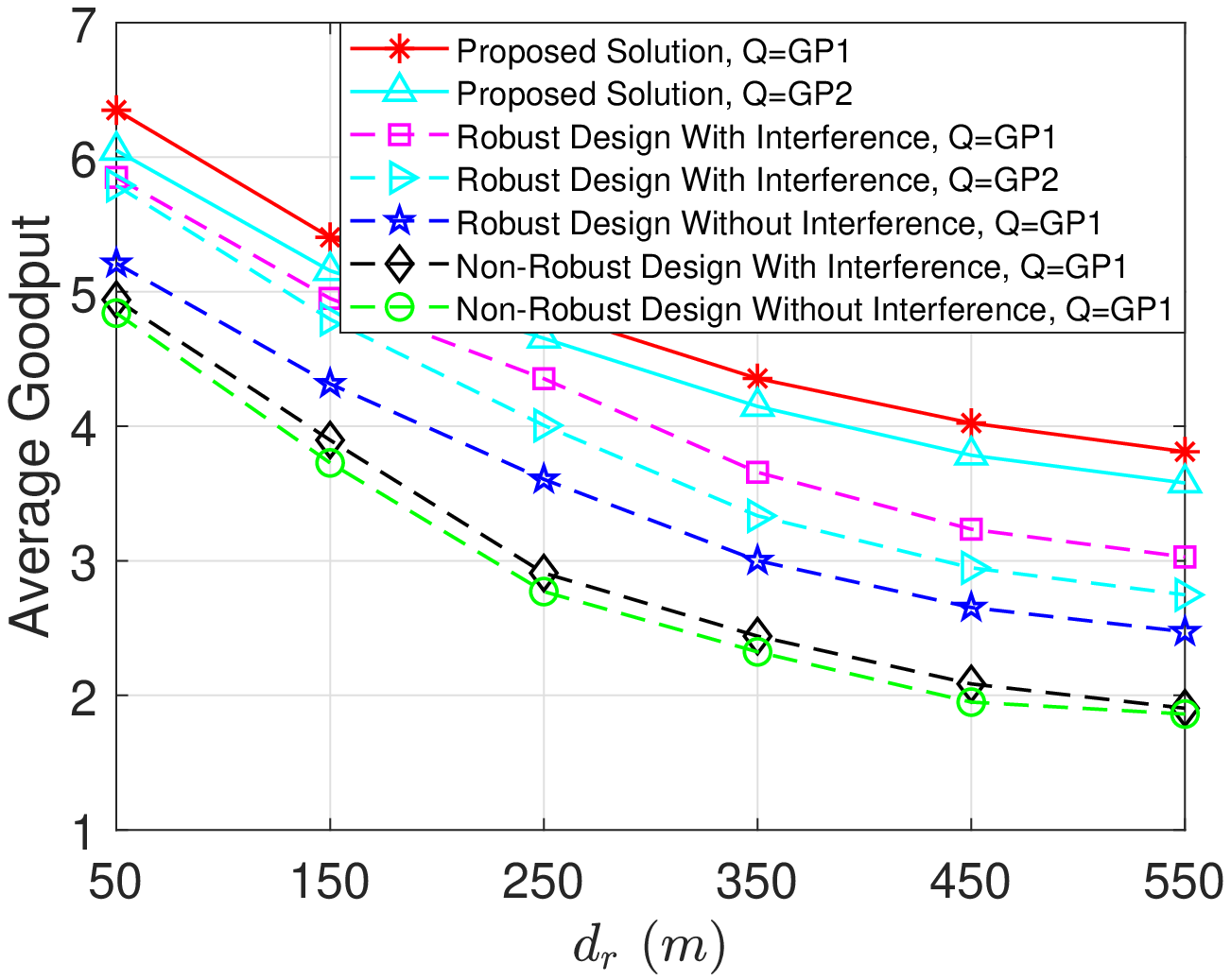}}}
\end{center}
\vspace{-2mm}
\caption{\small{Average goodput maximization.}}
\vspace{-2mm}
\label{fig:nonErgodic}
\end{figure}
In the scenario of ergodic rate maximization, we evaluate the ergodic rates of all schemes. Fig.~\ref{fig:ergodic} illustrates the ergodic rate versus $M_r$ $(=N_r)$, $K_{0,r}$ $(=K_{r,0})$, $\delta_1(=\delta_2)$, $d_{0,0}$, \blue{and $d_{r}$,}  respectively. Fig.~\ref{fig:ergodic} shows that \blue{$C^{(ER)}(\mathbf{v}^{\dag},\mathbf{w}^{\dag}_0)$ given by \eqref{eq:sinr} and its upper bound
$$
  \log_2\left(1+
 \frac{P_0\mathbb{E}\left[{g}_0
 ^{(ER)}\left(\mathbf{v}^{\dag},\mathbf{w}^{\dag}_0
\left(\hat{\mathbf{H}}_0\right),\hat{\mathbf{H}}_0\right)\right]}
{\sum\limits_{k \in \mathcal{K}\backslash \{0\}}P_k{g}_{k}\left(\mathbf{v}^{\dag}\right)+\sigma^2}\right)
$$}are very close to each other, indicating that the upper bound in Theorem~\ref{lem:ergodic Case 1 with reflector} is a good approximation of $C^{(ER)}(\mathbf{v},\mathbf{w}_0)$. In the scenario of average goodput maximization, we evaluate the \blue{average goodputs} of all schemes. Fig.~\ref{fig:nonErgodic} illustrates the average goodput versus $M_r$ $(=N_r)$, $K_{0,r}$ $(=K_{r,0})$, $\delta_1(=\delta_2)$, $d_{0,0}$, \blue{and $d_r$}, respectively.
Fig.~\ref{fig:nonErgodic} shows that the two proposed solutions for average goodput maximization have different preferable system parameters and complement each other to provide a higher average goodput.

In both scenarios, we can draw the following conclusions. \blue{The gain of the proposed solution over {\em Robust Design With Interference} stems from the joint design of instantaneous CSI-adaptive beamforming and quasi-static phase shift design; the} gain of the proposed solution over {\em Robust Design Without Interference} comes from the effective utilization of inter-cell interference; the gain of the proposed solution over {\em Non-Robust Design With Interference} comes from the effective utilization of imperfect CSIT; the gain of the proposed solution over {\em Non-Robust Design Without Interference} comes from the effective utilization of imperfect CSIT and inter-cell interference. The \blue{three} robust designs outperform the two non-robust designs, demonstrating the advantage of explicitly considering CSI \blue{estimation} errors.
\blue{The proposed solutions and {\em Robust Design With Interference} outperform {\em Robust Design Without Interference}, and {\em Non-Robust Design With Interference} outperforms {\em Non-Robust Design Without Interference},} indicating the advantage of explicitly considering inter-cell interference. Fig.~\ref{fig:Ergodic_r} and Fig.~\ref{fig:nonErgodic_r} show that the ergodic rate and average goodput of each scheme increase with $M_r$ $(=N_r)$, mainly due to the increment of reflecting signal power. Fig.~\ref{fig:Ergodic_K} and Fig.~\ref{fig:nonErgodic_K} show that the ergodic rate and average goodput of each scheme increase with $K_{0,r}$ $(=K_{r,0})$, mainly due to the increment of the channel power of each LoS component. Fig.~\ref{fig:Ergodic_d_r} and Fig.~\ref{fig:nonErgodic_d_r} show that the ergodic rate and average goodput of each scheme decrease with $\delta_1(=\delta_2)$, due to the increment of channel estimation error, and the ergodic rates and average goodputs of the two non-robust designs decrease faster than those of the two robust ones. Fig.~\ref{fig:Ergodic_d_00}\blue{,} Fig.~\ref{fig:nonErgodic_d_00}\blue{, Fig.~\ref{fig:Ergodic_d_r_r}, and Fig.~\ref{fig:nonErgodic_d_r_r}} show that the ergodic rate and average goodput of each scheme decrease with $d_{0,0}$ \blue{and $d_r$}, mainly due to the increment of interference.
\section{Conclusion}
This paper investigated the robust optimization of instantaneous CSI-adaptive beamforming and quasi-static phase shifts in an IRS-assisted system with inter-cell interference. In the scenario of coding over many slots, we considered the robust maximization of the user's ergodic rate. In the scenario of coding within each slot, we considered the robust optimization of the user's average goodput under \blue{the} successful transmission probability constraints. In both scenarios, we obtained closed-form robust \blue{instantaneous CSI-adaptive} beamforming designs that can promptly adapt to \blue{fast CSI} changes \blue{over slots} and robust quasi-static phase shift designs of low computation and phase adjustment costs. Numerical results further demonstrate notable gains of the proposed robust joint designs over existing designs.
\section*{Appendix A: Proof of Theorem~\ref{lem:ergodic Case 1 with reflector}}\label{app:theorem 1}
By Jensen's inequality from \eqref{eq:sinr}, we have: $$ C^{(ER)}\left(\mathbf{v},\mathbf{w}_0\right)\leq\log_2\left(1+
\frac{{P_0}\mathbb{E}_{\hat{\mathbf{H}}_0,\Delta\mathbf{H}_0}\left[\left\lvert\left(\mathbf{v}^H\mathbf{G}_{0,0}+ \mathbf{h}^H_{0,0}\right)\mathbf{w}_{0}\left(\hat{\mathbf{H}}_0\right)  \right\rvert^2\right]}{\sum\limits_{k \in \mathcal{K}\backslash\{0\}} {P_{k}}\mathbb{E} \left[ {\left\lvert \left(\mathbf{v}^H\mathbf{G}_{k,0} + \mathbf{h}^H_{k,0}\right)  \frac{{\mathbf{h}}_{k,k}}{\left\lVert {\mathbf{h}}_{k,k} \right\rVert_2}\right\rvert}^2 \right]+ {\sigma}^2}\right).$$
Following the proof in Appendix B of \cite{YuhangJia}, we have: \blue{\begin{align*} &\mathbb{E}\left[\left\lvert\left(\mathbf{v}^H\mathbf{G}_{0,0}+ \mathbf{h}^H_{0,0}\right)\mathbf{w}_{0}\left(\hat{\mathbf{H}}_0\right)  \right\rvert^2\right] \\
= &\mathbb{E}\left[\mathbb{E}\left[\left\lvert\left(
\mathbf{v}^H\hat{\mathbf{G}}_{0,0}+\hat{\mathbf{h}}^H_{0,0}\right)\mathbf{w}_0\left(\hat{\mathbf{H}}_0\right)
+\mathbf{v}^H\Delta{\mathbf{G}}_{0,0}\mathbf{w}_0\left(\hat{\mathbf{H}}_0\right)
+\Delta{\mathbf{h}}^H_{0,0}\mathbf{w}_0\left(\hat{\mathbf{H}}_0\right)\right\rvert^2
\bigg|\hat{\mathbf{H}}_0\right]\right]\\
\eqla & \mathbb{E}\left[\left\lvert\left(\mathbf{v}^H\hat{\mathbf{G}}_{0,0}+\hat{\mathbf{h}}^H_{0,0}\right)
\mathbf{w}_{0}\left(\hat{\mathbf{H}}_0\right)\right\rvert^2
+ \mathbb{E}\left[\left\lvert\mathbf{v}^H\Delta{\mathbf{G}}_{0,0}\mathbf{w}_{0}\left(\hat{\mathbf{H}}_0\right)  \right\rvert^2\bigg|\hat{\mathbf{H}}_0\right] \right.\\ & \left.+
\mathbb{E}\left[\left\lvert\Delta{\mathbf{h}}^H_{0,0}
\mathbf{w}_{0}\left(\hat{\mathbf{H}}_0\right)\right\rvert^2
\bigg|\hat{\mathbf{H}}_0\right]\right] \\ = & \mathbb{E}\bigg[\left\lvert\left(\mathbf{v}^H\hat{\mathbf{G}}_{0,0}+\hat{\mathbf{h}}^H_{0,0}\right)
\mathbf{w}_{0}\left(\hat{\mathbf{H}}_0\right)\right\rvert^2
+\mathbf{w}_{0}^H\left(\hat{\mathbf{H}}_0\right)\mathbb{E}\left[
\Delta{\mathbf{G}}^H_{0,0}\mathbf{v}\mathbf{v}^H
\Delta{\mathbf{G}}_{0,0}\right]\mathbf{w}_{0}\left(\hat{\mathbf{H}}_0\right) \\ & +
\mathbf{w}_{0}^H\left(\hat{\mathbf{H}}_0\right)
 \mathbb{E}\left[\Delta{\mathbf{h}}_{0,0}\Delta{\mathbf{h}}^H_{0,0}\right]
\mathbf{w}_{0}\left(\hat{\mathbf{H}}_0\right)
\bigg]
\\ \eqlb &\mathbb{E}\left[{g}_0^{(ER)}\left(\mathbf{v},\mathbf{w}_0
\left(\hat{\mathbf{H}}_0\right),\hat{\mathbf{H}}_0\right)\right],
\end{align*}
where $(a)$ is due to the independence between $\Delta{\mathbf{G}}_{0,0}$ and $\Delta\mathbf{h}_{0,0}$, $\mathbb{E}[\Delta\mathbf{G}_{0,0}]=0$, and $\mathbb{E}[\Delta\mathbf{h}_{0,0}]=0$, $(b)$ is due to the fact that $\mathbb{E}\left[
\Delta{\mathbf{G}}^H_{0,0}\mathbf{v}\mathbf{v}^H
\Delta{\mathbf{G}}_{0,0}\right] = M_rN_r\delta_1^2\mathbf{I}_{M_0N_0}$ (as the elements of $\mathbf{v}^H\Delta\mathbf{G}_{0,0}$ are i.i.d. according to $\mathcal{CN}(0,M_rN_r\delta_1^2)$), $\mathbb{E}\left[\Delta{\mathbf{h}}_{0,0}\Delta{\mathbf{h}}^H_{0,0}\right]
=\delta_2^2\mathbf{I}_{M_0N_0}$ (as the elements of $\Delta{\mathbf{h}}_{0,0}$ are i.i.d. according to $\mathcal{CN}(0,\delta_2^2)$), and $\left\lVert
\mathbf{w}_{0}\left(\hat{\mathbf{H}}_0\right)\right\rVert_2^2=1$. Similarly, we have:
\begin{align*}
  & \mathbb{E} \left[ {\left\lvert \left(\mathbf{v}^H\mathbf{G}_{k,0} + \mathbf{h}^H_{k,0}\right)  \frac{{\mathbf{h}}_{k,k}}{\left\lVert {\mathbf{h}}_{k,k} \right\rVert_2}\right\rvert}^2 \right]
=  \mathbb{E} \left[ \left\lvert \left(\mathbf{v}^H\mathbf{G}_{k,0} + \mathbf{h}^H_{k,0}\right)  \frac{{\mathbf{h}}_{k,k}
  {\mathbf{h}}_{k,k}^H}{\left\lVert {\mathbf{h}}_{k,k} \right\rVert^2_2} \left(\mathbf{v}^H\mathbf{G}_{k,0} + \mathbf{h}^H_{k,0}\right)^H\right\rvert \right] \\
  \eqla &\mathbb{E} \left[ \left\lvert \left(\mathbf{v}^H\mathbf{G}_{k,0} + \mathbf{h}^H_{k,0}\right) \mathbb{E}\left[ \frac{{\mathbf{h}}_{k,k}
  {\mathbf{h}}_{k,k}^H}{\left\lVert {\mathbf{h}}_{k,k} \right\rVert^2_2}\right] \left(\mathbf{v}^H\mathbf{G}_{k,0} + \mathbf{h}^H_{k,0}\right)^H\right\rvert \right] \\ \eqlb & \frac{1}{M_kN_k}\mathbb{E} \left[\left\lVert \mathbf{v}^H\mathbf{G}_{k,0} + \mathbf{h}^H_{k,0}\right\rVert_2^2 \right]
 \eqlc \frac{1}{M_kN_k} \left(\mathbb{E} \left[\left\lVert \mathbf{v}^H\mathbf{G}_{k,0}\right\rVert_2^2 \right]+2\mathbf{v}^H\mathbb{E} \left[ \mathbf{G}_{k,0}\right] \mathbb{E} \left[\mathbf{h}_{k,0} \right]+\mathbb{E} \left[\left\lVert \mathbf{h}^H_{k,0}\right\rVert_2^2 \right] \right) \\
  \eqld & \frac{1}{M_kN_k} \left(\mathbb{E} \left[\left\lVert \mathbf{v}^H\mathbf{G}_{k,0}\right\rVert_2^2 \right]+\alpha_{k,0}\right)
  \end{align*}
  \begin{align*}
    \eqle & \frac{1}{M_kN_k} \left(\left\lVert \mathbf{v}^H\bar{\mathbf{G}}_{k,0}\right\rVert_2^2 + \alpha_{r,0}\alpha_{k,r}\left( \frac{K_{r,0}\mathbb{E}\left[\left\lVert\mathbf{v}^H\text{diag}
\left(\bar{\mathbf{h}}_{r,0}^H\right)\tilde{\mathbf{H}}_{k,r}\right\rVert_2^2\right]
}{(K_{k,r}+1)(K_{r,0}+1)}\right.\right. \\ & \left.\left. +
\frac{K_{k,r}\mathbb{E}\left[\left\lVert\mathbf{v}^H\text{diag}
\left(\tilde{\mathbf{h}}_{r,0}^H\right)\bar{\mathbf{H}}_{k,r}\right\rVert_2^2\right]
}{(K_{k,r}+1)(K_{r,0}+1)}+
\frac{\mathbb{E} \left[\left\lVert \mathbf{v}^H\text{diag}
\left(\tilde{\mathbf{h}}_{r,0}^H\right)\tilde{\mathbf{H}}_{k,r}\right\rVert_2^2 \right]}{(K_{k,r}+1)(K_{r,0}+1)} + \alpha_{k,0}\right) \right)
\\
= & \frac{1}{M_kN_k} \left(\left\lVert \mathbf{v}^H\bar{\mathbf{G}}_{k,0}\right\rVert_2^2 + \alpha_{r,0}\alpha_{k,r}\left(   \frac{K_{r,0}
\mathbf{v}^H\text{diag}
\left(\bar{\mathbf{h}}_{r,0}^H\right)
\mathbb{E}\left[\tilde{\mathbf{H}}_{k,r}\tilde{\mathbf{H}}^H_{k,r}\right]
\text{diag}\left(\bar{\mathbf{h}}_{r,0}\right)\mathbf{v}
}{(K_{k,r}+1)(K_{r,0}+1)} \right. \right. \\ & \left. \left. + \frac{\mathbf{v}^H\mathbb{E} \left[\text{diag}
\left(\tilde{\mathbf{h}}_{r,0}^H\right)\bar{\mathbf{H}}_{k,r}
\bar{\mathbf{H}}_{k,r}^H\text{diag}
\left(\tilde{\mathbf{h}}_{r,0}\right)\right]\mathbf{v}}
{(K_{k,r}+1)(K_{r,0}+1)}+ \frac{\mathbf{v}^H\mathbb{E} \left[\text{diag}
\left(\tilde{\mathbf{h}}_{r,0}^H\right)\tilde{\mathbf{H}}_{k,r}
\tilde{\mathbf{H}}_{k,r}^H\text{diag}
\left(\tilde{\mathbf{h}}_{r,0}\right)\right]\mathbf{v}}
{(K_{k,r}+1)(K_{r,0}+1)} \right)\right)  \\
 \eqlf &
  {g}_{k}\left(\mathbf{v}\right), k \in \mathcal{K}\backslash\{0\},
\end{align*}
where $(a)$ is due to the fact that $\mathbf{h}_{k,k}$ is independent from $\mathbf{h}_{k,0}$  and $\mathbf{G}_{k,0}$, $(b)$ is due to the fact that  $\frac{{\mathbf{h}}_{k,k}}{\left\lVert {\mathbf{h}}_{k,k} \right\rVert_2}$ follows the uniform spherical distribution and hence satisfies $\mathbb{E}\left[ \frac{{\mathbf{h}}_{k,k}
  {\mathbf{h}}_{k,k}^H}{\left\lVert {\mathbf{h}}_{k,k} \right\rVert^2_2}\right] = \frac{1}{M_kN_k}\mathbf{I}_{M_kN_k}$, $(c)$ is
  due to the independence between $\mathbf{G}_{k,0}$ and $\mathbf{h}_{k,0}$, $(d)$ is due to $\mathbb{E} \left[\mathbf{G}_{k,0}\right]=0$ and $\mathbb{E} \left[\mathbf{h}_{k,0} \right] = 0$, $(e)$ is due to \eqref{eq:observedG}, and $(f)$ is due to the fact that the elements of $\tilde{\mathbf{h}}_{k,r}$  and $\tilde{\mathbf{h}}_{r,0}$ are i.i.d. according to $\mathcal{CN}(0,1)$.} Thus, we complete the proof of Theorem~\ref{lem:ergodic Case 1 with reflector}.
\section*{Appendix B: Proof of Theorem~\ref{theorem:eq}}\label{app:theorem 2}
First, we equivalently transform Problem~\ref{prob:approximation} to the following problem:
\begin{equation}\label{prob:ins}
\begin{split}
\mathop{\max}_{\mathbf{v}}\quad &
\frac{\mathbb{E}\left[
 \gamma_{0}^{(ER)*}\left(\mathbf{v}, \hat{\mathbf{H}}_0\right)
\right]}{\sum\limits_{k \in \mathcal{K}\backslash \{0\}}P_k{g}_{k}\left(\mathbf{v}\right)+\sigma^2} \\
s.t. \quad
& \eqref{eq:phi},
\end{split}
\end{equation}
where
\begin{equation}\label{prob:ins_w}
\begin{split}
\gamma_{0}^{(ER)*}\left(\mathbf{v}, \hat{\mathbf{H}}_0\right)\triangleq \mathop{\max}_{\mathbf{w}_0\left(\hat{\mathbf{H}}_0\right)} \quad & {g}_0^{(ER)}\left(\mathbf{v},\mathbf{w}_0\left(\hat{\mathbf{H}}_0\right),\hat{\mathbf{H}}_0\right) \\
s.t. \quad & \eqref{eq:w}.
\end{split}
\end{equation}
Next, by Cauchy-Schwartz inequality, we can show that $\mathbf{w}_{0}^{(ER)*}\left(\mathbf{v},
\hat{\mathbf{H}}_0\right)$ in \eqref{eq:wequivalence} is an optimal solution of the problem in \eqref{prob:ins_w} and $\gamma_{0}^{(ER)*}\left(\mathbf{v}, \hat{\mathbf{H}}_0\right)={g}_0^{(ER)}\left(\mathbf{v},\mathbf{w}_{0}^{(ER)*}\left(\mathbf{v},
\hat{\mathbf{H}}_0\right),\hat{\mathbf{H}}_0\right)$. Substituting $\gamma_{0}^{(ER)*}\left(\mathbf{v}, \hat{\mathbf{H}}_0\right)$ into the objective function of the problem in \eqref{prob:ins}, we can obtain Problem~\ref{prob:theta}.
Therefore, we complete the proof of Theorem~\ref{theorem:eq}.
\section*{Appendix C: Proof of Lemma~\ref{lem:lemma1}}\label{app:lemma 1}
By total probability theorem, we have: \begin{align}
\begin{split}
&\text{Pr}\left[C\left(\mathbf{v},
\mathbf{w}_0\left(
\hat{\mathbf{H}}_0\right),\hat{\mathbf{H}}_0,\Delta{\mathbf{H}}_0\right)
\geq r\left(\hat{\mathbf{H}}_0\right)\right] \\
=&\text{Pr}\left[\Delta\mathbf{H}_0\in \mathcal{E}\right]\text{Pr}\left[C\left(\mathbf{v},
\mathbf{w}_0\left(
\hat{\mathbf{H}}_0\right),\hat{\mathbf{H}}_0,\Delta{\mathbf{H}}_0\right)
\geq r\left(\hat{\mathbf{H}}_0\right) \big|\Delta\mathbf{H}_0\in \mathcal{E}\right]\\& +
\left(1-\text{Pr}\left[\Delta\mathbf{H}_0\in \mathcal{E}\right]\right)\text{Pr}\left[C\left(\mathbf{v},
\mathbf{w}_0\left(
\hat{\mathbf{H}}_0\right),\hat{\mathbf{H}}_0,\Delta{\mathbf{H}}_0\right)
\geq r\left(\hat{\mathbf{H}}_0\right) \big|\Delta\mathbf{H}_0\notin \mathcal{E}\right].
\end{split}\label{eq:preq}
\end{align}
By \eqref{eq:proconstraints} and \eqref{eq:preq}, we have:
\begin{align}
&\text{Pr}\left[C\left(\mathbf{v},
\mathbf{w}_0\left(
\hat{\mathbf{H}}_0\right),\hat{\mathbf{H}}_0,\Delta{\mathbf{H}}_0\right)
\geq r\left(\hat{\mathbf{H}}_0\right) \big|\Delta\mathbf{H}_0\in \mathcal{E}\right]\nonumber \\ \geq& \frac{\rho}{\text{Pr}\left[\Delta\mathbf{H}_0\in \mathcal{E}\right]}-\frac{1-\text{Pr}\left[\Delta\mathbf{H}_0\in \mathcal{E}\right]}{\text{Pr}\left[\Delta\mathbf{H}_0\in \mathcal{E}\right]}\text{Pr}\left[C\left(\mathbf{v},
\mathbf{w}_0\left(
\hat{\mathbf{H}}_0\right),\hat{\mathbf{H}}_0,\Delta{\mathbf{H}}_0\right)
\geq r\left(\hat{\mathbf{H}}_0\right) \big|\Delta\mathbf{H}_0\notin \mathcal{E}\right] \nonumber \\
\geq& \frac{\rho}{\text{Pr}\left[\Delta\mathbf{H}_0\in \mathcal{E}\right]}.
\label{eq:preqapp}
\end{align}
In addition, by \eqref{eq:bound}, we have:
\begin{align}
\text{Pr}\left[\Delta\mathbf{H}_0\in \mathcal{E}\right]=
F_{2M_0N_0M_rN_r}\left(\frac{2\varepsilon_1^2}{\delta_1^2}\right)=\rho.\label{eq:rho}
\end{align}
By \eqref{eq:preqapp} and \eqref{eq:rho}, we
can show Lemma~\ref{lem:lemma1}.
\section*{Appendix D: Proof of Lemma~\ref{theorem:GPeq}}\label{app:lemma 2}
Note that \eqref{eq:probability} is equivalent to:
\begin{align}
r\left(\hat{\mathbf{H}}_0\right)=
\mathop{\min}\limits_{\Delta\mathbf{H}_0\in\mathcal{E}}
C\left(\mathbf{v},\mathbf{w}_0
\left(\hat{\mathbf{H}}_0\right),\hat{\mathbf{H}}_0
,\Delta\mathbf{H}_0\right), \quad \hat{\mathbf{H}}_0 \in \mathbb{C}^{M_0N_0\times(M_rN_r+1)}.\label{eq::probmin}
\end{align}
In addition, we have:
\begin{align}
&\left\lvert\left(\mathbf{v}^H\mathbf{G}_{0,0}
+\mathbf{h}_{0,0}^H\right)\mathbf{w}_0
\left(\hat{\mathbf{H}}_0\right)\right\rvert
=\left\lvert\left(\mathbf{v}^H\left(\hat{\mathbf{G}}_{0,0}+
\Delta{\mathbf{G}}_{0,0}\right)
+\hat{\mathbf{h}}_{0,0}^H\right)\mathbf{w}_0
\left(\hat{\mathbf{H}}_0\right)+\Delta{\mathbf{h}}_{0,0}^H\mathbf{w}_0
\left(\hat{\mathbf{H}}_0\right)\right\rvert \nonumber \\
\overset{\text{(a)}}{\geq}& \left\lvert\left(\mathbf{v}^H\left(\hat{\mathbf{G}}_{0,0}+
\Delta{\mathbf{G}}_{0,0}\right)
+\hat{\mathbf{h}}_{0,0}^H\right)\mathbf{w}_0
\left(\hat{\mathbf{H}}_0\right)\right\rvert-\varepsilon_{2} \nonumber \\
\overset{\text{(b)}}{\geq} & \left\lvert\left(\mathbf{v}^H
\hat{\mathbf{G}}_{0,0}+\hat{\mathbf{h}}_{0,0}^H\right)
\mathbf{w}_0\left(\hat{\mathbf{H}}_0\right)\right\rvert
-\varepsilon_{1}\sqrt{M_rN_r}-\varepsilon_{2}, \label{ineq:b}
\end{align}
where $(a)$ is due to triangle inequality, and Cauchy-Schwartz inequality and $(b)$ is due to triangle inequality and Cauchy-Schwarz inequality. Note that the equality in $(a)$ holds when $\Delta\mathbf{h}_{0,0}=-\mathbf{w}_0
\left(\hat{\mathbf{H}}_0\right)\varepsilon_{2}e^{-j\angle{\left(\left(
\mathbf{v}^H\left(\hat{\mathbf{G}}_{0,0}+\Delta{
\mathbf{G}}_{0,0}\right)+\hat{\mathbf{h}}^H_{0,0}
\right)\mathbf{w}_0\left(\hat{\mathbf{H}}_0\right)\right)}}$, and the equality in $(b)$ holds when $\Delta{G}_{0,0}= -\frac{\mathbf{v}\mathbf{w}_0^H\left(\hat{\mathbf{H}}_0\right)\varepsilon_1}{\sqrt{M_rN_r}}
e^{-j\angle{\left(\left(\mathbf{v}^H\hat{\mathbf{G}}_{0,0}
+\hat{\mathbf{h}}^H_{0,0}\right)\mathbf{w}_0\left(\hat{\mathbf{H}}_0\right)\right)}}$. By \eqref{ineq:b} and the fact that $C\left(\mathbf{v},\mathbf{w}_0
\left(\hat{\mathbf{H}}_0\right),\hat{\mathbf{H}}_0
,\Delta\mathbf{H}_0\right)$ increases with $\left\lvert\left(\mathbf{v}^H\mathbf{G}_{0,0}
+\mathbf{h}_{0,0}^H\right)\mathbf{w}_0
\left(\hat{\mathbf{H}}_0\right)\right\rvert$, we have
$
r\left(\hat{\mathbf{H}}_0\right) = \log_2\left(1+
\frac{P_0g_0^{(GP)}\left(\mathbf{v},\mathbf{w}_0
\left(\hat{\mathbf{H}}_0\right),\hat{\mathbf{H}}_0\right)}
{\sum\limits_{k \in \mathcal{K}\backslash \{0\}}P_kg_k\left(\mathbf{v}\right)+\sigma^2}\right)
$.
Thus, Problem~\ref{prob:eq_sta_determin} can be equivalently converted to:
\begin{equation}\label{prob:ins_GP1}
\begin{split}
\mathop{\max}_{\mathbf{v}}\quad &
\frac{\mathbb{E}\left[
 \gamma_{0}^{(GP1)*}\left(\mathbf{v}, \hat{\mathbf{H}}_0\right)
\right]}{\sum\limits_{k \in \mathcal{K}\backslash \{0\}}P_k{g}_{k}\left(\mathbf{v}\right)+\sigma^2} \\
s.t. \quad
& \eqref{eq:phi},
\end{split}
\end{equation}
where
\begin{equation}\label{prob:ins_w_GP1}
\begin{split}
\gamma_{0}^{(GP1)*}\left(\mathbf{v}, \hat{\mathbf{H}}_0\right)\triangleq \mathop{\max}_{\mathbf{w}_0\left(\hat{\mathbf{H}}_0\right)} \quad & {g}_0^{(GP1)}\left(\mathbf{v},\mathbf{w}_0\left(\hat{\mathbf{H}}_0\right),\hat{\mathbf{H}}_0\right) \\
s.t. \quad & \eqref{eq:w}.
\end{split}
\end{equation}
Next, by Cauchy-Schwartz inequality, we can show that $\tilde{\mathbf{w}}_{0}^{(GP1)*}\left(\mathbf{v},
\hat{\mathbf{H}}_0\right)$ in \eqref{eq:GP1tildew} is an optimal solution of the problem in \eqref{prob:ins_w_GP1} and $\gamma_{0}^{(GP1)*}\left(\mathbf{v}, \hat{\mathbf{H}}_0\right)={g}_0^{(GP1)}\left(\mathbf{v},\tilde{\mathbf{w}}_{0}^{(GP1)*}\left(\mathbf{v},
\hat{\mathbf{H}}_0\right),\hat{\mathbf{H}}_0\right)$. Substituting $\gamma_{0}^{(GP1)*}\left(\mathbf{v}, \hat{\mathbf{H}}_0\right)$ into the objective function of the problem in \eqref{prob:ins_GP1}, we can obtain Problem~\ref{prob:2}.
Therefore, we complete the proof of Lemma~\ref{theorem:GPeq}.
\section*{Appendix E: Proof of Theorem~\ref{theorem:2}}
As $\mathbf{v}^{(GP1)*}$ is an optimal solution of  Problem~\ref{prob:2}, $\left(\mathbf{v}^{(GP1)*},\mathbf{w}_0^{(GP1)*},
r^{(GP1)*}\right)$ is a feasible solution of  Problem~\ref{prob:eq_sta_determin}. Thus, based on Lemma~\ref{lem:lemma1}, $\left(\mathbf{v}^{(GP1)*},\mathbf{w}_0^{(GP1)*},
r^{(GP1)*}\right)$ is an optimal solution of Problem~\ref{prob:eq_sta_determin}. Therefore, we complete the proof of Theorem~\ref{theorem:2}.
\blue{\section*{Appendix F: Approximation Based on Bernstein-type Inequality}
We have:
\begin{align}
&\text{Pr}\left[C\left(\mathbf{v},
\mathbf{w}_0\left(
\hat{\mathbf{H}}_0\right),\hat{\mathbf{H}}_0,\Delta{\mathbf{H}}_0\right)
\geq r\left(\hat{\mathbf{H}}_0\right) \right] \notag \\
\eqla &
\text{Pr}\left[\left\lvert\left(
\hat{\mathbf{h}}_{0,0}^H+\Delta\mathbf{h}_{0,0}^H+
\mathbf{v}^H\left(\hat{\mathbf{G}}_{0,0}+
\Delta\mathbf{G}_{0,0}\right)\right)\mathbf{w}_0
(\hat{\mathbf{H}}_0)\right\rvert^2
-q\geq0\right] \notag \\
= & \text{Pr}\left[\left\lvert\left(
\Delta\mathbf{h}_{0,0}^H+
\mathbf{v}^H\Delta\mathbf{G}_{0,0}\right)\mathbf{w}_0
(\hat{\mathbf{H}}_0)\right\rvert^2
+ 2Re\left\{\left(
\Delta\mathbf{h}_{0,0}^H+
\mathbf{v}^H\Delta\mathbf{G}_{0,0}\right)\mathbf{w}_0(\hat{\mathbf{H}}_0)
\mathbf{w}_0^H(\hat{\mathbf{H}}_0)\left(
\hat{\mathbf{h}}_{0,0}^H+
\mathbf{v}^H\hat{\mathbf{G}}_{0,0}\right)^H\right\}  \right. \notag \\  &  \left.+\left\lvert\left(
\hat{\mathbf{h}}_{0,0}^H+\mathbf{v}^H\hat{\mathbf{G}}_{0,0}
\right)\mathbf{w}_0\right\rvert^2-q \geq 0 \right] \notag \\ \eqlb & \text{Pr}\left[\mathbf{z}^H\mathbf{U}\mathbf{z}
+2Re\{\mathbf{u}^H\mathbf{z}\}+u\geq 0 \right],\label{eq:rewritten}
\end{align}
where $(a)$ is due to the expression of $C\left(\mathbf{v},
\mathbf{w}_0\left(
\hat{\mathbf{H}}_0\right),\hat{\mathbf{H}}_0,\Delta{\mathbf{H}}_0\right)$,
$(b)$ is due to $\Delta\mathbf{h}_{0,0} \in \mathbb{C}^{ M_0N_0 \times 1} \dis \delta_2\mathbf{z}_1$ with $\mathbf{z}_1 \sim \mathcal{CN}(0,\mathbf{I}_{M_0N_0})$, $\text{vec}(\Delta\mathbf{G}_{0,0}) \in \mathbb{C}^{ M_0N_0M_rN_r \times 1} \dis \delta_1\mathbf{z}_2$ with $\mathbf{z}_2 \sim \mathcal{CN}(0,\mathbf{I}_{M_0N_0M_rN_r})$, and
$\mathbf{z} \triangleq [\mathbf{z}_1^T,\mathbf{z}_2^T]^T\in\mathbb{C}^{M_0N_0(M_rN_r+1)} $.
Here, we let
\begin{align}\mathbf{U}  \triangleq & \begin{bmatrix}
\delta_2^2\mathbf{w}_0(\hat{\mathbf{H}}_0)
\mathbf{w}^H_0(\hat{\mathbf{H}}_0) & \delta_1\delta_2(\mathbf{w}_0(\hat{\mathbf{H}}_0)
\mathbf{w}^H_0(\hat{\mathbf{H}}_0) \otimes \mathbf{v}^T) \nonumber \\
\delta_1\delta_2(\mathbf{w}_0(\hat{\mathbf{H}}_0)
\mathbf{w}^H_0(\hat{\mathbf{H}}_0) \otimes \mathbf{v}^*) & \delta_1^2(\mathbf{w}_0(\hat{\mathbf{H}}_0)
\mathbf{w}^H_0(\hat{\mathbf{H}}_0) \otimes \text{diag}(\mathbf{v}^T))\\
\end{bmatrix} \in\mathbb{C}^{M_0N_0(M_rN_r+1)\times M_0N_0(M_rN_r+1)},\nonumber \\
\mathbf{u} \triangleq &
\begin{bmatrix}
\delta_2\mathbf{w}_0(\hat{\mathbf{H}}_0)
\mathbf{w}^H_0(\hat{\mathbf{H}}_0)(\hat{\mathbf{h}}_{0,0}
+\hat{\mathbf{G}}_{0,0}^H\mathbf{v})\\
\delta_1\text{vec}^*(\mathbf{v}(\hat{\mathbf{h}}_{0,0}
+\hat{\mathbf{G}}_{0,0}^H\mathbf{v})\mathbf{w}_0(\hat{\mathbf{H}}_0)
\mathbf{w}^H_0(\hat{\mathbf{H}}_0)(\hat{\mathbf{h}}_{0,0}) \\
\end{bmatrix} \in\mathbb{C}^{M_0N_0(M_rN_r+1)},\nonumber \\
u  \triangleq & \left\lvert\left(
\hat{\mathbf{h}}_{0,0}^H+\mathbf{v}^H\hat{\mathbf{G}}_{0,0}
\right)\mathbf{w}_0\right\rvert^2-\frac{1}{P_0}\left(2^{r(\hat{\mathbf{H}}_0)}-1\right)
\left(\sum\limits_{k \in \mathcal{K}\backslash\{0\}}P_kg_k\left(\mathbf{v}\right)
+\sigma^2\right). \label{eq:xxxxxxs}
\end{align}
By \eqref{eq:rewritten} and the Bernstein-type inequality [Method II, R1], the constraint in (11) of the previous draft can be approximated as:
\begin{align}
  &\text{tr}\{\mathbf{U}\} - \sqrt{2\ln\frac{1}{1-\rho}}x+
  \ln(1-\rho)y +u \geq 0, \label{ieq:final} \\
  & \sqrt{\left\lVert\mathbf{U}\right\rVert_F^2
  +2\left\lVert \mathbf{u}\right\rVert^2}\leq x, \label{ieq:x} \\
  & y\mathbf{I}_{M_0N_0(M_rN_r+1)}+\mathbf{U} \succeq \mathbf{0}_{M_0N_0(M_rN_r+1)},\quad  y\geq 0. \label{ieq:y}
\end{align}
Substituting \eqref{eq:xxxxxxs} into \eqref{ieq:final}, we have:
\begin{align}\label{eq:r_function}
  r(\hat{\mathbf{H}}_0) \leq \log_2\left(1+\frac{P_0\left(\left\lvert\left(
\hat{\mathbf{h}}_{0,0}^H+\mathbf{v}^H\hat{\mathbf{G}}_{0,0}
\right)\mathbf{w}_0\right\rvert^2+\text{tr}\{\mathbf{U}\} - \sqrt{2\ln\frac{1}{1-\rho}}x+
  \ln(1-\rho)y \right)}{\sum\limits_{k \in \mathcal{K}\backslash\{0\}}P_kg_k\left(\mathbf{v}\right)
+\sigma^2}\right).
\end{align}
Note that the upper bound on $r(\hat{\mathbf{H}}_0)$ in \eqref{eq:r_function} decreases with $x$ and $y$, and the objective function of the maximization problem in Problem 8 increases with $r(\hat{\mathbf{H}}_0)$. Thus, the constraints in \eqref{ieq:x} and \eqref{ieq:y} are active at the optimal solution of Problem 8. Substituting
$$ x = \sqrt{\left(\delta_1^2M_rN_r+\delta_{2}^2\right)^2+2\left(\delta_1^2M_rN_r+\delta_{2}^2\right)\left\lvert
\left(\hat{\mathbf{h}}_{0,0}^H+\mathbf{v}^H\hat{\mathbf{G}}_{0,0}\right)\mathbf{w}_0
\left(\hat{\mathbf{H}}_0\right)\right\rvert^2},\quad y = 0$$
into \eqref{eq:r_function}, we have \eqref{eq:constraint1}.}
\section*{Appendix G: Proof of Lemma~\ref{lem:equivalencesta}}
It is obvious that any feasible solution of Problem~\ref{prob:appBernstein} is also feasible for Problem~\ref{prob:appBernstein}. It remains to show that $\left(\mathbf{v}^{(GP2)},\mathbf{w}_0^{(GP2)},r^{(GP2)}\right)$ is feasible for Problem~\ref{prob:eq_sta_ergodic}. From Lemma~\ref{theorem:GPeq}, we know that Problem~\ref{prob:eq_sta_determin_2} is equivalent to Problem~\ref{prob:eq_sta_determin}. Thus, $\left(\mathbf{v}^{(GP2)},\mathbf{w}_0^{(GP2)},\hat{r}^{(GP2)}\right)$ is a feasible solution for Problem~\ref{prob:eq_sta_determin}. By Lemma~\ref{lem:lemma1}, we know that $\left(\mathbf{v}^{(GP2)},\mathbf{w}_0^{(GP2)},\hat{r}^{(GP2)}\right)$ is also a feasible solution for Problem~\ref{prob:eq_sta_ergodic}. Therefore, we complete the proof of Lemma~\ref{lem:equivalencesta}.
\section*{Appendix H: Proof of Lemma~\ref{theorem:GPeqxx}}
By noting that \eqref{eq:constraint1} is equivalent to $r\left(\hat{\mathbf{H}}_0\right)\leq\log_2\left(1+\frac{g_0^{(GP2)}\left(\mathbf{v},\mathbf{w}_0
\left(\hat{\mathbf{H}}_0\right),
\hat{\mathbf{H}}_0\right)}{\sum\limits_{k \in \mathcal{K}\backslash \{0\}}P_k{g}_{k}\left(\mathbf{v}\right)+\sigma^2}\right)$,
Problem~\ref{prob:appBernstein} is equivalent to:
\begin{align*}
\mathop{\max}_{\mathbf{v}}\quad &
\log_2\!\left(1+\frac{P_0g_0^{(GP2)}\left(\mathbf{v},\mathbf{w}_0\left(\hat{\mathbf{H}}_0\right)
,\hat{\mathbf{H}}_0\right)}{\sum\limits_{k \in \mathcal{K}\backslash\{0\}}P_kg_k\left(\mathbf{v}\right)
+\sigma^2}\right), \\
s.t. \quad & \eqref{eq:phi},\quad \eqref{eq:w},
\end{align*}
which can be equivalently converted to:
\begin{equation}\label{prob:ins_GP2}
\begin{split}
\mathop{\max}_{\mathbf{v}}\quad &
\frac{\mathbb{E}\left[
 \gamma_{0}^{(GP2)*}\left(\mathbf{v}, \hat{\mathbf{H}}_0\right)
\right]}{\sum\limits_{k \in \mathcal{K}\backslash \{0\}}P_k{g}_{k}\left(\mathbf{v}\right)+\sigma^2} \\
s.t. \quad
& \eqref{eq:phi},
\end{split}
\end{equation}
where
\begin{equation}\label{prob:ins_w_GP2}
\begin{split}
\gamma_{0}^{(GP2)*}\left(\mathbf{v}, \hat{\mathbf{H}}_0\right)\triangleq \mathop{\max}_{\mathbf{w}_0\left(\hat{\mathbf{H}}_0\right)} \quad & {g}_0^{(GP2)}\left(\mathbf{v},\mathbf{w}_0\left(\hat{\mathbf{H}}_0\right),\hat{\mathbf{H}}_0\right) \\
s.t. \quad & \eqref{eq:w}.
\end{split}
\end{equation}
Next, by Cauchy-Schwartz inequality, we can show that $\tilde{\mathbf{w}}_{0}^{(GP2)*}\left(\mathbf{v},
\hat{\mathbf{H}}_0\right)$ in \eqref{eq:GP2tildew} is an optimal solution of the problem in \eqref{prob:ins_w_GP2} and $\gamma_{0}^{(GP2)*}\left(\mathbf{v}, \hat{\mathbf{H}}_0\right)={g}_0^{(GP2)}\left(\mathbf{v},\tilde{\mathbf{w}}_{0}^{(GP2)*}\left(\mathbf{v},
\hat{\mathbf{H}}_0\right),\hat{\mathbf{H}}_0\right)$. Substituting $\gamma_{0}^{(GP2)*}\left(\mathbf{v}, \hat{\mathbf{H}}_0\right)$ into the objective function of the problem in \eqref{prob:ins_GP2}, we can obtain Problem~\ref{prob:3}. Therefore, we complete the proof of Lemma~\ref{theorem:GPeqxx}.
\section*{Appendix I: Proof of Theorem~\ref{theorem:3}}
As $\mathbf{v}^{(GP2)*}$ is an optimal solution of  Problem~\ref{prob:3}, $\left(\mathbf{v}^{(GP2)*},\mathbf{w}_0^{(GP2)*},
r^{(GP2)*}\right)$ is a feasible solution of  Problem~\ref{prob:appBernstein}. Thus, based on Lemma~\ref{lem:equivalencesta}, $\left(\mathbf{v}^{(GP2)*},\mathbf{w}_0^{(GP2)*},
r^{(GP2)*}\right)$ is an optimal solution of Problem~\ref{prob:appBernstein}. Therefore, we complete the proof of Theorem~\ref{theorem:3}.
\section*{Appendix J: Proof of Theorem~\ref{theoren:phi}}
Problem~\ref{prob:ins_surrogate} can be equivalently decoupled to $M_rN_r$ convex problems with the $n$-th problem given by  (ignoring constant terms):
\begin{align}
\begin{split}
\mathop{\min}_{v_n}\quad &
\tau\left\lvert v_n-v_n^{(Q,t-1)} \right\rvert^2-2Re
\left\{\sum\limits_{n=1}^{M_rN_r}c_{1,n}^{(Q,t)}v_n\right\}\\
s.t.\quad & \eqref{eq:cvxphi}.
\end{split}\label{prob:subv}
\end{align}
Obviously, the convex problem in \eqref{prob:subv} is strictly feasible, and hence satisfies the Slater's condition.
Thus, its KKT conditions:
\begin{align}
\begin{split}
&\lambda \geq 0, \quad \left\lvert v_n \right\rvert^2 \geq 1, \quad
 \lambda\left(\left\lvert v_n \right\rvert^2-1\right)=0, \quad v_n(\tau+\lambda)
-\left(\tau v_n^{(t-1)}+f_n^t\right)=0
\end{split}\label{prob:subvkkt}
\end{align}
provide necessary and sufficient conditions for optimality. It is clear that $v_n=\frac{\tau v_n^{(t-1)}+f_n^t}
{\tau+\lambda}$ and $\lambda=\left\lvert\tau v_n^{(t-1)}+f_n^t\right\rvert-\tau$ satisfy \eqref{prob:subvkkt}, implying that $\bar{v}_n^{(Q,t)}$ given in \eqref{eq:solutionv} is an optimal solution of Problem~\ref{prob:ins_surrogate}. Therefore, we complete the proof of Theorem~\ref{theoren:phi}.

\vspace{12pt}


\begin{thebibliography}{00}
\bibitem{QingqingWu1}
Q.~{Wu} and R.~{Zhang}, ``{Towards} {Smart} and {Reconfigurable} {Environment}:
  {Intelligent} {Reflecting} {Surface} {Aided} {Wireless} {Network},''
  \emph{{IEEE} Commun. Mag.}, vol.~58, no.~1, pp. 106--112, Nov., 2020.

\bibitem{EBasar}
E.~{Basar}, M.~{Di Renzo}, J.~{De Rosny}, M.~{Debbah}, M.~{Alouini}, and
  R.~{Zhang}, ``{Wireless} {Communications} {Through} {Reconfigurable}
  {Intelligent} {Surfaces},'' \emph{IEEE Access}, vol.~7, pp.
  116\,753--116\,773, Aug., 2019.
\blue{\bibitem{R7}
M.~{Di}~{Renzo} et al., “{Smart} radio environments empowered by reconfigurable intelligent
surfaces: {How} it works, state of research, and the road ahead”, \emph{{IEEE} J. Sel. Areas Commun.}, vol.~38, no.~11, pp.~2450-2525, Nov.~2021.}


\bibitem{SJin1}
W.~{Tang}, M.~Z. {Chen}, X.~{Chen}, J.~Y. {Dai}, Y.~{Han}, M.~{Di Renzo},
  Y.~{Zeng}, S.~{Jin}, Q.~{Cheng}, and T.~J. {Cui}, ``{Wireless}
  {Communications} with {Reconfigurable} {Intelligent} {Surface}: {Path} {Loss}
  {Modeling} and {Experimental} {Measurement},'' \emph{to appear in IEEE Trans.
  Wireless Commun.}, Sept., 2020.

\bibitem{HGuo1}
H.~{Guo}, Y.~{Liang}, J.~{Chen}, and E.~G. {Larsson}, ``{Weighted} {Sum}-{Rate}
  {Maximization} for {Reconfigurable} {Intelligent} {Surface} {Aided}
  {Wireless} {Networks},'' \emph{{IEEE} Trans. Wireless Commun.}, vol.~19,
  no.~5, pp. 3064--3076, Feb., 2020.

\bibitem{HShen1}
H.~{Shen}, W.~{Xu}, S.~{Gong}, Z.~{He}, and C.~{Zhao}, ``{Secrecy} {Rate}
  {Maximization} for {Intelligent} {Reflecting} {Surface} {Assisted}
  {Multi}-{Antenna} {Communications},'' \emph{{IEEE} Commun. Lett.}, vol.~23,
  no.~9, pp. 1488--1492, Jun., 2019.

\bibitem{XYu}
X.~{Yu}, D.~{Xu}, and R.~{Schober}, ``{Enabling} {Secure} {Wireless}
  {Communications} via {Intelligent} {Reflecting} {Surfaces},'' in \emph{Proc.
  of 2019 IEEE GLOBECOM}, Feb., 2019, pp. 1--6.

\bibitem{CHuang}
C.~{Huang}, A.~{Zappone}, G.~C. {Alexandropoulos}, M.~{Debbah}, and C.~{Yuen},
  ``{Reconfigurable} {Intelligent} {Surfaces} for {Energy} {Efficiency} in
  {Wireless} {Communication},'' \emph{{IEEE} Trans. Wireless Commun.}, vol.~18,
  no.~8, pp. 4157--4170, Jun., 2019.

\bibitem{8855810}
X.~{Yu}, D.~{Xu}, and R.~{Schober}, ``{MISO} {Wireless} {Communication}
  {Systems} via {Intelligent} {Reflecting} {Surfaces} : ({Invited} {Paper}),''
  in \emph{Proc. of 2019 IEEE/CIC ICCC}, Oct., 2019, pp. 735--740.

\bibitem{QingqingWu2}
Q.~{Wu} and R.~{Zhang}, ``{Intelligent} {Reflecting} {Surface} {Enhanced}
  {Wireless} {Network} via {Joint} {Active} and {Passive} {Beamforming},''
  \emph{{IEEE} Trans. Wireless Commun.}, vol.~18, no.~11, pp. 5394--5409, Aug.,
  2019.

\bibitem{SJin2}
Y.~{Han}, W.~{Tang}, S.~{Jin}, C.~{Wen}, and X.~{Ma}, ``{Large} {Intelligent}
  {Surface}-{Assisted} {Wireless} {Communication} {Exploiting} {Statistical}
  {CSI},'' \emph{{IEEE} Trans. Veh. Technol.}, vol.~68, no.~8, pp. 8238--8242,
  Jun., 2019.
\blue{\bibitem{hu2020statistical}
X.~{Hu}, J.~{Wang}, and C.~{Zhong}, ``{Statistical} {CSI} based {Design} for
  {Intelligent} {Reflecting} {Surface} {Assisted} {MISO} {Systems}.''
  \emph{Sci. China Inf. Sci.}, vol.~63, no.~222303, Aug. 2020.}

\bibitem{YuhangJia}
Y.~{Jia}, C.~{Ye}, and Y.~{Cui}, ``{Analysis} and {Optimization} of an
  {Intelligent} {Reflecting} {Surface}-{Assisted} {System} {With}
  {Interference},'' \emph{{IEEE} Trans. Wireless Commun.}, vol.~19, no.~12, pp.
  8068--8082, 2020.

\bibitem{zhao2019intelligent}
M.~M. {Zhao}, Q.~{Wu}, M.~J. {Zhao}, and R.~{Zhang}, ``{Intelligent}
  {Reflecting} {Surface} {Enhanced} {Wireless} {Network}: {Two}-{Timescale}
  {Beamforming} {Optimization},'' \emph{to appear in IEEE Trans. Wireless
  Commun.}, Sept., 2020.

\bibitem{CGuo}
C.~{Guo}, Y.~{Cui}, F.~{Yang}, and L.~{Ding}, ``{Outage} {Probability}
  {Analysis} and {Minimization} in {Intelligent} {Reflecting}
  {Surface}-{Assisted} {MISO} {Systems},'' \emph{{IEEE} Commun. Lett.},
  vol.~24, no.~7, pp. 1563--1567, Feb., 2020.

\bibitem{QTao}
Q.~{Tao}, J.~{Wang}, and C.~{Zhong}, ``{Performance} {Analysis} {of}
  {Intelligent} {Reflecting} {Surface} {Aided} {Communication} {Systems},''
  \emph{{IEEE} Commun. Lett.}, vol.~24, no.~11, pp. 2464--2468, Jul., 2020.
\blue{\bibitem{R8}
F.~{Alavi}, K.~{Cumanan}, Z.~{Ding}, and A. G. Burr, “{Robust} {Beamforming} {Techniques}
for {Non}-{Orthogonal} {Multiple} {Access} {Systems} with {Bounded} {Channel} {Uncertainties},” \emph{{IEEE}
Commun. Lett.}, vol.~21, no.~9, pp. 2033--2036, Sept. 2017.
\bibitem{R9}
 F.~{Alavi}, K.~{Cumanan}, M.~{Fozooni}, Z.~{Ding}, S.~{Lambotharan} and O.~A.~{Dobre}, “{Robust} {Energy}-{Efficient} {Design} for {MISO} {Non}-{Orthogonal} {Multiple} {Access} {Systems},” \emph{{IEEE}
Trans. Commun.}, vol.~67, no.~11, pp. 7937--7949, Nov. 2019.}

\bibitem{xu2020resource}
D.~{Xu}, X.~{Yu}, Y.~{Sun}, D.~W.~K. {Ng}, and R.~{Schober}, ``{Resource}
  {Allocation} for {IRS}-{Assisted} {Full}-{Duplex} {Cognitive} {Radio}
  {Systems},'' \emph{{IEEE} Trans. Commun.}, vol.~68, no.~12, pp. 7376--7394,
  2020.

\bibitem{yu2020robust}
X.~{Yu}, D.~{Xu}, Y.~{Sun}, D.~W.~K. {Ng}, and R.~{Schober}, ``{Robust} and
  {Secure} {Wireless} {Communications} via {Intelligent} {Reflecting}
  {Surfaces},'' vol.~38, no.~11, pp. 2637--2652, Jul., 2020.
\blue{\bibitem{hong2020robust}
S.~{Hong}, C.~{Pan}, H.~{Ren}, K.~{Wang}, K.~K. {Chai}, and A.~{Nallanathan},
  ``{Robust} {Transmission} {Design} for {Intelligent} {Reflecting} {Surface}
  {Aided} {Secure} {Communication} {Systems} with {Imperfect} {Cascaded}
  {CSI},'' \emph{{IEEE} Trans. Wireless Commun.}, vol. 20, no. 4, pp. 2487-2501, Apr. 2021.}

\bibitem{ZhouGui}
G.~{Zhou}, C.~{Pan}, H.~{Ren}, K.~{Wang}, and A.~{Nallanathan}, ``{A}
  {Framework} of {Robust} {Transmission} {Design} for {IRS}-{Aided} {MISO}
  {Communications} {With} {Imperfect} {Cascaded} {Channels},'' \emph{{IEEE}
  Trans. Signal Process.}, vol.~68, pp. 5092--5106, Aug., 2020.

\bibitem{JWang}
J.~{Wang}, Y.~{Liang}, S.~{Han}, and Y.~{Pei}, ``{Robust} {Beamforming} and
  {Phase} {Shift} {Design} for {IRS}-{Enhanced} {Multi}-{User} {MISO}
  {Downlink} {Communication},'' in \emph{Proc. of IEEE ICC 2020}, Jul., 2020.

\bibitem{TALe}
T.~A. {Le}, T.~{Van Chien}, and M.~{Di Renzo}, ``{Robust}
  {Probabilistic}-{Constrained} {Optimization} for {IRS}-{Aided} {MISO}
  {Communication} {Systems},'' \emph{to appear in IEEE Wireless Commun. Lett.},
  Aug., 2020.
\blue{\bibitem{deng2020robust}
Y.~{Deng}, Y.~{Zou}, S.~{Gong}, B.~{Lyu}, D.~T. {Hoang}, and D.~{Niyato},
  ``{Robust} {Beamforming} for {IRS}-assisted {Wireless} {Communications} under
  {Channel} {Uncertainty},'' in \emph{Proc. of IEEE WCNC 2021}, pp. 1-6, Apr., 2021.}

\bibitem{Czhong1}
J.~{Zhang}, Y.~{Zhang}, C.~{Zhong}, and Z.~{Zhang}, ``{Robust} {Design} for
  {Intelligent} {Reflecting} {Surfaces} {Assisted} {MISO} {Systems},''
  \emph{{IEEE} Commun. Lett.}, vol.~24, no.~10, pp. 2353--2357, Jun., 2020.

\bibitem{YongjunXu}
Y.~Xu, Z.~Qin, Y.~Zhao, G.~Li, G.~Gui, and H.~Sari, ``{{Resource} {Allocation}
  for {Intelligent} {Reflecting} {Surface} {Enabled} {Heterogeneous}
  {Networks}},'' Feb., 2020. [Online]. Available:
  \url{https://www.techrxiv.org/articles/preprint/Resource_Allocation_for_Intelligent_Reflecting_Surface_Enabled_Heterogeneous_Networks/11697666}

\bibitem{CPan1}
C.~{Pan}, H.~{Ren}, K.~{Wang}, W.~{Xu}, M.~{Elkashlan}, A.~{Nallanathan}, and
  L.~{Hanzo}, ``{Multicell} {MIMO} {Communications} {Relying} on {Intelligent}
  {Reflecting} {Surfaces},'' \emph{{IEEE} Trans. Wireless Commun.}, vol.~19,
  no.~8, pp. 5218--5233, May, 2020.

\bibitem{WangQun}
Q.~Wang, F.~Zhou, R.~Q. Hu, and Y.~Qian, ``{Energy}-{Efficient} {Beamforming}
  and {Cooperative} {Jamming} in {IRS}-{Assisted} {MISO} {Networks},'' in
  \emph{Proc. of IEEE ICC 2020}, Aug., 2020.
\blue{\bibitem{QingqingWu3}
M.~{Hua}, Q.~{Wu}, D.~W.~K. {Ng}, J.~{Zhao}, and L.~{Yang}, ``{Intelligent}
  {Reflecting} {Surface}-{Aided} {Joint} {Processing} {Coordinated}
  {Multipoint} {Transmission}.'' \emph{{IEEE} Trans. Commun.}, vol.~69,
  no.~3, pp. 1650-1665, Mar., 2021.}
\blue{\bibitem{YuanweiLiu}
W.~{Ni}, X.~{Liu}, Y.~{Liu}, H.~{Tian}, and Y.~{Chen}, ``{Resource}
  {Allocation} for {Multi}-{Cell} {IRS}-{Aided} {NOMA} {Networks}.''
  \emph{to appear in IEEE Wireless Commun.}, 2021.}
\blue{\bibitem{jia2020reconfigurable}
S.~{Jia}, X.~{Yuan}, and Y.-C. {Liang}, ``{Reconfigurable} {Intelligent}
  {Surfaces} for {Energy} {Efficiency} in {D2D} {Communication} {Network}.''
  \emph{{IEEE} Trans. Wireless Commun. Lett.}, vol.~10,
  no.~3, pp. 683--687, Mar., 2021.}
\blue{\bibitem{DMishra}
D.~{Mishra} and H.~{Johansson}, ``{Channel} estimation and low-complexity beamforming design for passive intelligent surface assisted {MISO} wireless energy transfer,” \emph{Proc. of {{IEEE} ICASSP}}, Apr., 2019, pp. 4659–4663.
\bibitem{Que}
 Q.~{Nadeem}, H.~{Alwazani}, A.~{Kammoun}, A.~{Chaaban}, M.~{Debbah}, and M.~{Alouini},
“{Intelligent} reflecting surface{-}assisted multi-user MISO communication: Channel estimation and
beamforming design,” \emph{{IEEE} Open J. Commun. Society}, vol. 1, pp. 661– 680,
May, 2020.}

\bibitem{IEEEexample:van2004optimum}
H.~L. Van~Trees, \emph{{Optimum} array processing: {Part} {IV} of detection,
  estimation, and modulation theory}.\hskip 1em plus 0.5em minus 0.4em\relax
  John Wiley \& Sons, 2004.

\bibitem{A_Liu}
A.~{Liu}, V.~K.~N. {Lau}, and M.~{Zhao}, ``{Online} {Successive} {Convex}
  {Approximation} for {Two}-{Stage} {Stochastic} {Nonconvex} {Optimization},''
  \emph{{IEEE} Trans. Signal Process.}, vol.~66, no.~22, pp. 5941--5955, Sept.,
  2018.

\bibitem{cover1991elements}
T.~M. {Cover} and J.~A. {Thomas}, \emph{Elements of information theory}, 1991.

\bibitem{KWang}
K.~{Wang}, A.~M. {So}, T.~{Chang}, W.~{Ma}, and C.~{Chi}, ``{Outage}
  {Constrained} {Robust} {Transmit} {Optimization} for {Multiuser} {MISO}
  {Downlinks}: {Tractable} {Approximations} by {Conic} {Optimization},''
  \emph{{IEEE} Trans. Signal Process.}, vol.~62, no.~21, pp. 5690--5705, Sept.,
  2014.

\bibitem{7412752}
Y.~{Yang}, G.~{Scutari}, D.~P. {Palomar}, and M.~{Pesavento}, ``{A} {Parallel}
  {Decomposition} {Method} for {Nonconvex} {Stochastic} {Multi}-{Agent}
  {Optimization} {Problems},'' \emph{{IEEE} Trans. Signal Process.}, vol.~64,
  no.~11, pp. 2949--2964, Feb., 2016.
\blue{\bibitem{Ergodic_baseline}
A.~{Gründinger}, D.~{Leiner}, M.~{Joham}, C.~{Hellings} and W.~{Utschick}, ``{Ergodic} robust rate
balancing for rank-one vector broadcast channels via sequential approximations,” \emph{Proc. of {{IEEE} ICASSP}}, May, 2013, pp. 4744-4748.
\bibitem{nonErgodic_baseline}
K.~{Wang}, A.~M.~{So}, T.~{Chang}, W.~{Ma}, and C.~{Chi},
``{Outage} {Constrained} {Robust} {Transmit} {Optimization} for {Multiuser} {MISO} {Downlinks}: {Tractable} {Approximations} by {Conic} {Optimization},” \emph{{IEEE} Trans. Signal Process.}, vol.~62, no.~21, pp.~5690–5705, Sept., 2014.
}
\end{thebibliography}
\end{document}